\documentclass[12pt]{report}
\usepackage[utf8]{inputenc}
\usepackage{amsmath}
\usepackage{setspace}
\usepackage[explicit]{titlesec}
\usepackage{type1cm}
\usepackage{xcolor}
\usepackage{latexsym}
\usepackage{amssymb}
\usepackage[Sonny]{fncychap}
\usepackage{mathtools}
\usepackage{geometry}
\usepackage{braket}
\usepackage{graphicx}
\usepackage{dsfont}
\usepackage{float}
\usepackage{amsfonts}
\usepackage{times}
\usepackage{suffix}
\usepackage[toc,page]{appendix}
\usepackage{MnSymbol}
\usepackage{tikz}
\usepackage{lipsum}  
\usepackage[square,sort,comma,numbers,sort&compress]{natbib}
\usepackage[colorlinks=true,linktoc=page,linkcolor=blue,citecolor=blue,urlcolor=blue]{hyperref}
\usepackage{braket}

\usepackage{multirow}
\usepackage{xcolor}
\usepackage{slashed}
\allowdisplaybreaks

\begin{document}

\thispagestyle{empty}
%
%
%
%
%

\begin{titlepage}
	\begin{center}
		
		\huge{\textbf{Probing Anomalous Higgs Boson Couplings at Colliders}}
		
		\vspace{0.5cm}
		\Large 
		by\\
		\vspace{0.3cm}
		\LARGE
		{\textbf {Biswajit Das}}\\
		\vspace{0.2cm}
		\large
		PHYS07201404011\\
		\vspace{0.4cm}
		\Large
		\textbf{Institute of Physics, Bhubaneswar}\\ 
		\vspace{1cm} 
		\normalsize       
		\textit{A thesis submitted to the\\
			Board of Studies in Physical Sciences\\
			In partial fulfillment of requirements \\ for the Degree of}\\
		\vspace{1cm}
		\Large
		\textbf{DOCTOR OF PHILOSOPHY}\\
		\textit{of}\\
		\Large
		\textbf{HOMI BHABHA NATIONAL INSTITUTE}\\
		\vspace{6cm}
		\includegraphics[width=0.2\textwidth]{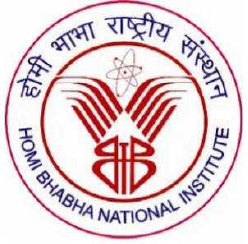}\\
		\vspace{0.5cm}
		May, 2023
	\end{center}
\end{titlepage}
\pagenumbering{gobble}
\thispagestyle{empty}

\newpage

\doublespacing

\phantomsection
\tableofcontents\clearpage
\newpage
\pagenumbering{roman}
\addcontentsline{toc}{chapter}{Summary}
\chapter*{Summary}
\label{chap_summery}
  In the present day, the standard model is known for its illustrious success in describing the fundamental building blocks of nature. Despite its great success, the standard model has a few limitations in describing nature fully. 
  This motivates physicists to consider beyond-the-standard model scenarios, which may address the limitations of the standard model. Despite the elegance of some of the beyond-the-standard model theories, no significant departure from the standard model predictions has been found in colliders at current energy scales.
  After the discovery of the Higgs boson at the Large Hadron Collider, its properties have been studied extensively. They are found to be consistent with the standard model.
  However, some of the Higgs boson couplings are still not well determined. Self-couplings of the Higgs boson and the couplings with  some of the standard model particles still do not have stringent bounds.
  
  In this thesis, our main focus is on the $HHH$ and $VVHH$ couplings which are loosely bound. The $HHH$ coupling will determine the shape the Higgs potential and $VVHH$ couplings will tell us how the gauge bosons are coupled to the Higgs fields. We consider a few processes of Higgs boson production and decay to study these couplings.
   We investigate these processes in the context of the $\kappa$-framework to study the anomalous behavior of $HHH$ and $VVHH$ couplings. We calculate one-loop QCD and electroweak correction for these processes.
   In this thesis, we discuss spinor helicity formalism, which has been used in computing Feynman amplitudes for QCD and electroweak correction to these processes. We discuss one-loop Feynman integrals and one-loop electroweak renormalization in this thesis.
    We compute all possible self-energy contributions at one-loop for electroweak renormalization. We adopt Catani-Saymour dipole subtraction technique to have IR-safe amplitudes.
    
    In this thesis, in the first work, we consider the process $b\bar{b}\rightarrow W^+W^-H$. Due to the non-negligible coupling of the Higgs boson with the bottom quarks, there is a dependence on the $WWHH$ coupling in this process.
    This process receives the largest contribution when the W bosons are longitudinally polarized. We compute one-loop QCD corrections to various ﬁnal states with polarized $W$ bosons. We ﬁnd that the corrections to the ﬁnal state with the longitudinally polarized $W$ bosons are large.
     It is shown that the measurement of the polarization of the W bosons can be used as a tool to probe the $WWHH$ coupling in this process. We also examine the effect of varying $WWHH$ coupling in the $\kappa$-framework. The variation in the cross-section is significantly high when we consider longitudinally polarized $W$ bosons.
     In the second work, we study one-loop electroweak correction to $H\rightarrow \nu_e\bar{\nu_e}\nu_\mu\bar{\nu_\mu}$. 
     We discuss $\gamma^5$-anomaly, complex mass scheme and input parameter schemes which are relevant for one-loop electroweak correction.
     The corrections depend on the $HHH$ and $ZZHH$ couplings. We investigate this dependence in $\kappa$-framework. We find that the width depends on $HHH$ coupling significantly.
      The dependence on $ZZHH$ coupling is marginal in the $\alpha(M_Z)$ scheme but significant in the $G_F$ scheme. We also study the dependence on $ZZW W$ coupling. The scaling of $HHH$ coupling does not violate gauge invariance but the scaling of $ZZHH$ and $ZZWW$ couplings violate gauge invariance.
      In the third work, we study one-loop electroweak correction to the $H\rightarrow e^+e^-\mu^+\mu^-$ process. The corrections depend on $HHH$ and $ZZHH$ couplings.
      This process has IR divergences which we handle by adopting Catani-Saymour dipole subtraction for QED.
      We observe the same behavior in the relative change of the decay width for this process as the process $H\rightarrow \nu_e\bar{\nu_e}\nu_\mu\bar{\nu_\mu}$ when we vary $HHH$ and $ZZHH$ couplings.
  

\newpage
\phantomsection
\addcontentsline{toc}{chapter}{\listfigurename}
\listoffigures

\newpage
\phantomsection
\addcontentsline{toc}{chapter}{\listtablename}
\listoftables


\newpage
\pagenumbering{arabic}

\chapter{Introduction}
 There has always been a great curiosity in human society to know the basic building blocks of the nature. In modern times, we are more or less satisfied that the thirst for this curiosity has been fulfilled via a model called Standard Model (SM). The SM has a zoo of elementary particles and describe their interaction through a spontaneous broken gauge symmetry.
 Starting from the discovery of the electron in $1897$ by J.J. Thomson, the physics community started building the SM. Then slowly, every other element of SM, except the Higgs boson, has  been discovered over the past century. Finally,  after a long struggle and immense dedication, the last particle in the zoo was discovered in $2012$ at Large Hadron Collider (LHC).

 The physical theory of the SM has a certain symmetry group structure which tells us about the structure and principle of elementary particles and their interactions. The gauge-group structure of the SM is $SU_C(3)\times SU_L(2)\times U_Y(1)$.  The $SU_C(3)$ represents the color-charge dependent interaction of the quarks and gluons. Other elementary particles do not carry any color charge. The force associated with the $SU_C(3)$ gauge group is the strong force. This force is very strong at low energy; as a result, the quarks and gluons are not seen as free particles in nature. They always form bound states. At very high energy, when this force becomes weak, the quarks and gluons become free, which are called asymptotic free states.
   The gauge group $SU_L(2)\times U_Y(1)$ describes weak and electromagnetic interaction among the fermions and gauge bosons. There are four gauge bosons related to this gauge group. The four gauge bosons are $W^\pm$ boson, $Z$ boson and photon.
   The $W^\pm$ bosons are massive charged bosons and are responsible for charged current interactions, whereas $Z$ boson is a massive neutral gauge boson and is responsible for weak neutral interaction. The fourth gauge boson photon is massless and is responsible for electromagnetic interaction.
     The fermions and gauge bosons are  massless with unbroken SM gauge group. As one tries to put the mass terms, it will lead to the violation of gauge symmetries. One needs a better mechanism to introduce the mass terms in the SM. In the late sixties, Weinberg and Salam showed that the fermions and gauge bosons could acquire masses via the Higgs mechanism without spoiling the gauge symmetries  at the level of action.
      In this mechanism, a doublet complex scalar field has been introduced. The gauge symmetry has been broken spontaneously by giving a vacuum expectation value to the doublet. This is known as spontaneous symmetry breaking (SSB), and as a result, all leptons, quarks, $W^\pm$ and $Z$ bosons acquire masses.
       The Higgs mechanism became more acceptable when 't Hooft showed the renormalizability of the Glashow, Weinberg and Salam (GWS) model of electroweak interaction in $1971$.

  Despite the illustrious success, still, the SM is not a complete theory to analyze nature. There are certain questions that can not be addressed within the framework of the SM. A few of them are as follows. The SM does not include the gravitational force for the unification as the scale for the gravitation is very high ($\sim 10^{34}$).
 The mass term for neutrinos are not included and the neutrino oscillation cannot be explained by the SM. It fails to explain the matter-antimatter asymmetry in the universe. The strong CP problem, dark matter, etc., also can not be explained by the SM.
  In this regard, there was a need for new theories which may cast light on the shaded regions of the SM. There are many beyond the standard model (BSM) scenarios which address a few of the aforementioned issues.
   The BSM models like 2HDM, Seesaw model, SMEFT, HEFT, SUSY, etc., have attracted the physics community because of their elegance to describe the loopholes of the SM. Despite the elegance of these theories, no significant departure from the SM predictions have been found. There is no hint of any specific BSM model in the experimental data.
    After the Higgs boson discovery at the LHC, its properties have been studied extensively. They are found to be consistent with the SM. However, some of the couplings of the Higgs bosons are still to be determined fully.
    The Higgs sector  of the model is not yet fully explored. This has left open the question of the shape of the Higgs potential.
  The Higgs potential can
still have many allowed shapes \cite{Agrawal:2019bpm}. Self-couplings of the Higgs boson and its couplings with some of the standard model particles are
still loosely bound. The more precise measurement of the couplings can also lead to hints to beyond the standard model scenarios.
  In the standard model, the $VVH$ and $VVHH$ couplings are related.
The experimental verification of this relationship is important
to put the standard model on a firm footing. There are scenarios beyond the SM,
where these couplings are either not related or have a different relationship \cite{Bishara_2017}.
The ATLAS collaboration has put a bound on this $VVHH$ coupling at 
 the LHC. Using the vector-boson fusion (VBF) production mechanism of a pair of Higgs bosons and using 126 fb$^{-1}$ of data at 13
    TeV, there is a bound of $ -0.43 < \kappa_{V_2H_2} < 2.56$
 at 95$\%$ confidence level \cite{Aad_2021}. Here $ \kappa_{V_2H_2}$ is the scaling factor for the $VVHH$ coupling. However, in this process, bound
  on $WWHH$ and $ZZHH$ couplings cannot be separated. The 
    process $p p \to HHV$, where a
   pair of Higgs bosons are produced in association with a $W$ or a $Z$ boson, allows us to separately measure $HHWW$ and $HHZZ$ couplings. Gluon-gluon fusion would contribute to
    $HHZ$ production. This mechanism is important at HE-LHC and FCC-hh.
    However, dependence on the scaling of $HHVV$ coupling is weak. The expected bound
    from the $WHH$ production 
  at the HL-LHC is $ -10.6 < \kappa_{V_2H_2} < 11$ \cite{Nordstrom:2018ceg}, which is quite loose.
  It is important to measure all the properties of the Higgs boson with good enough precision to demonstrate the complete validity of the standard model.
  The theoretical precision computation is necessary to fit the experimental data in order to find the proper shape of the Higgs potential and couplings with the gauge bosons. There are a few processes where one can look for the Higgs production and decays where these properties can be probed.
   We used both Higgs production and Higgs decay to probe the $HHH$ and $VVHH$ couplings through a few processes. The $HHH$ coupling will shape the Higgs potential and $VVHH$ couplings will tell how the gauge bosons are coupled to the Higgs field.
   
   Our motivation in this thesis is to study a few processes in the context of $\kappa$-framework~\cite{LHCHiggsCrossSectionWorkingGroup:2012nn,Ghezzi:2015vva} to study the anomalous  behavior of $HHH$ and $VVHH$ couplings. 
   In this framework, a naive scaling of a particular coupling has been done consistently for a given process. The idea is to consider all possible new physics effects in that particular coupling without worrying about a particular BSM model. 
   This framework works in an intuitive way that may face issues beyond leading order correction. It may violate the gauge invariance and unitarity.
   Our primary goal is to maintain the gauge invariance and rescale the Higgs boson couplings. Also, the arbitrary scaling may violate the unitarity. We vary the $\kappa$-factor within the allowed experimental bounds.
   Most of the current LHC data for the Higgs boson searches are formulated in this framework. A similar approach also may be fruitful in analyzing future collider data. This framework is useful for side-by-side comparison with the well-motivated new physics models to analyze the current and future collider data.
   As it is intuitive and may violate gauge invariance and unitarity, this framework is not the ultimate framework to analyze the Higgs sector or any other results. Our future goal is to adopt certain EFT theories to probe the anomalous coupling effects, which will be consistent in the sense of not violating the gauge invariance and unitarity.

In this thesis, we have focused on the precision computation of cross-section for Higgs production and decay; and how the scaling of $HHH$ and $VVHH$ coupling affect the results.
 We calculate the one-loop QCD and electroweak (EW) corrections to a few processes which are sensitive to $HHH$ and $VVHH$ couplings.
 In the perturbative computation, we calculate cross-section and decay width, which involve computation of matrix element (S-matrix) followed by the phase space integration for the final state particles.
 To calculate matrix elements, one needs to compute the Feynman diagrams with the given set of Feynman rules. These Feynman diagrams have been generated through a {\tt {Mathematica}} package called {\tt {FeynArts}}~\cite{Hahn:2000kx}. We calculate the Feynman amplitude for a particular diagram using spinor helicity formalism~\cite{Peskin:2011in}.
  We treat all fermions except $t$-quarks as massless particles and the spinor helicity formalism is used for massless fermions.
  We classify the diagrams for a given process in a set of generic classes and compute complete amplitude with the help of these generic diagrams.  
 We use the symbolic manipulation program {\tt FORM}~\cite{Vermaseren:2000nd}, to calculate the helicity amplitudes. The spinor helicity formalism can be implementable in {\tt FORM} with the helicity identities.  One-loop (OL) Feynman amplitudes also contain loop integrals.
    These loop integrals are conventionally computed in $d$-dimension. Depending upon the process, pentagon, box, triangle, bubble and tadpole-type integrals may appear. The loop-integrals with loop-momentum in the numerator will give tensor integrals and loop-integrals without loop-momentum in the numerator will give scalar integrals.
     The scalar integrals are computed using a package called OneLOop~\cite{vanHameren:2010cp} . There are many techniques to compute tensor integrals. These tensor integrals are computed via Oldenborg-Vermaseren technique~\cite{vanNeerven:1983vr}. To compute the tensor integrals, we use an in-house reduction code {\tt OVReduce}~\cite{Agrawal:2012df,Agrawal:1998ch} where this technique has been used.
      The $d$-dimensional amplitude has been computed with different dimensionality. We use four-dimensional helicity scheme (FDH)\cite{BERN1992451,Gnendiger:2017pys} as well as 't Hooft-Veltman (HV)\cite{THOOFT1972189} dimensional scheme to compute the one-loop integrals. Finally, the phase space integrals have been done with the advanced Monte-Carlo integration ({\tt AMCI})~\cite{Veseli:1997hr} package. In {\tt AMCI}, the {\tt VEGAS}\cite{Lepage:1977sw} algorithm is implemented using the parallel virtual machine ({\tt PVM})~\cite{10.7551/mitpress/5712.001.0001} package .
      The phase space points are generated with random numbers from {\tt AMCI} with our in-house phase space routine. Then the matrix element square i.e., integrand, is evaluated at the phase space points. {\tt AMCI} tries to shift the grid in the relevant areas in each iteration and compute stable results. 

In this thesis, firstly, we studied the process $b\bar{b}\rightarrow W^+W^-H$. The QCD correction at the next-to-leading order (NLO) has been computed in this work. This process is relevant in order to probe $HHH$ and  $VVHH$ couplings as these couplings appear in this process. These couplings do not arise in the light quark channels.
 The contribution of this channel is also sizable compared to the $gg$ channel. We also get a significant contribution at a higher center-of-mass energies. As we calculate helicity amplitudes, it helps to see the polarization dependence of cross-sections.
  Our finding is that the longitudinal polarization  states of external $W^\pm$ bosons give a larger contribution compared to other polarization modes. This longitudinal mode can help to reduce the backgrounds; hence this is relevant for the background study for the $W^+W^-H$ production.
   Our other main finding is that this process does not depend significantly on the anomalous $HHH$ coupling when we vary corresponding $\kappa_{HHH}$. But we have a significant change in cross-section, in particular in longitudinal mode, when we vary $VVHH$ coupling. From the kinematic distributions, we see that the negative $\kappa_{VVHH}$ plays a significant role in putting a more stringent bound.
   In the second work in this thesis, we studied Higgs boson decay width in the $H\rightarrow \nu_e\bar{\nu}_e\nu_\mu\bar{\nu}_\mu$ channel. We calculate one-loop electroweak correction to the decay width. This process again also includes the anomalous Higgs couplings.
 The $HHH$ and $VVHH$ couplings are involved in the loop-level diagrams. Here we again vary these couplings and study their effects. We find that the decay width of the Higgs in this channel depends on anomalous $HHH$ coupling significantly, whereas the effect of the anomalous $VVHH$ coupling is marginal. We also scale $ZZWW$ coupling in this process and find the decay width depends significantly on $\kappa_{ZZWW}$ in this channel.
 In the third work in this thesis, we studied the Higgs boson decay width in $H\rightarrow e^+e^-\mu^+\mu^-$. This is known as the golden channel. The collider signature is very relevant for this process as final state particles are charged leptons.
 We calculate one-loop EW correction to this process. Again, this process is sensitive to $HHH$ and $ZZHH$ couplings. We can study the  effects of anomalous $HHH$ and $ZZHH$ couplings on partial decay widths of the Higgs boson. We find the same anomalous effects on partial decay as the previous process. 

This thesis is organized as follows. In the chapter~\ref{chap_shf}, we discuss the formalism of spinor helicity techniques and the derivation of different identities, which help to compute the amplitudes.
Then we discuss the functional form of the Lorentz-invariant spinor products and vector currents, which are needed to calculate the numerical results.
  In the chapter~\ref{chap_ren_ol}, we discuss the electroweak one-loop renormalization. In this chapter, we have computed all self-energy diagrams which are needed to calculate the counterterms. We also mentioned Feynman rules for a few counterterms diagrams, which are needed for the processes mentioned above.
  In the chapter~\ref{chap_ir_div_dp_sub}, we discuss the IR singularities in virtual and real emission diagrams. In this chapter, we discuss the dipole subtraction procedure, which has been implemented to get the IR-safe amplitudes for the above-mentioned processes.
  In the chapter~\ref{chap_wwh_bb_fus}, we discuss process $b\bar{b}\rightarrow W^+W^-H$ in detail. In the chapter~\ref{chap_h24nu}, we have discussed the process $H\rightarrow \nu_e\bar{\nu}_e\nu_\mu\bar{\nu}_\mu$ and in the chapter~\ref{chap_h22e2m}, we discussed the process $H\rightarrow e^+e^-\mu^+\mu^-$. In the final chapter, we summarize the works being discussed in this thesis. We also mention our future goals in the conclusion chapter.

\chapter{Spinor helicity formalism}
\label{chap_shf}
   The computation of Feynman amplitudes for a given process is an essential and vital part of calculating the physical observables. Computation of traces with a series of $\gamma$-matrices that appear in computing the cross sections and decay widths, is a very cumbersome job. 
   The calculation is even more tedious for loop amplitudes. One has to compute these traces for finding the matrix element squares for a given process. To avoid this, one can numerically evaluate amplitudes, in particular helicity amplitudes.
   One gets the helicity amplitudes by computing so. The helicity amplitudes are less cumbersome to compute. Also, the helicity amplitudes are important to probe many physical observables when $W/Z$ bosons are present in the process.
   For massless particle, considerably simple amplitudes can be determined using the spinor helicity formalism. The one-loop amplitudes can also be computed with the spinor helicity formalism which is less tedious.
    There are many reviews~\cite{Berger:2009zb,ELLIS2012141,Peskin:2011in} where this formalism has been discussed in detail. In this chapter, we discuss spinor helicity formalism techniques: representation, helicity identities, functional form of Lorentz-invariant spinor products and vector current, etc.

\section{Spinor products and Vector currents}
\label{sec:chap_shf_rspvc}
  Spinor helicity formalism has been developed for the massless fermions. The massless fermion with momentum $p$ satisfies the Dirac equation
\begin{eqnarray}
\slashed{p}\:U(p)=0\:.
\label{eq_shf_de}
\end{eqnarray}
  One gets two solutions for the above equation; one is for the right-handed spinor, and the other one is for the left-handed spinor. In the chiral basis, the $\gamma$-matrices can be represented as 
\begin{gather}
\gamma^\mu=
\begin{pmatrix} 0 & \sigma^\mu \\ \bar{\sigma}^\mu & 0
\end{pmatrix}
 ,\quad
 \gamma^5=
  \begin{pmatrix}
   1 & 0 \\ 0 & -1
   \end{pmatrix}\:,
   \label{equ_shf_gmtx}
\end{gather}
 where $\sigma^\mu =(1,\vec{\sigma})$ and $\bar{\sigma}^\mu = (1,-\vec{\sigma})$.
 In this basis, the spinor solutions take the form
 \begin{gather}
U_R(p)=
\begin{pmatrix}
0 \\
u_R(p)
\end{pmatrix}
 ,\quad
 U_L(p)=
  \begin{pmatrix}
u_L(p) \\
0
   \end{pmatrix}
   ,
   \label{equ_shf_urspn}
\end{gather}
where the two-component spinors $u_{R/L}$ satisfy the equations
\begin{equation}
p\:.\:\sigma \:\:u_R = 0\:,\quad p\:.\:\bar{\sigma}\:\: u_L = 0\:\: .
\label{eq_shf_tse}
\end{equation}

There are unique solutions corresponding to each equation of~\ref{eq_shf_tse}. These two-component spinor solutions are related by the transformation 
\begin{equation}
u_R(p)=i\sigma^2u_L^*(p)\:.
\label{equ_shf_url}
\end{equation}

  One also needs antifermion spinor solutions $V(p)$, which satisfy the same equation as Eq.~\ref{eq_shf_de}. The same equations as in Eq.~\ref{eq_shf_tse} can be used to get the solutions for $V(p)$. 
  The spinors $V_R(p)$ and $V_L(p)$ are used for the creation of left-handed and right-handed antifermions. As fermion and antifermion satisfy the same equation, we will use $U(p)$ spinor for fermion and antifermion spinor solutions with proper helicity index. 
  The outgoing left- and right-handed fermions are represented by the spinors $\overline{U}_L(p)$ and $\overline{U}_R(p)$ respectively; and outgoing left- and right-handed antifermions are represented by the spinors $U_R(p)$ and $U_L(p)$ respectively.
  
  We use the popular notation for compactness to represent the spinors as
  \begin{equation}
  \overline{U}_L(p)=\langle p\:,\quad \overline{U}_R(p)=[ p\:,\quad U_L(p)= p]\:,\quad U_R(p) = p\rangle\:.
  \label{equ_shf_snr_rep}
  \end{equation}
  With this notation, the Lorentz-invariant spinor product can be written  as 
  \begin{equation}
  \overline{U}_L(p)\:U_R(q) = \langle pq\rangle\:,\quad \overline{U}_R(p)\:U_L(q) = [pq]\:.
  \label{equ_shf_lspnrp}
  \end{equation}
  In a similar way, the vector currents are written as
  \begin{equation}
  \overline{U}_L(p)\gamma^\mu U_L(q)=\langle p \gamma^\mu q]\:,\quad
  \overline{U}_R(p)\gamma^\mu U_R(q) = [p\gamma^\mu q\rangle\:.
  \label{equ_shf_lvecrnt}
  \end{equation}
 
 The helicity amplitudes at the tree level can be solely written in terms of Lorentz-invariant spinor products $\langle pq\rangle$ or $[pq]$. The vector currents are used in one-loop amplitudes as it contracts with the loop-momenta and the contraction can not be written in terms of spinor products.

\section{Helicity identities}
\label{sec_shf_hi}
Spinor helicity formalism is a very elegant technique to compute the amplitudes with the identities. From the properties of Dirac spinor solutions, $\gamma$-matrices, a set of helicity identities can be derived which will be used to simplify the matrix elements in terms of spinor products and vector currents.

  Taking complex conjugate of Eq.~\ref{equ_shf_lspnrp} one can easily get the  relation 
  \begin{equation}
  \langle pq \rangle = [qp]^*\:.
  \label{equ_shf_anid}
  \end{equation}
  
  Exploiting the Eq.~\ref{equ_shf_url} and properties of $\sigma^2$ matrix, we can obtain the following relations 
  \begin{equation}
  \langle pq \rangle = -\langle qp \rangle\:,\quad [pq] = -[qp]\:.
  \label{equ_shf_ansqid}
  \end{equation}
  This also tells us
  \begin{equation}
  \langle pp \rangle = [pp]=0\:.
  \label{equ_shf_pp_as_id}
  \end{equation}
  From Eq.~\ref{equ_shf_urspn} and Eq.~\ref{equ_shf_lvecrnt} the vector currents can be written in terms of two-component spinors as follows.
  \begin{gather}
  \langle p \gamma^\mu q]=\overline{U}_L(p)\gamma^\mu U_L(q)=u_L^\dag(p)\overline{\sigma}^\mu u_L(q)\nonumber \\
  [p\gamma^\mu q\rangle =\overline{U}_R(p)\gamma^\mu U_R(q)=u_R^\dag(p)\sigma^\mu u_R(q)
  \label{equ_shf_lrspexp}
  \end{gather}
  
  With the property given in Eq.~\ref{equ_shf_url}, we can derive the relation 
  \begin{equation}
  \langle p \gamma^\mu q] = [q \gamma^\mu p\rangle\:.
  \label{equ_shf_lrrlid}
  \end{equation}
  We can also show using Eq.~\ref{equ_shf_gmtx} and~\ref{equ_shf_urspn} that 
  \begin{equation}
  \langle p \gamma^\mu q\rangle = [ p \gamma^\mu q] = 0\:.
  \label{equ_shf_vec_zr_id}
  \end{equation}
  Similar identities can be found with the chain of $\gamma$-matrices sandwiched between two spinors with possible helicities. The identities are 
  \begin{gather}
  \langle p \gamma^{\mu_1}...\gamma^{\mu_{2n+1}} q]=[q \gamma^{\mu_{2n+1}}...\gamma^{\mu_1} p\rangle\nonumber\\
  \langle p \gamma^{\mu_1}...\gamma^{\mu_{2n}} q\rangle=-\langle q \gamma^{\mu_{2n}}...\gamma^{\mu_1} p\rangle,\quad [ p \gamma^{\mu_1}...\gamma^{\mu_{2n}} q]=-[ q \gamma^{\mu_{2n}}...\gamma^{\mu_1} p]\nonumber\\
  \langle p \gamma^{\mu_1}...\gamma^{\mu_{2n+1}} q\rangle=[p \gamma^{\mu_1}...\gamma^{\mu_{2n+1}} q]=0\nonumber\\
  \langle p \gamma^{\mu_1}...\gamma^{\mu_{2n}} q]=[ p \gamma^{\mu_1}...\gamma^{\mu_{2n}} q\rangle = 0\:.
  \end{gather}
  
  The Fiertz identity for sigma matrices is given by
  \begin{equation}
  (\overline{\sigma}^\mu)_{ab}(\overline{\sigma}_\mu)_{cd}=2\:(i\sigma^2)_{ac}(i\sigma^2)_{bd}\:.
  \end{equation}
  
  With the help of this identity and the Eq.~\ref{equ_shf_lrspexp}, one can derive the below identity from the contraction of two vector currents.
  \begin{equation}
  \langle p \gamma^\mu q]\langle k \gamma_\mu l] = 2\:\langle p k \rangle [lq]\:,\quad \langle p \gamma^\mu q]\langle k \gamma_\mu l] = 2\:\langle p l\rangle[kq]
  \label{equ_shf_vec_con_id}
  \end{equation}
  This identity is very useful for calculating the helicity amplitudes. The spin sum of the spinor outer product for massless fermion is given by
  \begin{equation}
   \sum_{s=1}^{2} U_s(p)\overline{U}_s(p) = \slashed{p}\:.
  \end{equation}   
  In the chiral representation, the spin basis of spinors can be converted into the helicity basis of spinors and the above equation can be written as 
  \begin{equation}
  \slashed p = p\rangle [p + p]\langle p \:. 
  \label{equ_shf_slpid}
  \end{equation}
  The dot product of two momenta can also be written in terms of spinor products given in Eq.~\ref{equ_shf_lspnrp} with the help of the relation given in Eq.~\ref{equ_shf_ansqid} and \ref{equ_shf_slpid}. The dot product of two momenta can be written as 
  \begin{equation}
  2\:p.q = \frac{1}{2} \text{Tr}(\slashed p \slashed q)=\langle pq\rangle[qp]=|\langle pq\rangle|^2=|[pq]|^2\:.
  \end{equation}
  From the anticommutation relation of $\gamma$-matrices, one very useful identity can be derived, which has been used extensively to compute helicity amplitudes in the spinor helicity method.
  We can write with the anticommutation relation of $\gamma$-matrix as follow.
  \begin{align}
  &\{\gamma_\mu,\gamma_\nu\}=2\:\eta_{\mu\nu}\nonumber\\
  &\implies l^\mu_1l^\nu_2\{\gamma_\mu,\gamma_\nu\}=2\:\eta_{\mu\nu}l^\mu_1l^\nu_2\nonumber\\
  &\implies \frac{\slashed {l_1}\slashed {l_2}+\slashed {l_2}\slashed{l_1}}{2\:l_1.l_2}=1
  \label{equ_shf_antcmt_id}
\end{align}   

  We can take reference momenta $l_1$ and $l_2$ lightlike and unequal (obvious). This identity can be inserted in between two $\gamma$-matrices for a tensor current with a chain of $\gamma$-matrices sandwiched between spinors.
  Then we can use the identity given in Eq.~\ref{equ_shf_slpid} and convert the chain of $\gamma$-matrices with the spinors into the vector currents $\langle p \gamma^\mu q]$ or $[p \gamma^\mu q\rangle$. This identity is quite useful for the computation of one-loop amplitude.
  
  Below, we summarize the helicity identities and the properties of spinors for massless fermions
  \begin{align}
  &\langle pq \rangle = [qp]^*\:,\nonumber\\
  &\langle pq \rangle = -\langle qp \rangle\:,\quad [pq] = -[qp]\nonumber\\
  &\langle pp \rangle = [pp] = 0\:,\nonumber\\
  &\langle p \gamma^\mu q] = [q \gamma^\mu p\rangle\:,\nonumber\\
  &\langle p \gamma^\mu q\rangle = [q \gamma^\mu p]=0\:,\nonumber\\
  &\langle p \gamma^{\mu_1}...\gamma^{\mu_{2n+1}} q]=[q \gamma^{\mu_{2n+1}}...\gamma^{\mu_1} p\rangle\:,\nonumber\\
  &\langle p \gamma^{\mu_1}...\gamma^{\mu_{2n}} q\rangle=-\langle q \gamma^{\mu_{2n}}...\gamma^{\mu_1} p\rangle,\quad [ p \gamma^{\mu_1}...\gamma^{\mu_{2n}} q]=-[ q \gamma^{\mu_{2n}}...\gamma^{\mu_1} p]\:,\nonumber\\
  &\langle p \gamma^{\mu_1}...\gamma^{\mu_{2n+1}} q\rangle=[p \gamma^{\mu_1}...\gamma^{\mu_{2n+1}} q]=0\:,\nonumber\\
  &\langle p \gamma^{\mu_1}...\gamma^{\mu_{2n}} q]=[ p \gamma^{\mu_1}...\gamma^{\mu_{2n}} q\rangle = 0\:,\nonumber\\
  &\langle p \gamma^\mu q]\langle k \gamma_\mu l] = 2\:\langle p k \rangle [lq]\:,\quad \langle p \gamma^\mu q]\langle k \gamma_\mu l] = 2\:\langle p l\rangle[kq]\:,\nonumber\\
  &\slashed p = p\rangle [p + p]\langle p\:,\nonumber\\
   &2\:p.q =\langle pq\rangle[qp]=|\langle pq\rangle|^2=|[pq]|^2\:,\nonumber\\
   &\slashed p = p\rangle [p + p]\langle p\:.
   \label{equ_shf_shids}
  \end{align}

There are other identities, such as Schouten, Fiertz with charge conjugation, etc., that can be derived in spinor helicity formalism~\cite{ELLIS2012141}. We only use the above spinor helicity identities and the identity given in Eq.~\ref{equ_shf_antcmt_id} to calculate the Feynman amplitudes for the scattering and decay process at one loop.

\section{Polarization vector}
\label{sec_shf_pol_rep}

The polarization vectors for a massless gauge boson of definite helicity can be represented as
\begin{gather}
\epsilon^{*\mu}_R(k)=\frac{1}{\sqrt{2}}\frac{\langle r \gamma^\mu k]}{\langle rk\rangle}\:,\quad \epsilon^{*\mu}_L(k)=-\frac{1}{\sqrt{2}}\frac{[r \gamma^\mu k\rangle}{[rk]}\:.
\label{equ_shf_eps_rep}
\end{gather}
  Here $\epsilon_L$ and $\epsilon_R$ are the left- and right-handed polarization respectively.
  $k$ is the momentum of the vector boson and $r$ is some lightlike reference momenta, which should not be equal to $k$.
  This representation can easily be understood from a fermionic current with definite helicity of massless fermions to which a massless gauge boson can decay.
  This representation follows the properties and identities of the polarization vector of a massless vector boson. It can be easily checked from the Eq.~\ref{equ_shf_eps_rep} that 
  \begin{equation}
  [\epsilon_R^*(k)]^*=\epsilon_L^*(k)\:,\quad k_\mu \epsilon_{R,L}^{*\mu}=0\:.
  \end{equation}
  The last relation is derived with the identities given in Eq.~\ref{equ_shf_pp_as_id} and~\ref{equ_shf_slpid}. Helicity vectors also satisfy 
  \begin{equation}
  \epsilon_R^*(k).[\epsilon_L^*(k)]^*=\epsilon_R^*(k).\epsilon_R^*(k)=0\:. 
  \end{equation}
The normalization and orthogonality properties can also be checked. We can write
\begin{equation}
|\epsilon^{*}_R(k)|^2=\frac{1}{2}\frac{\langle r\gamma^\mu k] \langle k\gamma_\mu r]}{\langle rk \rangle[kr]}=-1\:.
\end{equation}
  Using the relations given in Eq.~\ref{equ_shf_lrrlid} and~\ref{equ_shf_vec_con_id}, we can find the above property. Similarly, we can find 
  \begin{equation}
  |\epsilon^{*}_L(k)|^2=-1\:,\quad \epsilon^{*}_R(k).[\epsilon^*_L(k)]^*=0\:.
  \end{equation}
  The basic properties of the helicity vector for a massless gauge boson  are satisfied by the representation given in Eq.~\ref{equ_shf_eps_rep}. The lightlike reference momenta in Eq.~\ref{equ_shf_eps_rep} can be chosen from the same process to reduce the length of the amplitude.
   In the tree-level amplitude, the polarization vector is contracted with other polarization vectors, momenta and with vector currents; which can be written in terms of spinor products.
   In one-loop amplitudes, the polarization vectors can be contracted with the loop momenta. So, in one-loop amplitudes, not only the spinor product but also the vector current given in Eq.~\ref{equ_shf_lvecrnt} is required to compute the helicity amplitudes.
\section{Functional form of Spinor products and Vector currents}
\label{sec_shf_ff_sp_vec}
  We have discussed about the spinor helicity formalism in the above sections. Next, we need the functional form of both Lorentz-invariant spinor product and vector current to compute one-loop amplitudes.
  In this section, we derive the functional form of the spinor product and vector current following Ref.~\cite{Kleiss:1985yh,Kleiss:1986qc}. Here we will denote right- and left-handed spinors with $U_{\pm}(p)$, unlike $U_{R/L}(p)$. 
  For a massless fermion with momentum $p$ and helicity $\lambda=\pm 1$, we have the relation 
  \begin{equation}
  U_\lambda(p)\:\overline{U}_\lambda(p)=\omega_\lambda\slashed p\:,
  \label{equ_shf_uub_id}
  \end{equation}
  where $\omega_\lambda=\frac{1}{2}(1+\lambda\gamma^5)$. We consider two momenta $k_0$ and $k_1$, with the following properties
  \begin{equation}
  k_0.k_0=0\:,\quad k_1.k_1=-1\:,\quad k_0.k_1=0.
  \label{equ_shf_k0k1_condition}
  \end{equation}
  We choose
  \begin{equation}
  U_{+}(k_0)=\slashed k_1 U_{-}(k_0)\:,
  \label{equ_shf_uk0_slk1}
  \end{equation}
    which satisfies Eq.~\ref{equ_shf_uub_id} with the condition given in Eq.~\ref{equ_shf_k0k1_condition}. 
  Now for any lightlike momentum we can construct spinors as 
  \begin{equation}
  U_\lambda(p)=\frac{\slashed p U_{-\lambda}(k_0)}{\sqrt{2p.k_0}}\:.
  \label{equ_shf_spnr_ltlk}
  \end{equation}
  This spinor again satisfies the relation given in Eq.~\ref{equ_shf_uub_id}.
  
  We label the Lorentz-invariant spinor products by 
  \begin{gather}
  s_1(p_1,p_2)=\overline{U}_+(p_1)U_-(p_2)=\overline{U}_R(p_1)U_L(p_2)=[p_1p_2]=-s_1(p_2,p_1)\nonumber\\
  s_2(p_1,p_2)=\overline{U}_-(p_1)U_+(p_2)=\overline{U}_L(p_1)U_R(p_2)=\langle p_1p_2\rangle=-s_2(p_2,p_1)
  \end{gather}
  Now we derive the functional form of the $s_1(p_1,p_2)$ and $s_2(p_1,p_2)$. Let's start with the first spinor product.
  \begin{align}
  s_1(p_1,p_2) &= [p_1,p_2]\nonumber\\
  &=\overline{U}_+(p_1)U_-(p_2)\nonumber\\
&=\frac{\overline{U}_-(k_0)\slashed p_1\slashed p_2U_+(k_0)}{2\sqrt{(p_1.k_0)(p_2.k_0)}}\quad\quad\quad\quad\quad\quad\quad\quad\quad\quad\quad\quad\quad{\text{(using Eq.~\ref{equ_shf_spnr_ltlk}})}\nonumber\\
 &=\frac{\overline{U}_-(k_0)\slashed p_1\slashed p_2\slashed k_1U_-(k_0)}{2\sqrt{(p_1.k_0)(p_2.k_0)}}\:\quad\quad\quad\quad\quad\quad\quad\quad\quad\quad\quad\quad{\text{(using Eq.~\ref{equ_shf_uk0_slk1}})}\nonumber\\
 &=\frac{\text{Tr}\Big[\overline{U}_-(k_0)\slashed p_1\slashed p_2\slashed k_1U_-(k_0)\Big]}{2\sqrt{(p_1.k_0)(p_2.k_0)}}\nonumber\\
 &=\frac{\text{Tr}\Big[\omega_-\slashed k_0\slashed p_1\slashed p_2\slashed k_1\Big]}{2\sqrt{(p_1.k_0)(p_2.k_0)}}
   \end{align}
   As the spinor products are scalar, one can take the trace. We compute the above trace and set the following choice of $k_0$ and $k_1$, which satisfy the condition given in Eq.~\ref{equ_shf_k0k1_condition}
   \begin{equation}
   k_0^\mu=(1,1,0,0)\:,\quad k_1^\mu=(0,0,1,0)\:.
   \label{equ_shf_k0k1_choice}
   \end{equation}
    We  get the functional form of spinor product $s_1(p_1,p_2)$ in terms of four-momentum component of $p_1$ and $p_2$ as
\begin{equation}
   s_1(p_1,p_2)=(p_1^y+ip_1^z)\Big(\frac{p_2^0-p_2^x}{p_1^0-p_1^x}\Big)^{\frac{1}{2}}-(p_2^y+ip_2^z)\Big(\frac{p_1^0-p_1^x}{p_2^0-p_2^x}\Big)^{\frac{1}{2}}\:.
   \label{equ_shf_s1_exp}
\end{equation}
Here superscripts $0$, $x$, $y$ and $z$ denote the energy and spatial components of momenta. Similarly, one can derive the functional form of $s_2(p_1,p_2)$. The spinor product $s_2(p_1,p_2)$ is derived as 
\begin{align}
s_2(p_1,p_2) &= \langle p_1,p_2\rangle\nonumber\\
&=\overline{U}_-(p_1)U_+(p_2)\nonumber\\
&=\frac{\overline{U}_-(k_0)\slashed k_1\slashed p_1\slashed p_2U_-(k_0)}{2\sqrt{(p_1.k_0)(p_2.k_0)}}\nonumber\\
&=\frac{\text{Tr}\Big[\omega_-\slashed k_0\slashed k_1\slashed p_1\slashed p_2\Big]}{2\sqrt{(p_1.k_0)(p_2.k_0)}}\:\:.
\end{align}
Again we calculate this trace and set the choice given in Eq.~\ref{equ_shf_k0k1_choice}. We get the functional form of $s_2(p_1,p_2)$ as
\begin{equation}
s_2(p_1,p_2)=-(p_1^y-ip_1^z)\Big(\frac{p_2^0-p_2^x}{p_1^0-p_1^x}\Big)^{\frac{1}{2}}+(p_2^y-ip_2^z)\Big(\frac{p_1^0-p_1^x}{p_2^0-p_2^x}\Big)^{\frac{1}{2}}\:.
\label{equ_shf_s2_exp}
\end{equation}
  It can be easily seen that $s_1(p_1,p_2)$ and $s_2(p_1,p_2)$ follow the identity given in Eq.~\ref{equ_shf_anid},~\ref{equ_shf_ansqid} and 
~\ref{equ_shf_pp_as_id}.
  Now following similar steps as in the computation of $s_1(p_1,p_2)$ and $s_2(p_1,p_2)$, we derive the functional form of the vector current $\langle p \gamma^\mu q]$ and $[p\gamma^\mu q\rangle$.
  We denote the vector current as  
  
  \begin{equation}
  t^\mu(p_1,p_2)= \langle p_1 \gamma^\mu p_2]
  \end{equation}
  and it follows $t^\mu(p_1,p_2)= [p_2\gamma^\mu p_1\rangle$ from the identity given in Eq.~\ref{equ_shf_lrrlid}. We calculate $t^\mu(p_1,p_2)$ as 
  \begin{align}
  t^\mu(p_1,p_2)&= \overline{U}_-(p_1)\gamma^\mu U_-(p_2)\nonumber\\
  &=\frac{\overline{U}_+(k_0)\slashed p_1\gamma^\mu\slashed p_2 U_+(k_0)}{2\sqrt{(p_1.k_0)(p_2.k_0)}}\nonumber\\
  &=\frac{\text{Tr}(\omega_+\slashed k_0\slashed p_1\gamma^\mu \slashed p_2)}{2\sqrt{(p_1.k_0)(p_2.k_0)}}\nonumber\\
  &=\frac{((k_0.p_1)p_2^\mu -(p_1.p_2)k_0^\mu +(k_0.p_2)p_1^\mu +i\epsilon^{\mu\alpha\beta\rho}k_{0\alpha}p_{1\beta}p_{2\rho})}{2\sqrt{(p_1.k_0)(p_2.k_0)}}\:.
  \label{equ_shf_tmu_exp}
  \end{align}
  
  Here $\epsilon^{\mu\alpha\beta\rho}$ is the Levi-Civita symbol and $\epsilon^{0123}=1$. We can now compute the four components of $t^\mu(p_1,p_2)$ function from the Eq.~\ref{equ_shf_tmu_exp}. They are
  \begin{align}
  t^0(p_1,p_2)&=\frac{(p_1^y-ip_1^z)(p_2^y+ip_2^z)+(p_1^0-p_1^x)(p_2^0-p_2^x)}{(\sqrt{(p_1^0-p_1^x)(p_2^0-p_2^x)})}\:,\nonumber\\
  t^x(p_1,p_2)&=\frac{(p_1^y-ip_1^z)(p_2^y+ip_2^z)-(p_1^0-p_1^x)(p_2^0-p_2^x)}{(\sqrt{(p_1^0-p_1^x)(p_2^0-p_2^x)})}\:,\nonumber\\
  t^y(p_1,p_2)&=(p_1^y-ip_1^z)\Big(\frac{p_2^0-p_2^x}{p_1^0-p_1^x}\Big)^\frac{1}{2}+(p_2^y+ip_2^z)\Big(\frac{p_1^0-p_1^x}{p_2^0-p_2^x}\Big)^\frac{1}{2}\:,\nonumber\\
  t^z(p_1,p_2)&=(p_1^z+ip_1^y)\Big(\frac{p_2^0-p_2^x}{p_1^0-p_1^x}\Big)^\frac{1}{2}+(p_2^z-ip_2^y)\Big(\frac{p_1^0-p_1^x}{p_2^0-p_2^x}\Big)^\frac{1}{2}\:.
  \label{equ_shf_t_exp}
  \end{align}
  
  These are four components of the vector current $\langle p_1 \gamma^\mu p_2]$ as a function of $p_1$ and $p_2$, and it can be numerically evaluated. Numerically we have verified the identities  given in Eq.~\ref{equ_shf_vec_con_id}  with the functions given in Eq.~\ref{equ_shf_s1_exp},~\ref{equ_shf_s2_exp} and~\ref{equ_shf_t_exp}.
 
  In this chapter, we have discussed the spinor helicity formalism. As we have seen, the tree-level amplitudes can be calculated in terms of spinor products. 
  The Feynman amplitudes at the one-loop can be written in terms of dot products of momenta, polarization vectors and the vector currents along with the spinor products.
   We use the symbolic manipulation program {\tt FORM}~\cite{Vermaseren:2000nd} to calculate the helicity amplitudes in spinor helicity formalism. The spinor helicity identities given in Eq.~\ref{equ_shf_shids} have been implemented in a {\tt FORM} code. 
   Especially, the identity given in Eq.~\ref{equ_shf_antcmt_id} is very useful when we calculate loop amplitudes. The proper choice of reference momenta in this identity and in polarization vector representation in Eq.~\ref{equ_shf_eps_rep} can reduce the length of the amplitudes significantly.
   We can also make {\it do loop} in {\tt FORM} code for the choice of reference momenta and minimize the size of a amplitude. The spinor helicity formalism has been discussed in $4$-dimension. This formalism is very elegant in computing the helicity amplitudes at one loop.
\chapter{Renormalization at one-loop}
\label{chap_ren_ol}
  In this chapter, we will discuss the one-loop Feynman integrals that appear in the one-loop diagrams for a quantum field theoretical scattering and decay process. We will see that these integrals are divergent at large loop momenta ($k\rightarrow \infty$) regime. 
   This type of divergence is called \textit{ultraviolet} (UV) divergence. The standard model is a completely renormalizable theory. In the standard model, the divergences that appear at any order in perturbative computation can be absorbed in the bare parameters of the standard model.
   The renormalized perturbation theory tells us how the bare parameters can be written in terms of renormalized parameters and corresponding counterterms. We will discuss one-loop renormalization for electroweak theory in the following sections.
   We adopt the on-shell renormalization scheme for massive particles and the $\overline{\text{MS}}$ scheme for massless particles.
\section{One-loop integrals}
\label{sec_ren_ol_int}
\begin{figure}[h!]
 \begin{center}
\includegraphics [angle=0,width=0.4\linewidth]{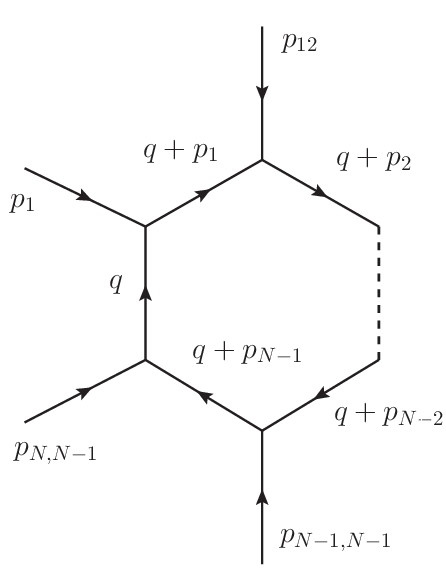}\\
	\caption{Generic one-loop $N$-point integral with loop and external momenta.}
	\label{fig_ren_ol_gnrc}
	\end{center}
\end{figure}

A general one-loop integral in $D$ dimension with $N$ number of propagators and $P$ number of loop momenta in the numerator can be written as 
\begin{align}
 T^{(N)}_{\mu_1...\mu_P}(p_0,...,p_{N-1},m_0,...,m_{N-1})=\frac{(2\pi\mu)^{(4-D)}}{i\pi^2}\int d^Dq\frac{q_{\mu_1}...q_{\mu_P}}{N_0...N_{N-1}}\:,
 \label{equ_ren_ol_int}
 \end{align}
  with the denominator factors
 \begin{align}
 N_0=q^2-m_0^2+i\epsilon\:,\quad D_i=(q+p_i)^2-m_i^2+i\epsilon\:,\quad i=1,...,N-1\:.
 \end{align}
  These denominators originate from the propagators in the Feynman diagram. The schematic diagram for the one-loop integral given in Eq.~\ref{equ_ren_ol_int}, has been shown in Fig.~\ref{fig_ren_ol_gnrc}. The loop momenta for $i^{\text{th}}$ internal leg is $q+p_i$ and the external momenta are
  \begin{align}
   p_{0i}=p_i\quad {\text{and}}\quad p_{ij}=p_i-p_j.
  \end{align}    
  Conventionally, the integral $T^N$ is denoted by alphabets with the $N$ value as $T^1\equiv A$, $T^2\equiv B$, $T^2\equiv C$, ... and the scalar integrals are denoted by $A_0$, $B_0$, $C_0$, ... . Traditionally, $A$, $B$, $C$, $D$, ... -type integrals are called tadpole, bubble, triangle, box, ... -type integrals respectively.
  There are several reduction techniques~\cite{Passarino:1978jh,vanNeerven:1983vr} that help to reduce tensor integrals into scalar integrals. The basic property of these reductions is that any integral $T^N$ can be written as a linear combination of one-loop scalar integrals and a finite remnant $\mathcal{R}$ of dimensional regularization procedure. This can be expressed as
  \begin{align}
  T^N=c_{4;j}T^4_{0;j}+c_{3;j}T^3_{0;j}+c_{2;j}T^2_{0;j}+c_{1;j}T^1_{0;j}+\mathcal{R}+\mathcal{O}(D-4)\:.
  \label{equ_ren_Tn_decmp}
  \end{align}
  The coefficient $c_{N;j}\:(N=1,...,4)$ in Eq.~\ref{equ_ren_Tn_decmp} are evaluated in $D=4$ dimension i.e, they do not depend on $\epsilon$. They are a function of external momenta, masses and the different scalar integrals.
  The $T^N_{0;j}\:(N=0,...,4)$ denote the one-loop scalar integral of type $j$. The type $j$ specifies the external momenta that specify the denominators for corresponding scalar integral. This decomposition originates from the Lorentz covariance of decomposition of the tensor structure to invariant form factors and the four-dimensional nature of space-time, which allows the decomposition of higher scalar integrals to a sum of box-scalar integrals.
  The explicit Lorentz decomposition for bubble tensor integrals can be written as 
  \begin{align}
  T^2_{\mu}\equiv B_\mu= {p_1}_\mu\:B_1\:,\quad T^2_{\mu\nu}\equiv B_{\mu\nu}=g_{\mu\nu}\:B_{00}+{p_1}_\mu {p_1}_\nu\:B_{11}\:.  
  \label{equ_ren_b_int_dec}
  \end{align}
  In a similar way, one can write the decomposition of other tensor integrals too~\cite{Denner:1991kt}.
  
  The integral given in Eq.~\ref{equ_ren_ol_int} are UV divergent with the condition $P+D-2N \geq 0$. The UV divergences are regularized in dimensional regularization in which the one-loop integrals have been evaluated in the $D$ dimension. 
  The UV divergence comes only from the tadpole and bubble scalar integrals. In cut-off regularization, we can easily see the divergences in $D=4$ dimension as below
  \begin{align}
  A_0\sim\int^\Lambda \frac{d^4q}{N_0}\sim\int^\Lambda \frac{d^4q}{q^2}\sim \Lambda^2\:,\quad B_0\sim\int^\Lambda \frac{d^4q}{N_0N_1}\sim\int^\Lambda \frac{d^4q}{q^2.q^2}\sim \Lambda^0\:.
  \label{equ_ren_a0_b0_uv}
  \end{align}
  In Eq.~\ref{equ_ren_a0_b0_uv}, the $A_0$ is quadratic and $B_0$ is logarithmic divergent. In dimension regularization these divergences appear as $\frac{1}{\epsilon}$ poles. The UV poles for these two scalar integrals in dimensional regularization are
  \begin{align}
  A_0(m)\Big\vert_{UV} = -2m^2\:\frac{1}{D-4}\:\quad {\text{and}}\quad B_0(p_1,m_0,m_1)\Big\vert_{UV}=-2\:\frac{1}{D-4}\:.
  \end{align}
  One can remove the UV divergences by renormalizing the bare parameters and the fields for a given renormalizable theory. 
%
%
\section{EW renormalization at one-loop}
\label{sec_ren_ew_ol_ren}
 The parameters in the standard model have to be extracted with the experiments. The usual parameters of SM are couplings of different interactions and the masses of elementary particles. At the tree level, physical observables can be calculated with these parameters and they have direct relations with the experiments.
 This scenario becomes problematic in higher order corrections. As we have seen that the higher order corrections can be UV divergent. In a renormalizable theory, these UV divergences can be absorbed and one  can predict the physical observables with higher order accuracy.
 The procedure of renormalization for a renormalizable theory involves the redefinition of bare SM parameters and fields in terms of renormalized parameters and fields. The renormalized parameters are related to physical observables whereas the field renormalization makes Green functions finite.
 The heavy particles masses are known from the experiments, so on-shell renormalization scheme is a good choice for the renormalization associated with the heavy particles. For the particles whose masses are not known properly, the $\overline{\text{MS}}$ renormalization scheme will be convenient for them.
 There are several independent parameter sets with which one can use  in renormalization. We choose the parameter set: $e$, $M_W$, $M_Z$, $M_H$ and $m_f$. Here $e$ is the electric charge, $M_W$, $M_Z$, $M_H$ and $m_f$ are the masses of $W$, $Z$, $H$ boson and fermions respectively.
  We take the quark mixing CKM matrix ($V_{CKM}$) as the diagonal matrix, so no renormalization is needed for $V_{CKM}$. We adopt on-shell renormalization scheme for the electroweak standard model. Here we will only discuss the renormalizations which are relevant for  the process discussed in chapter~\ref{chap_h24nu} and~\ref{chap_h22e2m}.
  
  We express the bare parameters in terms of renormalized parameters and corresponding counterterms as below
  \begin{align}
  e_0=(1+\delta Z_e)e\:,\quad M^2_{W,0}=M^2_{W}+\delta M^2_W\:,\quad
  M^2_{Z,0}=M^2_Z+\delta M_Z^2\:.
  \label{equ_ren_ren_para}
  \end{align}
 As we have discussed that the fields are also needed to be renormalized. We write similar expressions as in Eq.~\ref{equ_ren_ren_para} for the fields renormalization as below
 \begin{align}
 Z_0=(1+\frac{1}{2}\delta Z_{ZZ})\: Z+\frac{1}{2}\delta Z_{ZA}\; A\:,&\quad A_0=(1+\frac{1}{2}\delta Z_{AA})\: A+\frac{1}{2}\delta Z_{AZ}\: Z\:,\nonumber \\
W_0^\pm=(1+\frac{1}{2}\delta Z_W)\: W^\pm\:,&\quad\:H_0=(1+\frac{1}{2}\delta Z_H)\: H\:,\nonumber \\
f_{i,0}^L=(\delta_{ij}+\frac{1}{2}\delta Z_{ij}^{f,L})\: f_j^L\:,&\quad f_{i,0}^R=(\delta_{ij}+\frac{1}{2}\delta Z_{ij}^{f,R})\: f_j^R\:.
\label{equ_ren_ren_wf}
 \end{align}
 We denote bare quantities by the `$0$' and the renormalized quantities in the normal fashion in Eq.~\ref{equ_ren_ren_para},~\ref{equ_ren_ren_wf}. The $\delta Z$  and $\delta M$ given in Eq.~\ref{equ_ren_ren_para},~\ref{equ_ren_ren_wf} are the renormalization constants.
 In renormalized perturbation theory with the expansion $Z=1+\delta Z$, the bare Lagrangian $\mathcal{L}_0$  can be written as
 \begin{align}
 \mathcal{L}_0=\mathcal{L}+\delta \mathcal{L}\:,
 \end{align}
 where $\mathcal{L}$ is the renormalized Lagrangian which depends on renormalized parameters and fields; and $\delta \mathcal{L}$ is the counterterm.
 One can set up the Feynman rules for counterterms from $\delta \mathcal{L}$ and draw all possible counterterm diagrams for a given process.
 The renormalization constants given in Eq.~\ref{equ_ren_ren_para},~\ref{equ_ren_ren_wf} can be determined by the suitable on-shell renormalization conditions. Following the Ref~\cite{Denner:1991kt}, we write the explicit expression form of the renormalization constants below
\begin{align}
\delta M_W^2={\text {Re}}\:\Sigma ^W_T(M_W^2)\:,\quad& \delta Z_W=-{\text{Re}}\frac{\partial}{\partial k^2}\Sigma^W_T(k^2)\bigg |_{k^2=M_W^2}\:,\nonumber\\
\delta M_Z^2={\text {Re}}\:\Sigma ^{ZZ}_T(M_Z^2)\:,\quad& \delta Z_{ZZ}=-{\text{Re}}\frac{\partial}{\partial k^2}\Sigma^{ZZ}_T(k^2)\bigg |_{k^2=M_Z^2}\:,\nonumber\\
\delta Z_{ZA}=2\frac{\Sigma^{AZ}_T(0)}{M_Z^2},\quad& \delta Z_{AA}=-\frac{\partial}{\partial k^2}\Sigma^{AA}_T(k^2)\bigg |_{k^2=0}\:,\quad\nonumber\\
\delta Z_e=-\frac{1}{2}\delta Z_{AA}-\frac{s_W}{c_W}\frac{1}{2}\delta Z_{ZA}\:,\quad&\delta Z_{H}=-{\text{Re}}\frac{\partial}{\partial k^2}\Sigma^H(k^2)\bigg |_{k^2=M_H^2}\:,\nonumber\\
\delta Z_{ii}^{f,L}=-{\text {Re}}\:\Sigma_{ii}^{f,L}(m_{f,i}^2)-m_{f,i}^2\frac{\partial}{\partial k^2}{\text {Re}}&\Big[\Sigma_{ii}^{f,L}(k^2)+\Sigma_{ii}^{f,R}(k^2)+2\Sigma_{ii}^{f,S}(k^2)\Big]\bigg |_{k^2=m_{f,i}^2}\:,\nonumber\\
\delta Z_{ii}^{f,R}=-{\text {Re}}\:\Sigma_{ii}^{f,R}(m_{f,i}^2)-m_{f,i}^2\frac{\partial}{\partial k^2}{\text {Re}}&\Big[\Sigma_{ii}^{f,L}(k^2)+\Sigma_{ii}^{f,R}(k^2)+2\Sigma_{ii}^{f,S}(k^2)\Big]\bigg |_{k^2=m_{f,i}^2}\:.
\label{equ_ren_ren_cnst}
\end{align}
 The $\sum$ in Eq.~\ref{equ_ren_ren_cnst} are the self-energies for different propagators. We denote sin and cosine of the Weinberg angle as $s_W$ and $c_W$. In the next section, we have written down the explicit expressions for various self-energies which are relevant for the process given in chapter~\ref{chap_h24nu},~\ref{chap_h22e2m}.
 \subsection{Self-energies}
 \label{equ_ren_slf_enrg}
 In the sections below, we have given the analytical expressions for self-energies of gauge bosons and Higgs boson. These self-energies are needed to calculate the renormalized constants given in Eq.~\ref{equ_ren_ren_cnst}. We denote Goldstone bosons by $G^{\pm,0}$ and ghost fields by $\eta^{\pm,Z,\gamma}$. All fermions loop diagrams have been shown with one generic diagram with the fermion line $f$ (and $f^\prime$). The momentum $p$ is the incoming and outgoing momenta for all self-energy diagrams in this section. We drop $\frac{\alpha}{2\pi}$ factor from each self-energy as it is a common factor for all. The self-energies are written in terms of tadpole, bubble scalar integrals and coefficients given in Eq.~\ref{equ_ren_b_int_dec}. In the self-energy expressions, we introduce the term $C^R_I$, which is equal to one in the 't Hooft-Veltman (HV) dimensional  scheme  and zero in the four-dimensional helicity scheme (FDH).
 \subsubsection{Photon self-energy :}
 \label{equ_ren_gam_se} 
\begin{figure}[h!]
 \begin{center}
\includegraphics [angle=0,width=0.7\linewidth]{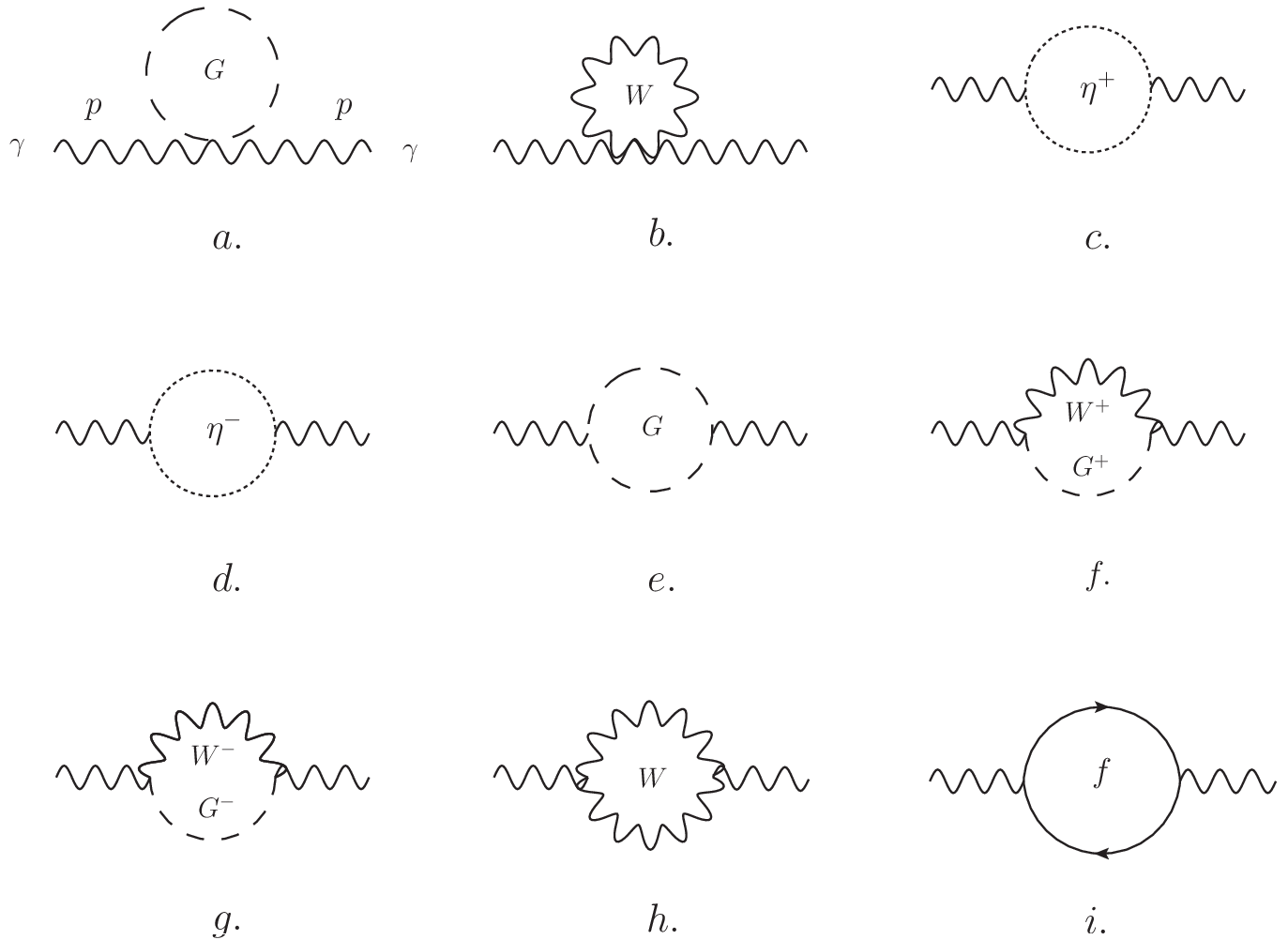}\\
	\caption{Photon self-energy diagrams at one loop.}
	\label{fig_ren_self_gam_gam}
	\end{center}
\end{figure}

   %
%
 We have listed the photon self-energy diagrams in Fig.~\ref{fig_ren_self_gam_gam}. The contributions for the self-energy diagram of photon from each diagram in Fig.~\ref{fig_ren_self_gam_gam} are listed below.
 %
 \begin{align}
 \Sigma_T^{AA}(p^2)=&\quad 2\times\Big[4A_0(M_W^2)-2M_W^2C_I^R\Big]\quad\quad\quad\quad\quad\quad\quad\quad\quad\quad\quad\:\:\:
 &&....\:({\text{dia. a and b}})\nonumber\\
 &+2\times B_{00}(p^2,M_W^2,M_W^2)&& ....\:({\text{dia. c and d}})\nonumber\\
 &-4B_{00}(p^2,M_W^2,M_W^2)&& ....\: ({\text{dia. e}})\nonumber\\
 &+2\times M_W^2B_{0}(p^2,M_W^2,M_W^2)&& .... \:({\text{dia. f and g}})\nonumber\\
&+\Big[-10B_{00}(p^2,M_W^2,M_W^2)-2A_0(M_W^2)-(2M_W^2+4p^2).\nonumber\\
 &\quad\quad\quad\quad B_0(p^2,M_W^2,M_W^2)+4(M_W^2-\frac{p^2}{6})C_I^R\Big]
&& ....\:({\text{dia. h}})\nonumber\\
 &+4Q_f^2\Big[\frac{1}{3}(p^2+2m_f^2)B_{0}(p^2,m_f^2,m_f^2)-\frac{p^2}{9}\nonumber\\
&\quad\quad\quad -\frac{2}{3}m_f^2 B_0(0,m_f^2,m_f^2)\Big]
&& ....\:({\text{dia. i}})\nonumber\\
 \end{align}
 \subsubsection{$\gamma\text{-} Z$ boson self-energy :}
 \label{equ_ren_self_a_z} 
\begin{figure}[h!]
 \begin{center}
\includegraphics [angle=0,width=0.7\linewidth]{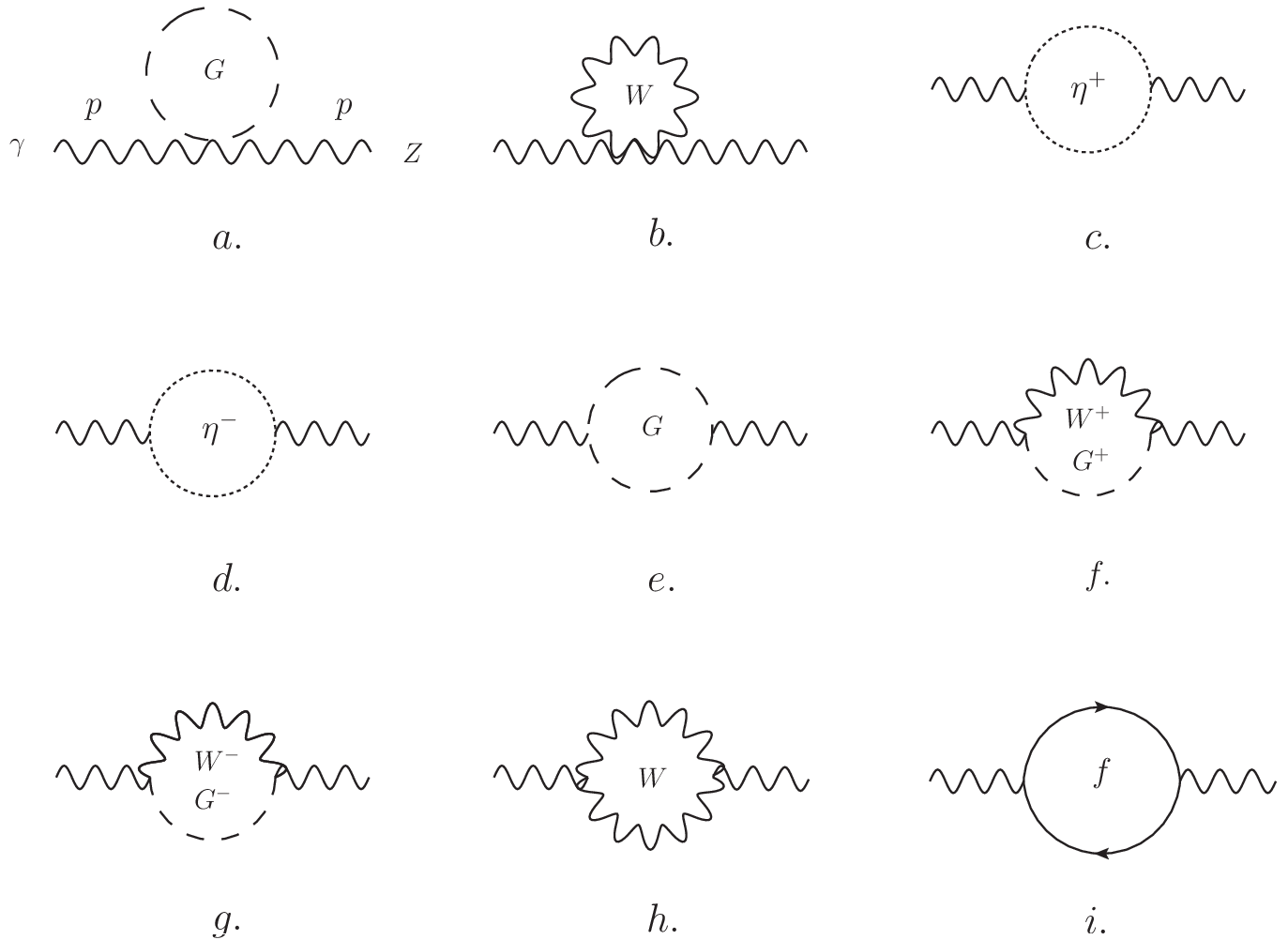}\\
	\caption{$\gamma{\text{-}Z}$ boson self-energy diagrams at one loop.}
	\label{fig_ren_self_gam_z}
	\end{center}
\end{figure}

 In Fig.~\ref{fig_ren_self_gam_z}, the $\gamma\text{-}Z$ boson type self-energy diagrams have been shown. The contributions from each diagram in Fig.~\ref{fig_ren_self_gam_z} have been listed below.
 \begin{align}
 \Sigma_T^{AZ}(p^2)=&-\frac{(2c_W^2-1)}{s_Wc_W}\:A_0(M_W^2)\quad\quad\quad\quad\quad\quad\quad\quad\quad\quad\quad\quad
 \quad\quad\quad\:\:&&
 ....\:({\text{dia. a}})\nonumber\\
 &-2\frac{c_W}{s_W} \Big(3A_0(M_W^2)-2M_W^2C_I^R\Big)&& ....\:({\text{dia. b}})\nonumber\\
 &-2\times \frac{c_W}{s_W}\:B_{00}(p^2,M_W^2,M_W^2)&& .... \:({\text{dia. c and d}})\nonumber\\
 &+2\frac{(2c_W^2-1)}{s_Wc_W}\:B_{00}(p^2,M_W^2,M_W^2)&& ....\: ({\text{dia. e}})\nonumber\\
 &+2\times M_W^2 \frac{s_W}{c_W}\:B_{0}(p^2,M_W^2,M_W^2)&& .... \:({\text{dia. f and g}})\nonumber\\
 &-\frac{c_W}{s_W}\Big[-10B_{00}(p^2,M_W^2,M_W^2)-2A_0(M_W^2)-(2M_W^2+4p^2).\nonumber\\
 &\quad\quad\quad\quad B_0(p^2,M_W^2,M_W^2)+4(M_W^2-\frac{p^2}{6})C_I^R\Big]
 && ....\:({\text{dia. h}})\nonumber\\
 &-\frac{2}{3}Q_f(g^+_f+g^-_f)\Big[(p^2+2m_f^2)B_{0}(p^2,m_f^2,m_f^2)-\frac{p^2}{3}\nonumber\\
&\quad\quad\quad -2m_f^2 B_0(0,m_f^2,m_f^2)\Big]
 && ....\:({\text{dia. i}})\nonumber\\
 \end{align}

  Values of $g_f^+$ and $g_f^-$ are given in the Sec.~\ref{subsec_ren_ct_dia_ew}.
\newpage
 \subsubsection{$Z$ boson self-energy :}
 \label{equ_ren_self_zz} 
   \begin{figure}[h!]
 \begin{center}
\includegraphics [angle=0,width=0.7\linewidth]{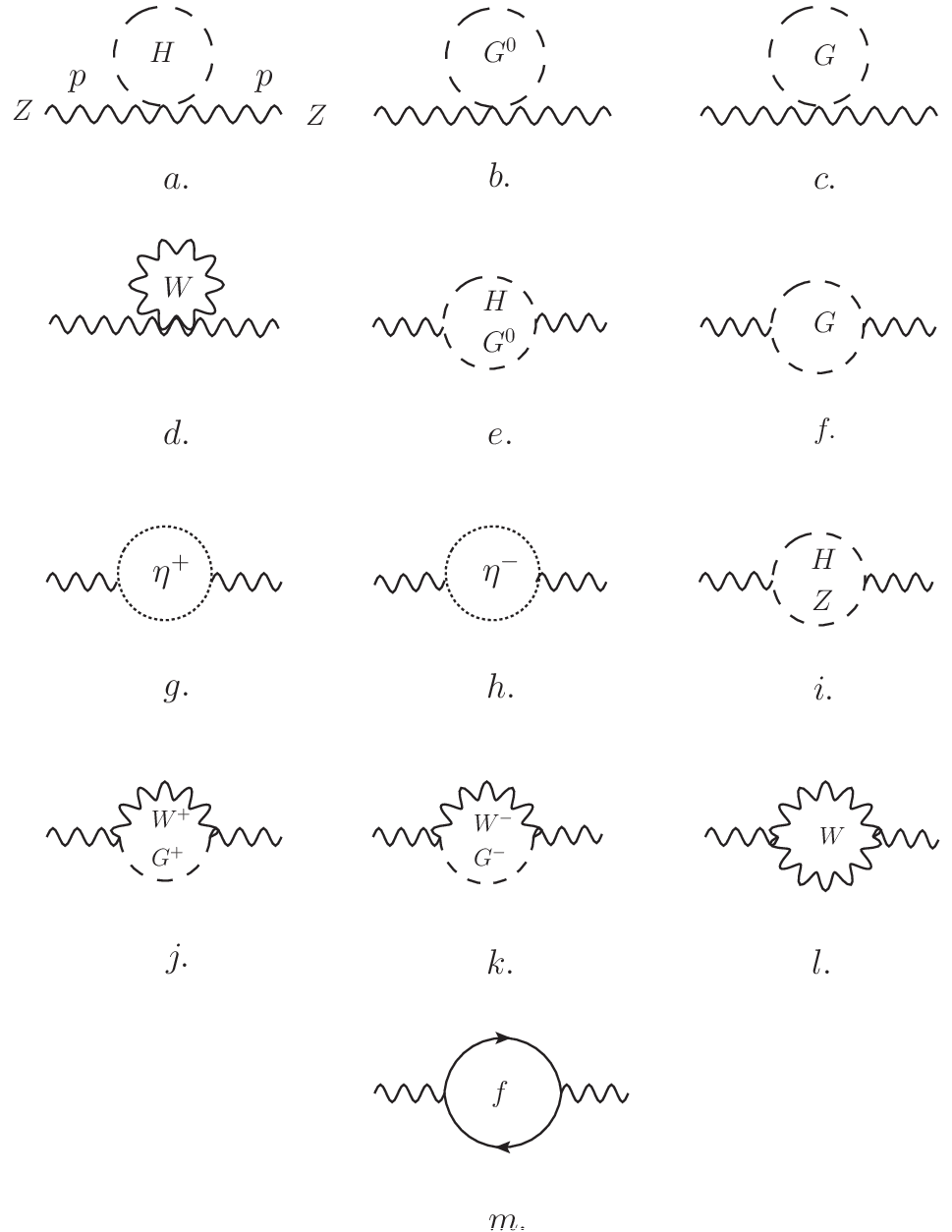}\\
	\caption{$Z$ boson self-energy diagrams at one loop.}
	\label{fig_ren_self_zz}
	\end{center}
\end{figure}
 We have listed the $Z$ boson self-energy diagrams in Fig.~\ref{fig_ren_self_zz}. The contributions for the $Z$ boson self-energy diagram from each diagram in Fig.~\ref{fig_ren_self_zz} are listed below 
 \begin{align}
 \Sigma_T^{ZZ}(p^2)=&\quad\frac{1}{4s_W^2c_W^2}\:A_0(M_H^2)\quad\quad\quad\quad\quad\quad\quad\quad\quad\quad\quad\quad
 \quad\quad\quad\quad\quad&&
 ....\:({\text{dia. a}})\nonumber\\
 &+\frac{1}{4s_W^2c_W^2}\:A_0(M_Z^2)&& ....\:({\text{dia. b}})\nonumber\\
 &+\frac{(2c_W^2-1)^2}{2s_W^2c_W^2}\:A_0(M_W^2)&& ....\: ({\text{dia. c}})\nonumber\\
 &+2\frac{c_W^2}{s_W^2}\:\Big(3A_0(M_W^2)-2M_W^2C_I^R\Big)&& .... \:({\text{dia. d}})\nonumber\\
 &-\frac{1}{s_W^2c_W^2}\:B_{00}(p^2,M_Z^2,M_H^2)&& .... \:({\text{dia. e}})\nonumber\\
  &-\frac{(2c_W^2-1)^2}{s_W^2c_W^2}\:B_{00}(p^2,M_W^2,M_W^2)&& .... \:({\text{dia. f}})\nonumber\\
 &+2\times \frac{c_W^2}{s_W^2}\:B_{00}(p^2,M_W^2,M_W^2)&& .... \:({\text{dia. g and h}})\nonumber\\
 &+\frac{M_Z^2}{s_W^2c_W^2}\:B_{0}(p^2,M_Z^2,M_H^2)&& .... \:({\text{dia. i}})\nonumber\\
 &+2\times M_Z^2s_W^2\:B_{0}(p^2,M_W^2,M_W^2)&& .... \:({\text{dia. j and k}})\nonumber\\
 &-\frac{c_W^2}{s_W^2}\Big[10B_{00}(p^2,M_W^2,M_W^2)+2A_0(M_W^2)+\Big(2M_W^2\nonumber\\
 &\quad\quad\quad +4p^2\Big)B_0(p^2,M_W^2,M_W^2)-4\Big(M_W^2-\frac{p^2}{6}\Big)C_I^R\Big]
 && ....\:({\text{dia. l}})\nonumber\\
 &-\Big[\frac{2}{3}\Big((g^+_f)^2+(g^-_f)^2\Big)\Big\{-\Big(p^2+2m_f^2\Big)B_{0}(p^2,m_f^2,m_f^2)+\frac{p^2}{3}\nonumber\\
&\quad\quad +2m_f^2 B_0(0,m_f^2,m_f^2)\Big\}+\frac{1}{2s_W^2c_W^2}m_f^2B_0(p^2,m_f^2,m_f^2)\Big]
 && ....\:({\text{dia. m}})\nonumber\\
 \end{align}
  \subsubsection{$W$ boson self-energy :}
 \label{equ_ren_self_ww} 
\begin{figure}[h!]
 \begin{center}
\includegraphics [angle=0,width=0.7\linewidth]{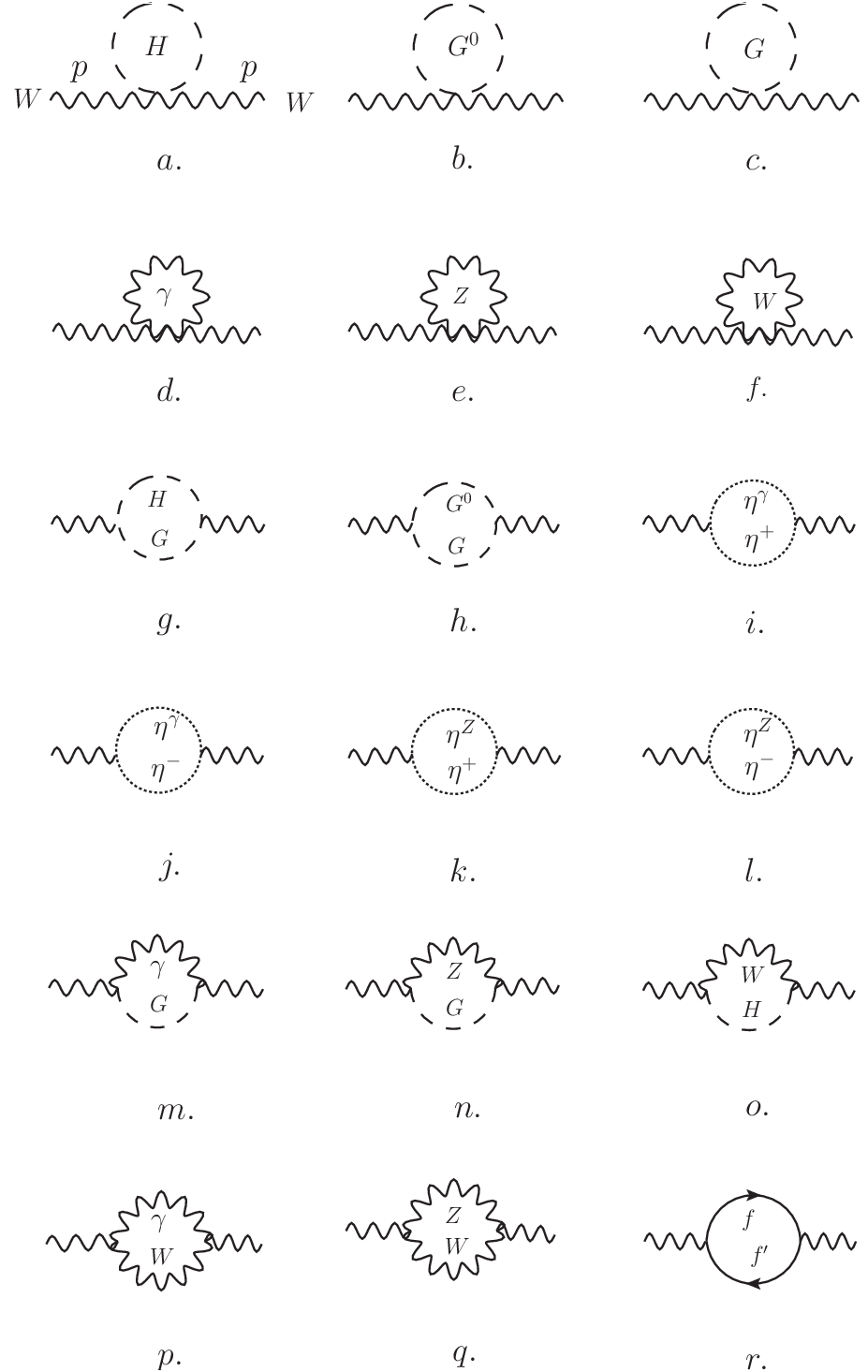}\\
	\caption{$W$ boson self-energy diagrams at one loop.}
	\label{fig_ren_self_ww}
	\end{center}
\end{figure}

 We have listed the $W$ boson self-energy diagrams in Fig.~\ref{fig_ren_self_ww}. The contributions from each diagram in Fig.~\ref{fig_ren_self_ww} are listed below.
 \newpage
 \begin{align}
 \Sigma_T^{W}(p^2)=&\quad\frac{1}{4s_W^2}\:A_0(M_H^2)\quad\quad\quad\quad\quad\quad\quad\quad\quad\quad\quad\quad\quad
 \quad\quad\quad\quad\quad\:&& ....\:({\text{dia. a}})\nonumber\\
 &+\frac{1}{4s_W^2}\:A_0(M_Z^2)&& ....\:({\text{dia. b}})\nonumber\\
 &+\frac{1}{2s_W^2}\:A_0(M_W^2)&& ....\: ({\text{dia. c}})\nonumber\\
 &+0&& ....\: ({\text{dia. d}})\nonumber\\
 &+\frac{c_W^2}{s_W^2}\:\Big(3A_0(M_Z^2)-2M_Z^2C_I^R\Big)&& .... \:({\text{dia. e}})\nonumber\\
 &+\frac{1}{s_W^2}\:\Big(3A_0(M_W^2)-2M_W^2C_I^R\Big)&& .... \:({\text{dia. f}})\nonumber\\
 &-\frac{1}{s_W^2}\:B_{00}(p^2,M_H^2,M_W^2)&& .... \:({\text{dia. g}})\nonumber\\
 &-\frac{1}{s_W^2}\:B_{00}(p^2,M_Z^2,M_W^2)&& .... \:({\text{dia. h}})\nonumber\\
  &+2\times B_{00}(p^2,M_W^2,0)&& .... \:({\text{dia. i and j}})\nonumber\\
 &+2\times \frac{c_W^2}{s_W^2}\:B_{00}(p^2,M_W^2,M_Z^2)&& .... \:({\text{dia. k and l}})\nonumber\\
 &+M_W^2\:B_{0}(p^2,0,M_W^2)&& .... \:({\text{dia. m}})\nonumber\\
 &+M_Z^2s_W^2\:B_{0}(p^2,M_Z^2,M_W^2)&& .... \:({\text{dia. n}})\nonumber\\
 &+\frac{M_W^2}{s_W^2}\:B_{0}(p^2,M_H^2,M_W^2)&& .... \:({\text{dia. o}})\nonumber\\
 &-\Big[10B_{00}(p^2,0,M_W^2)+2A_0(M_W^2)+2p^2B_1(p^2,0,M_W^2)\nonumber\\
 &\quad\quad\quad +5p^2B_0(p^2,0,M_W^2)-2\Big(M_W^2-\frac{p^2}{3}\Big)C_I^R\Big]
 && ....\:({\text{dia. p}})\nonumber\\
 &-\frac{c_W^2}{s_W^2}\Big[10B_{00}(p^2,M_W^2,M_Z^2)+2A_0(M_W^2)\nonumber\\
 &\quad\quad\quad +\Big(2M_Z^2+3p^2\Big)B_0(p^2,M_W^2,M_Z^2)-
 2p^2B_1(p^2,M_W^2,M_Z^2)\nonumber\\
 &\quad\quad\quad -\frac{2}{3}\Big(3M_W^2+3M_Z^2-p^2\Big)C_I^R\Big]
 && ....\:({\text{dia. q}})\nonumber\\
 &-\frac{1}{3s_W^2}\Big[-\Big(p^2-\frac{m_f^2+m_{f^\prime}^2}{2}\Big)B_{0}(p^2,m_f^2,m_{f^\prime}^2)+\frac{p^2}{3}\nonumber\\
&\quad\quad +\Big(m_f^2 B_0(0,m_f^2,m_f^2)+m_{f^\prime}^2 B_0(0,m_{f^\prime}^2,m_{f^\prime}^2)\Big)\nonumber\\
&+\frac{(m_f^2-m_{f^\prime}^2)}{2p^2}\Big(B_0(p^2,m_f^2,m_{f^\prime}^2)-B_0(0,m_f^2,m_{f^\prime}^2\Big)\Big]
 && ....\:({\text{dia. r}})\nonumber\\
 \end{align}
 \newpage
   \subsubsection{Higgs boson self-energy :}
 \label{subsec_ren_self_hh}  
\begin{figure}[h!]
 \begin{center}
\includegraphics [angle=0,width=0.7\linewidth]{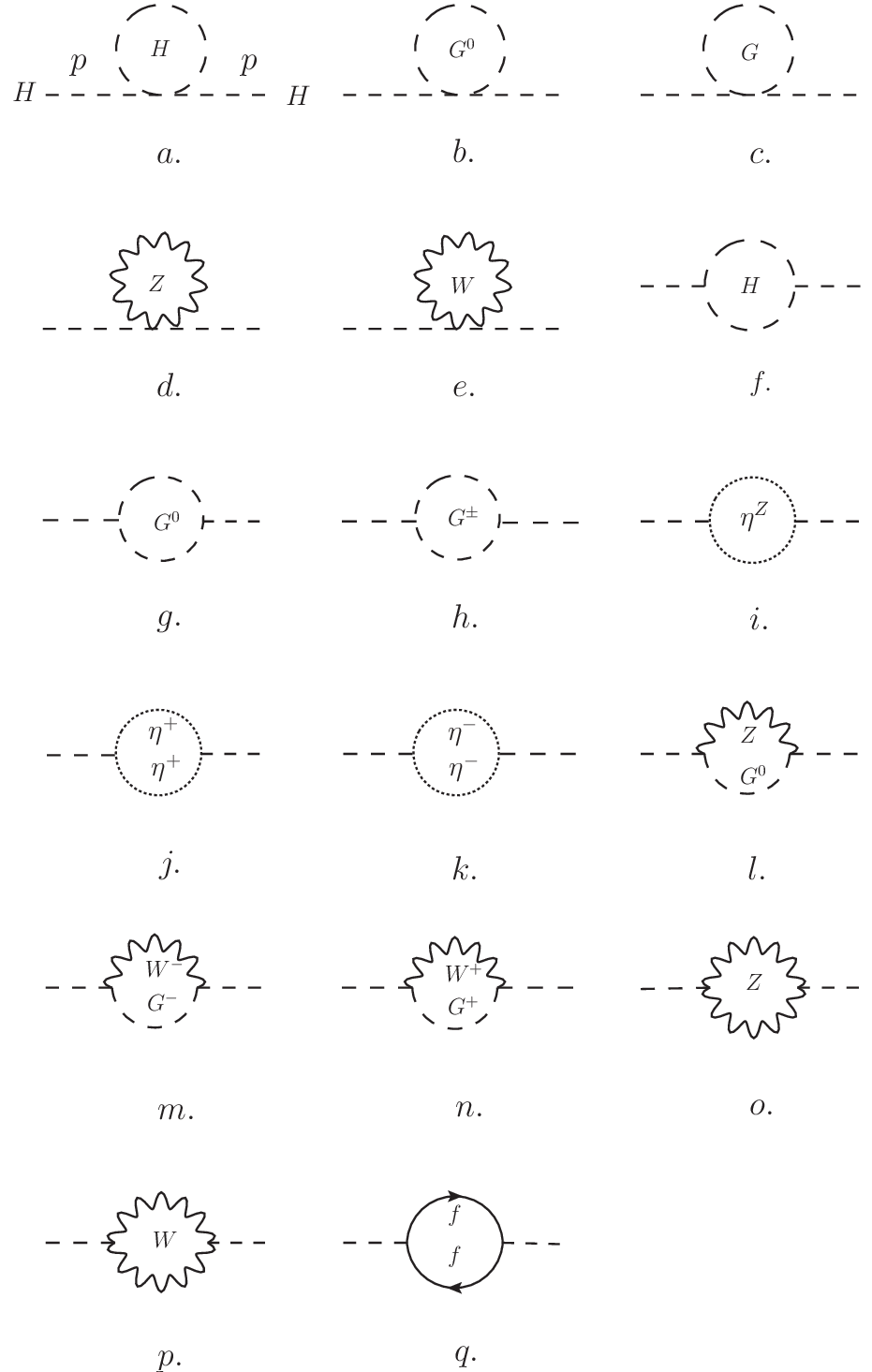}\\
	\caption{$H$ boson self-energy diagrams at one loop.}
	\label{fig_ren_self_hh}
	\end{center}
\end{figure}

  We have listed the $H$ boson self-energy diagrams in Fig.~\ref{fig_ren_self_hh}. The contributions from each diagram in Fig.~\ref{fig_ren_self_hh} are listed below.
 \begin{align}
 \Sigma^{H}(p^2)=&\quad\frac{3M_H^2}{8s_W^2M_W^2}\:A_0(M_H^2)\quad\quad\quad\quad\quad\quad\quad\quad\quad\quad\quad\quad\quad
 \quad\quad\quad\:\:&& ....\:({\text{dia. a}})\nonumber\\
 &+\frac{M_H^2}{8s_W^2M_W^2}\:A_0(M_Z^2)&& ....\:({\text{dia. b}})\nonumber\\
 &+\frac{M_H^2}{4s_W^2M_W^2}\:A_0(M_W^2)&& ....\: ({\text{dia. c}})\nonumber\\
 &+\frac{M_H^2}{4s_W^2c_W^2M_W^2}\:\Big(4A_0(M_Z^2)-2M_Z^2C_I^R\Big)&& .... \:({\text{dia. d}})\nonumber\\
 &+\frac{1}{2s_W^2}\:\Big(4A_0(M_W^2)-2M_W^2C_I^R\Big)&& .... \:({\text{dia. e}})\nonumber\\
 &+\frac{9M_H^4}{8s_W^2M_W^2}\:B_{0}(p^2,M_H^2,M_H^2)&& .... \:({\text{dia. f}})\nonumber\\
 &+\frac{M_H^4}{8s_W^2M_W^2}\:B_{0}(p^2,M_Z^2,M_Z^2)&& .... \:({\text{dia. g}})\nonumber\\
  &+\frac{M_H^4}{4s_W^2M_W^2}\:B_{0}(p^2,M_W^2,M_W^2)&& .... \:({\text{dia. h}})\nonumber\\
 &-\frac{M_Z^2}{4s_W^2c_W^2}\:B_{0}(p^2,M_Z^2,M_Z^2)&& .... \:({\text{dia. i}})\nonumber\\
  &-2\times\frac{M_W^2}{4s_W^2}\:B_{0}(p^2,M_W^2,M_W^2)&& .... \:({\text{dia. j and k}})\nonumber\\
 &-\frac{1}{4s_W^2c_W^2}\:\Big(2p^2B_{0}(p^2,M_Z^2,M_Z^2)+A_0(M_Z^2)\nonumber\\
 &\quad\quad\quad\quad +M_Z^2B_{0}(p^2,M_Z^2,M_Z^2)\Big)&& .... \:({\text{dia. l}})\nonumber\\
 &-2\times \frac{1}{4s_W^2}\:\Big(2p^2B_{0}(p^2,M_W^2,M_W^2)+A_0(M_W^2)\nonumber\\
 &\quad\quad\quad\quad +M_W^2B_{0}(p^2,M_W^2,M_W^2)\Big)&& .... \:({\text{dia. m and n}})\nonumber\\
 &+\frac{M_Z^2}{2s_W^2c_W^2}\:\Big(4B_0(p^2,M_Z^2,M_Z^2)-2C_I^R\Big)&& .... \:({\text{dia. o}})\nonumber\\
 &+\frac{M_W^2}{s_W^2}\:\Big(4B_0(p^2,M_W^2,M_W^2)-2C_I^R\Big)&& .... \:({\text{dia. p}})\nonumber\\
 &-\frac{N_cm_f^2}{2s_W^2M_W^2}\Big[2A_0(m_f^2)+(4m_f^2-p^2)B_0(p^2,m_f^2,m_f^2)\Big]
 && ....\:({\text{dia. q}})\nonumber\\
 \end{align}
 \newpage
 \subsubsection{Fermion self-energy :}
 \label{subsec_ren_self_fermion}  
 
   \begin{figure}[h!]
 \begin{center}
\includegraphics [angle=0,width=0.7\linewidth]{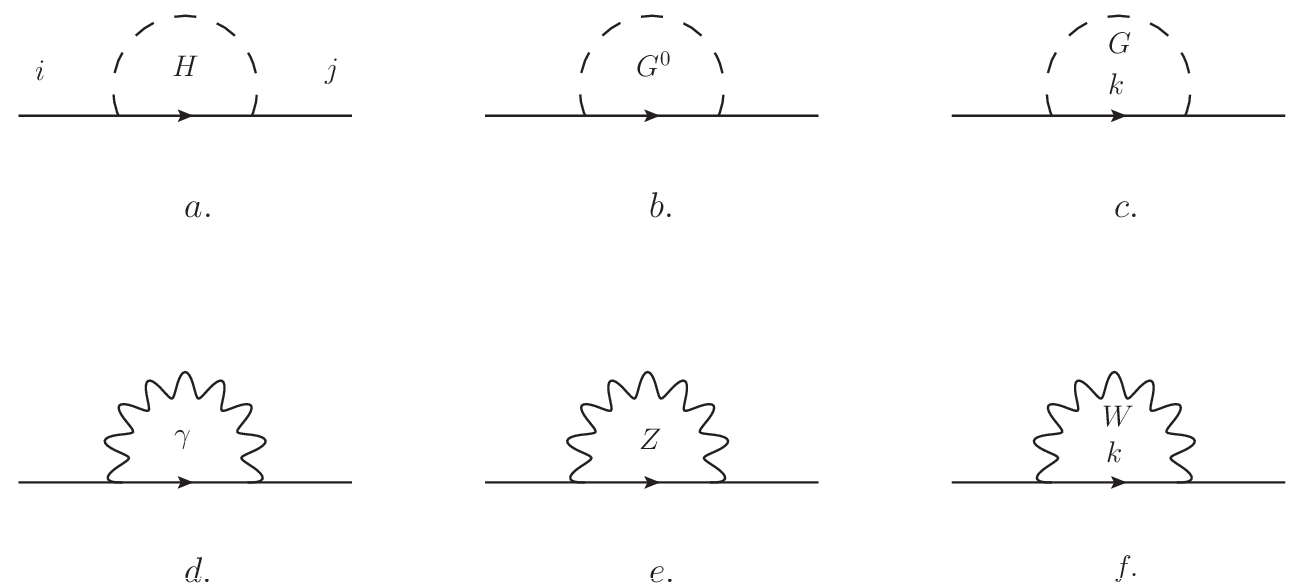}\\
	\caption{Fermion self-energy diagrams at one loop.}
	\label{fig_ren_self_fermion}
	\end{center}
\end{figure}
 We have listed the fermion self-energy diagrams in Fig.~\ref{fig_ren_self_fermion}. The fermion self-energy can be written in three components as 
 \begin{align}
 \Sigma^f_{ij}(p^2)=\slashed p L \Sigma^{f,L}_{i,j}(p^2)+\slashed p R \Sigma^{f,R}_{i,j}(p^2)+(m_{f,i}L+m_{f,j}R)\Sigma^{f,S}_{i,j}(p^2)\:,
 \label{equ_ren_self_fr_split}
 \end{align}
 where $L$ and $R$ are defined as $L=\frac{1-\gamma^5}{2}$, $R=\frac{1+\gamma^5}{2}$. The contributions from each diagram in Fig.~\ref{fig_ren_self_fermion} are listed below. 
\begin{align}
 \Sigma^{f}_{ij}(p^2)=&-\frac{m_f^2}{4s_W^2M_W^2}\:\delta_{ij}\:
 \Big[\slashed{p}LB_1(p^2,m_f^2,M_H^2)\nonumber\\
 &\quad\quad +\slashed{p}RB_1(p^2,m_f^2,M_H^2)-m_fB_0(p^2,m_f^2,m_H^2)\Big]
 \quad\quad\quad\quad\quad\quad\quad\quad\quad\:&& ....\:({\text{dia. a}})\nonumber\\
 &-\quad\frac{m_f^2}{4s_W^2M_W^2}\:\delta_{ij}\:
 \Big[\slashed{p}LB_1(p^2,m_f^2,M_Z^2)\nonumber\\
 &\quad\quad +\slashed{p}RB_1(p^2,m_f^2,M_Z^2)+m_fB_0(p^2,m_f^2,M_Z^2)\Big]
 && ....\:({\text{dia. b}})\nonumber\\
 &-\quad\frac{1}{2s_W^2M_W^2}\:\sum_k V_{ik}V^{*}_{kj}\:
 \Big[\slashed{p}\Big(L\:m_{f^\prime,k}^2+R\:m_{f,i}\:m_{f,j}\Big)B_1(p^2\nonumber\\
 &\quad\quad\quad ,m_{f^\prime,k}^2,M_W^2)+m_{f^\prime,k}^2B_0(p^2,m_{f^\prime,k}^2,M_W^2)\Big(m_{f,i}L+m_{f,j}R\Big)\Big]&& ....\:({\text{dia. c}})\nonumber\\
&-\quad Q_f^2\:\delta_{ij}\:
 \Big[\slashed{p}L\Big(2B_1(p^2,m_f^2,0)+C_I^R\Big)\nonumber\\
 &\quad\quad\quad +\slashed{p}R\Big(2B_1(p^2,m_f^2,0)+C_I^R\Big)+2m_f\Big(2B_0(p^2,m_f^2,0)-C_I^R\Big)\Big]
 && ....\:({\text{dia. d}})\nonumber\\
 &-\quad \delta_{ij}\:
 \Big[\slashed{p}L(g_f^-)^2\Big(2B_1(p^2,m_f^2,M_Z^2)+C_I^R\Big)+ \slashed{p}R(g_f^+)^2\Big(2B_1(p^2,m_f^2\nonumber\\
 &\quad\quad\quad ,M_Z^2)+C_I^R\Big)+g_f^- g_f^+ m_f\Big(4B_0(p^2,m_f^2,M_Z^2)-2C_I^R\Big)\Big]
 && ....\:({\text{dia. e}})\nonumber\\
 &-\quad\frac{1}{2s_W^2}\:\sum_k V_{ik}V^{*}_{kj}\:
 \slashed{p}L\Big(2B_1(p^2,m_{f^\prime,k}^2,M_W^2)+C_I^R\Big)
&& ....({\text{dia. f}})\nonumber\\
\label{equ_ren_self_fermion}
\end{align}
One can identify $\Sigma^{f,L}_{i,j}$, $\Sigma^{f,R}_{i,j}$ and $\Sigma^{f,S}_{i,j}$ in Eq.~\ref{equ_ren_self_fermion} with the help of the Eq.~\ref{equ_ren_self_fr_split}, which are relevant to calculate the renormalization constants given in Eq.~\ref{equ_ren_ren_cnst}.
 %
 \subsection{Counterterms}
 \label{subsec_ren_ct_dia_ew}  
 In this section, we have given a few counterterm diagrams with their Feynman rules. These counterterm diagrams are relevant for the process given in chapter~\ref{chap_h24nu} and~\ref{chap_h22e2m}.
   \begin{figure}[h!]
 \begin{center}
\includegraphics [angle=0,width=0.8\linewidth]{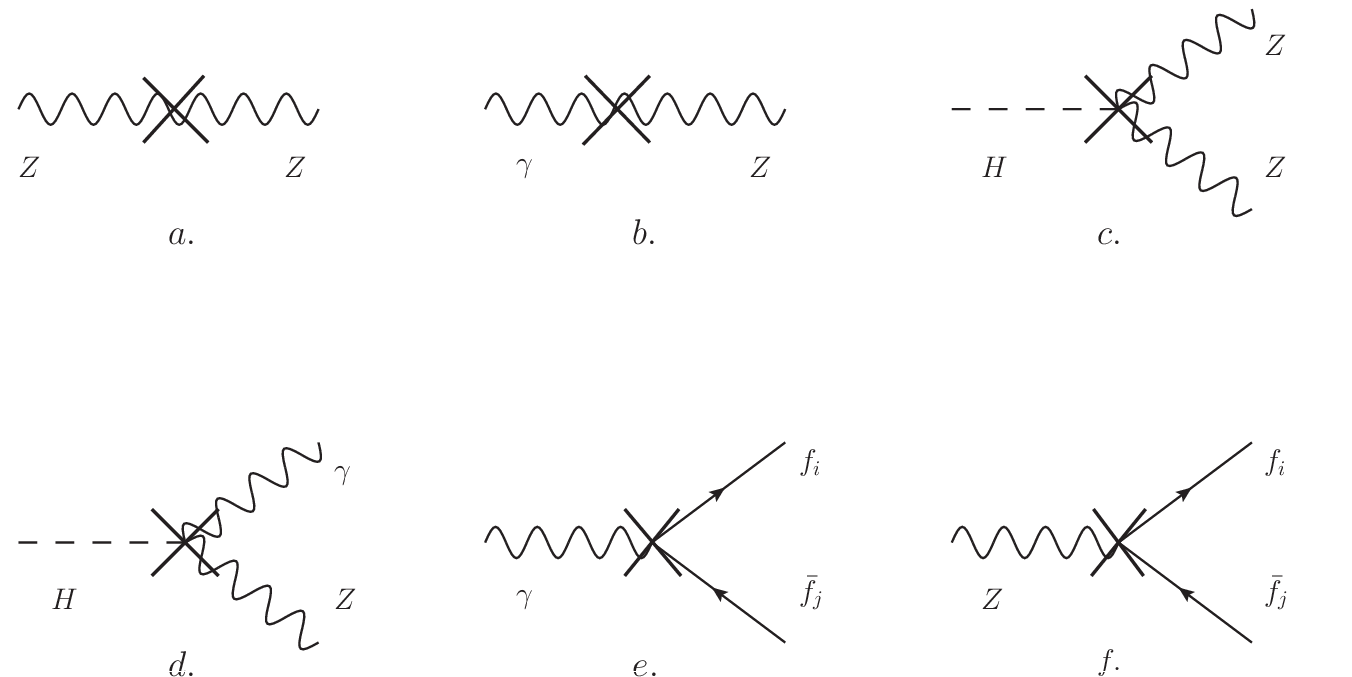}\\
	\caption{Counterterms for self-energy and vertex diagrams.}
	\label{fig_ren_ct_dia_ew}
	\end{center}
\end{figure}
\begin{align}
&{\text{Dia. a : }} SE^{CT}_{ZZ}\equiv -ig^{\mu\nu}\Big[p^2\delta Z_{ZZ}-\Big(M_Z^2\delta_{ZZ}+\delta M_Z^2\Big)\Big]\nonumber\\
&{\text{Dia. b : }} SE^{CT}_{\gamma Z}\equiv -ig^{\mu\nu}\frac{1}{2}\Big[p^2\Big(\delta Z_{AZ}+\delta Z_{ZA}\Big)-M_Z^2\delta Z_{ZA}\Big]\nonumber\\
&{\text{Dia. c : }} V^{CT}_{HZZ}\equiv ieg^{\mu\nu}\frac{M_W}{s_Wc_W^2}\Big[\delta Z_e+\frac{2s_W^2-c_W^2}{c_W^2}\frac{\delta s}{s}+\frac{1}{2}\frac{\delta M_W^2}{M_W^2}+\frac{1}{2}\delta Z_H+\delta Z_{ZZ}\Big]\nonumber\\
&{\text{Dia. d : }} V^{CT}_{HZ\gamma}\equiv ieg^{\mu\nu}\frac{M_W}{s_Wc_W^2}\frac{1}{2}\delta Z_{ZA}\nonumber\\
&{\text{Dia. e : }} V^{CT}_{\gamma f_i\bar{f}_{j}}\equiv -ie\gamma^\mu\Big[\omega_+\Big\{Q_f\Big[\delta_{ij}\Big(\delta Z_e+\frac{1}{2}\delta Z_{AA}\Big)+\frac{1}{2}\Big(\delta Z^{f,R}_{ij}+\delta Z^{f,R\dag}_{ij}\Big)\Big]+\delta_{ij}g_f^+\frac{1}{2}\delta Z_{ZA}\Big\}\nonumber\\
&\quad\quad\quad\quad\quad\quad
\quad\quad\quad+\omega_-\Big\{Q_f\Big[\delta_{ij}\Big(\delta Z_e+\frac{1}{2}\delta Z_{AA}\Big)+\frac{1}{2}\Big(\delta Z^{f,L}_{ij}+\delta Z^{f,L\dag}_{ij}\Big)\Big]+\delta_{ij}g_f^-\frac{1}{2}\delta Z_{ZA}\Big\}\Big]\nonumber\\
&{\text{Dia. f : }} V^{CT}_{Z f_i\bar{f}_{j}}\equiv ie\gamma^\mu\Big[\omega_+\Big\{g_f^+\Big[\delta_{ij}\Big(\frac{\delta g_f^+}{g_f^+}+\frac{1}{2}\delta Z_{ZZ}\Big)+\frac{1}{2}\Big(\delta Z^{f,R}_{ij}+\delta Z^{f,R\dag}_{ij}\Big)\Big]-\delta_{ij}Q_f\frac{1}{2}\delta Z_{AZ}\Big\}\nonumber\\
&\quad\quad\quad\quad\quad\quad
\quad\quad+\omega_-\Big\{g_f^-\Big[\delta_{ij}\Big(\frac{\delta g_f^-}{g_f^-}+\frac{1}{2}\delta Z_{ZZ}\Big)+\frac{1}{2}\Big(\delta Z^{f,L}_{ij}+\delta Z^{f,L\dag}_{ij}\Big)\Big]-\delta_{ij}Q_f\frac{1}{2}\delta Z_{AZ}\Big\}\Big]\nonumber\\
\end{align}
where
\begin{align}
g_f^+=-\frac{s_W}{c_W}Q_f\:,&\quad \delta g_f^+=-\frac{s_W}{c_W}Q_f\Big[\delta Z_e+\frac{1}{c_W^2}\frac{\delta s}{s}\Big]\:,\nonumber\\
g_f^-=-\frac{I^3_{W,f}-s^2Q_f}{s_Wc_W}\:,&\quad \delta g_f^-=-\frac{I^3_{W,f}}{s_Wc_W}\Big[\delta Z_e+\frac{s_W^2-c_W^2}{c_W^2}\frac{\delta s}{s}\Big]+\delta g_f^+\:
\end{align}
\chapter{IR divergences and Dipole subtraction}
\label{chap_ir_div_dp_sub}
 In the previous chapter, we have discussed the loop integrals that appear in the Feynman amplitudes at one loop. We have discussed the source of UV divergences in the loop integrals.
 In this chapter, we will discuss the infrared (IR) singularities that appear in a perturbative computation. For an inclusive process, the IR singularities appear in virtual as well as in real emission amplitudes in certain momenta regions.
 The IR singularities appear in a loop integral when a massless propagator becomes on-shell for certain loop momentum, whereas in real emission diagrams, the IR singularities appear in soft and collinear regions.
 These divergences are regularized by the small masses $m$ in mass regularization and they appear as $\text{ln}(m/Q)$, where $Q$ is a large scale. For the  massless case ($m=0$), the IR singularities appear as $\frac{1}{\epsilon}$ in dimensional regularization where $\epsilon=(4-D)/2$.
 According to Kinoshita-Lee-Nauenberg (KLN) theorem~\cite{Kinoshita:1962ur,Lee:1964is}, the singularities completely cancel order by order in a sufficiently inclusive process. As we see, the IR singularities from virtual diagrams cancel the singularities from real emission diagrams.
 For a Drell-Yan process, the phase space integral for the real emission diagram is not that challenging. Calculating the LHC processes, which lead to the final state with more than two particles, become very complicated to perform the phase space as well as the loop integrals. For such processes, many techniques~\cite{Harris:2001sx,Frederix:2009yq} have been developed to handle the IR singularities in a perturbative computation.
 We use {\it dipole subtraction} formalism to remove IR singularities from inclusive cross sections and decay widths. In this chapter, we discuss the sources of IR singularities in loop integrals and in real emission diagrams, and different aspects of the {\text dipole subtraction} procedure etc..
 
 \section{IR singularities in one-loop integrals}
 \label{sec_ir_loop_int}
 
\begin{figure}[h!]
 \begin{center}
\includegraphics [angle=0,width=0.5\linewidth]{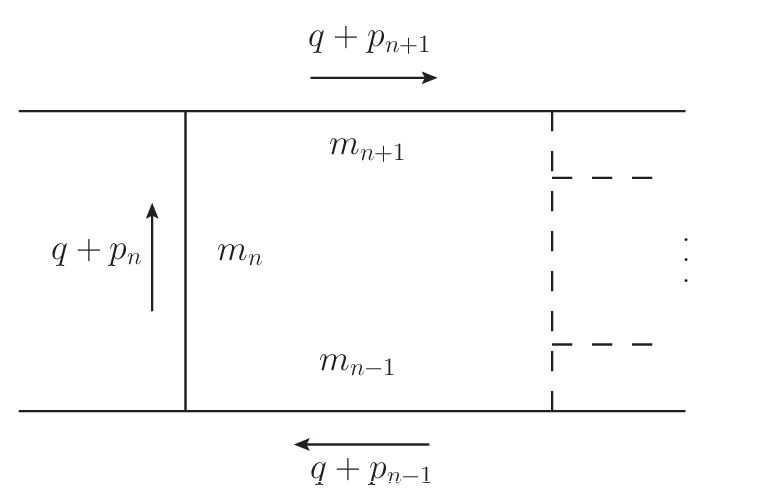}\\
	\caption{A generic one-loop $N$-point integral with masses and momenta}
	\label{fig_idd_ir_loop}
	\end{center}
\end{figure}

We have shown a diagrammatic representation of a generic one-loop integral in Fig.~\ref{fig_idd_ir_loop}. The generic one-loop $N$-point integral in $D$-dimension is given by
 \begin{align}
 T^{(N)}_{\mu_1...\mu_P}(p_0, ... ,p_{N-1},m_0, ... ,m_{N-1})=\frac{(2\pi\mu)^{(4-D)}}{i\pi^2}\int d^Dq\frac{q_{\mu_1} ... q_{\mu_P}}{N_0 ... N_{N-1}}\:,
 \label{equ_idd_ol_int}
 \end{align}
 with the denominator factors
 \begin{equation}
 N_n=(q+p_n)^2-m_n^2+i0\:,\quad n=0,...,N-1\:.
 \end{equation}
 Here $q$ is the loop momentum, $p_n$ forms external momenta and $m_n$ are the masses of internal propagators. Here we have not set the $p_0$ to zero for the generic treatment of related integrals.

 ``Mass" singularities appear in a loop-integral when the internal masses and the external momenta $(p_{n+1}-p_n)^2$ become small. Here we are not interested in specific or isolated points in phase space where the singularities may appear from the threshold effect.
  The threshold effect is discussed in Sec.~\ref{subsec:renorm_cms}. The ``mass" singularities can appear in one-loop diagrams in two situations. These singularities have been discussed in detail in the following subsections. They are known as collinear and soft singularities in virtual diagrams.

\subsection{Collinear Singularity}
\label{sub_sec_idd_col_sng}
  When an external (on-shell) massless particle is attached to two massless propagators in the loop, then there will be a singularity that appears in the loop integral. There will be a $n$ for which we write this condition as 
  \begin{equation}
  (p_{n+1}-p_n)^2\rightarrow 0 \:,\quad m_{n+1}\rightarrow 0\:,\quad m_n\rightarrow 0\:.
  \label{equ_idd_col_cnd}
  \end{equation}

   As we will see, the singularity is logarithmic. The singularity originates from the loop integration momenta $q$ with
   \begin{equation}
   q\rightarrow -p_n+x_n(p_n-p_{n+1})\:, 
   \end{equation}
   where $x_n$ is an arbitrary real variable. As the loop momentum $(q+p_n)$ of $n^{\text {th}}$ line of the integral given in Eq.~\ref{equ_idd_ol_int} is proportional to the external line of momentum $(p_n-p_{n+1})$, such singularities are called collinear singularity.
   In this region (collinear), we can see the logarithmic singular behavior of the $N$-point scalar integral given in Eq.~\ref{equ_idd_ol_int}. 
   The $N$-point scalar integral can be written as 
   \begin{align}
 T^{(N)}_{0}(p_0,...,p_{N-1},m_0,...,m_{N-1})=&\:\frac{(2\pi\mu)^{(4-D)}}{i\pi^2}\int d^Dq\frac{1}{N_0...N_{N-1}}\nonumber\\
 =&\:\frac{(2\pi\mu)^{(4-D)}}{i\pi^2}\int d^Dq\frac{1}{N_0...N_{n}N_{n+1}...N_{N-1}}\:.
 \label{equ_idd_ol_int_sclr} 
   \end{align}
   We introduce a infinitesimal momentum $\epsilon_\perp$ perpendicular to the external momentum $(p_{n}-p_{n+1})$ in the loop momentum of $n^{\text {th}}$ line as 
   \begin{equation}
   q+p_n=x_n(p_n-p_{n+1})+\epsilon_\perp\:.
   \end{equation}
   
   In this regime, the propagators $N_n$ and $N_{n+1}$ become
   \begin{align}
   N_n&=(q+p_n)^2-m_n^2\nonumber\\
   &=\{x_n(p_n-p_{n+1})+\epsilon_\perp\}^2-m_n^2\nonumber\\
   &=x_n^2(p_n-p_{n+1})^2+2x_n(p_n-p_{n+1}).\epsilon_\perp+\epsilon_\perp^2-m_n^2\:,\nonumber\\
   N_{n+1}&=(q+p_{n+1})^2-m_{n+1}^2\nonumber\\
   &=\{(q+p_n)-(p_n-p_{n+1})\}^2-m_{n+1}^2\nonumber\\
   &=\{(x_n-1)(p_n-p_{n+1})+\epsilon_\perp\}^2-m_{n+1}^2\nonumber\\
   &=(x_n-1)^2(p_n-p_{n+1})^2+2(x_n-1)(p_n-p_{n+1}).\epsilon_\perp+\epsilon_\perp^2-m_{n+1}^2\:.
   \end{align}
   With the condition (collinear) given in Eq.~\ref{equ_idd_col_cnd} and $(p_n-p_{n+1}).\epsilon_\perp=0$ (perpendicular momenta), the propagators $N_n$ and $N_{n+1}$ become
   \begin{equation}
   N_n\simeq \epsilon_\perp^2\xrightarrow {\epsilon_\perp\rightarrow 0} 0\:,\quad N_{n+1}\simeq \epsilon_\perp^2\xrightarrow {\epsilon_\perp\rightarrow 0} 0\:.
   \end{equation}
Other propagators are non-singular in the collinear region. The $N$-point scalar integral in Eq.~\ref{equ_idd_ol_int_sclr} becomes
\begin{align}
 T^{(N)}_{0}(p_0,...,p_{N-1},m_0,...,m_{N-1})\simeq&\:\int d^Dq(\epsilon_\perp)\frac{1}{N_0...N_{n}(\epsilon_\perp)N_{n+1}(\epsilon_\perp)...N_{N-1}}\nonumber\\
 \simeq&\:\int d^D\epsilon_\perp\frac{1}{N_0...\epsilon_\perp^2\epsilon_\perp^2...N_{N-1}}\nonumber\\
 \simeq&\:\int d^D\epsilon_\perp\frac{1}{\epsilon_\perp^4}\:.
 \label{equ_idd_int_sclr_col} 
\end{align}
This integral is logarithmic divergent in $D=4$ dimension.
\subsection{Soft Singularity}
\label{sub_sec_idd_sft_sng}
   When a massless particle is exchanged between two on-shell particles, then there will be a singularity that appears in a loop integral. There will be a $n$ that satisfies this condition for which we can write
   \begin{equation}
   m_n\rightarrow 0\:,\quad (p_{n-1}-p_n)^2-m_{n-1}^2\rightarrow 0\:,\quad (p_{n+1}-p_n)^2-m_{n+1}^2\rightarrow 0\:.
   \label{equ_idd_sft_cnd}
   \end{equation}
This singularity is also logarithmic, like the collinear singularity. This singularity originates from the loop momentum region
\begin{equation}
q\rightarrow -p_n\:.
\end{equation}
  As the momentum transfer $(q+p_n)$ in the $n^{\text{th}}$ line in the loop diagram given in Fig.~\ref{fig_idd_ir_loop} is zero, the singularity is called {\it{soft singularity}}.
  In this region (soft), we can see the logarithmic singular behavior of the $N$-point scalar integral given in Eq.~\ref{equ_idd_ol_int}. 
  The $N$-point scalar integral can be written as 
   \begin{align}
 T^{(N)}_{0}(p_0,...,p_{N-1},m_0,...,m_{N-1})=&\:\frac{(2\pi\mu)^{(4-D)}}{i\pi^2}\int d^Dq\frac{1}{N_0...N_{N-1}}\nonumber\\
 =&\:\frac{(2\pi\mu)^{(4-D)}}{i\pi^2}\int d^Dq\frac{1}{N_0...N_{n-1}N_{n}N_{n+1}...N_{N-1}}\:.
 \label{equ_idd_ol_int_slr_soft} 
   \end{align}
   We introduce a infinitesimal momentum $\epsilon^\prime$ to the loop momentum $(q+p_n)$ in the soft region as
   \begin{equation}
   q+p_n=\epsilon^\prime\:.
   \end{equation}
   In the soft region, the propagators $N_{n-1}$, $N_n$ and $N_{n+1}$ become
   \begin{align}
   N_{n-1}&=(q+p_{n-1})^2-m_{n-1}^2\nonumber\\
   &=(p_{n-1}-p_n+\epsilon^\prime)^2-m_{n-1}^2\nonumber\\
   &=(p_{n-1}-p_n)^2+2(p_{n-1}-p_n).\epsilon^\prime+{\epsilon^\prime}^2-m_{n-1}^2\:,\nonumber\\
   N_{n}\quad&=(q+p_{n})^2-m_{n}^2\nonumber\\
   &={\epsilon^\prime}^2-m_{n}^2\:,\nonumber\\
   N_{n+1}&=(q+p_{n+1})^2-m_{n+1}^2\nonumber\\
   &=(p_{n+1}-p_n+\epsilon^\prime)^2-m_{n+1}^2\nonumber\\
   &=(p_{n+1}-p_n)^2+2(p_{n+1}-p_n).\epsilon^\prime+{\epsilon^\prime}^2-m_{n+1}^2\:.
   \end{align}

   With the condition (soft) given in Eq.~\ref{equ_idd_sft_cnd} and considering the leading power, these propagators become 
   \begin{equation}
   N_{n-1}\simeq \epsilon^\prime\xrightarrow {\epsilon^\prime\rightarrow 0} 0\:,\quad N_n\simeq {\epsilon^\prime}^2\xrightarrow {\epsilon^\prime\rightarrow 0} 0\:,\quad N_{n+1}\simeq \epsilon^\prime\xrightarrow {\epsilon^\prime\rightarrow 0} 0\:.
   \end{equation}
   All other propagators are non-singular in the soft region. The $N$-point scalar integral given in Eq.~\ref{equ_idd_ol_int_slr_soft} can be written as
   \begin{align}
 T^{(N)}_{0}(p_0,...,p_{N-1},m_0,...,m_{N-1})\simeq&\:\int d^Dq(\epsilon^\prime)\frac{1}{N_0...N_{n-1}(\epsilon^\prime)N_{n}(\epsilon^\prime)N_{n+1}(\epsilon^\prime)...N_{N-1}}\nonumber\\
 \simeq&\:\int d^D\epsilon^\prime\frac{1}{N_0...\epsilon^\prime.{\epsilon^\prime}^2.{\epsilon^\prime}...N_{N-1}}\nonumber\\
 \simeq&\:\int d^D\epsilon^\prime\frac{1}{{\epsilon^\prime}^4}\:.
 \label{equ_idd_int_sclr_col} 
\end{align}
This integral is logarithmic divergent in $D=4$ dimension.
  
  We have seen that depending on the scenario (Eq.~\ref{equ_idd_col_cnd},~\ref{equ_idd_sft_cnd}) one-loop integrals give rise to singularities that are logarithmic in nature. These singularities appear as $\frac{1}{\epsilon}$ in dimensional regularization, where $\epsilon=(4-D)/2$.
   If there is a common region where the collinear and soft conditions are satisfied, then one can get the singularities as $\frac{1}{\epsilon^2}$ in dimensional regularization.
   One can check the singular structure of any loop diagram for a given process with these conditions. 
   
\section{IR singularities in real emission diagrams}
\label{sec_idd_ir_sng_rlem_dia}
\begin{figure}[h!]
 \begin{center}
\includegraphics [angle=0,width=0.4\linewidth]{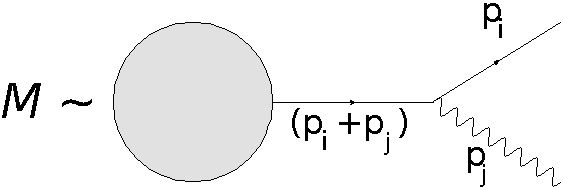}\\
	\caption{A generic real emission diagram where a fermion spit into a massless gauge boson and a fermion.}
	\label{fig_idd_ir_rem}
	\end{center}
\end{figure}

  From KLN theorem, the IR singularities must cancel order by order in a perturbative computation. For an inclusive process, the real emission diagrams must be included for fixed order computation to get the right prediction of physical observables. 
  In Fig.~\ref{fig_idd_ir_rem}, we have drawn a generic real emission diagram. In this real emission diagram, a massless fermion splits into a massless gauge boson and a massless fermion. The $p_i$ and $p_j$ are the momenta of the external fermion and gauge boson. 
  The blob in Fig.~\ref{fig_idd_ir_rem} represents the rest of the matrix element of the diagram. 
   
 The momentum of intermediate fermion propagator is $(p_i+p_j)$. In the massless limit, the intermediate propagator becomes 
 \begin{equation}
 \frac{1}{(p_i+p_j)^2-m_i^2}=\frac{1}{2p_i.p_j}\:.
 \end{equation}
  When the external gauge boson's momentum becomes parallel to the external fermion i.e, $p_j=xp_i$, where $x$ is a real variable; we get 
  \begin{align}
  p_i.p_j\sim p_i^2=m_i^2\rightarrow 0\:.
  \label{equ_idd_sft_prm}
  \end{align}
  This implies that the matrix element $\mathcal{M}$ is divergent in the massless limit. As this divergence comes from the collinear momenta regime i.e, $p_j\parallel p_i$ , it is called collinear divergence.
  When the external gauge boson momentum is very soft i.e, $p_j=\lambda q$, where $q$ is a lightlike momenta and $\lambda\rightarrow 0$; we get
  \begin{align}
  p_i.p_j\sim \lambda (q.p_i)\xrightarrow{\lambda\rightarrow 0} 0\:.
  \label{equ_idd_sft_prm}
  \end{align}
  In this region, matrix element $\mathcal{M}$ is again divergent. This divergence comes from the soft momentum regime, hence called soft divergence. 
  This is a very naive way to see the divergence in the collinear and soft regions. The phase space integration over radiated gluon or photon will give the divergence as a function of $\frac{1}{\epsilon}$ in the dimensional regularization.
  \section{Dipole subtraction}
  \label{sec_idd_dps}
  
   The computation of hadronic cross-sections in the perturbative approach is based on the parton model of hadrons. 
  In the parton model, the cross-section for a hard-scattering process can be  written as a convolution of the structure function of partons and a partonic level cross section. The structure function $f(x,Q^2)$ of partons (quarks and gluons) is known as parton distribution function (PDF).
  Here $x$ is the momentum fraction of hadron's momentum shared by a parton and $Q$ is the scale of the problem, typically large momentum transfer.
  The PDFs are non-perturbative and universal in nature, i.e., they do not depend on a process. The PDFs are determined by experimental results for a certain energy scale, but their evolution with the varying scale can be calculated from the Dokshitzer–Gribov–Lipatov–Altarelli–Parisi (DGLAP) evolution equation.
  The partonic cross section is perturbative in nature in the regime $Q\gg \Lambda_{QCD}$, where $\Lambda_{QCD}$ is the QCD scale. As we see from previous sections that the virtual and real amplitudes in the partonic cross section (and decay) are singular in low-momentum (soft) and small-angle (collinear) regions.
  The physical observables should be infrared safe (collinear and soft) i.e., the perturbative prediction has to be independent of the number of collinear and soft partons in the final states.
  The coherent sum over collinear and soft configurations in the final states in virtual and real amplitudes leads to the cancellation of soft singularities. The left-over collinear singularities are factorized and absorbed in the PDFs. This makes the physical observables infrared safe.
  Such complications make the perturbative computation fully inclusive. Thus one has to perform phase space integration over final state particles for real and virtual contributions in such a way that only the UV singularities will appear in the intermediate steps, which can be removed by renormalization.  
  The analytical calculation of phase space integral for multi-parton final states is very complicated, indeed next to impossible. In analytical calculation, the collinear and soft singularities that appear in the intermediate steps are regularized in $d$-dimension and calculated independently, which yield equal-and-opposite pole in $\epsilon$ and remove IR singularities from the inclusive process.
   The numerical methods are also not helpful for these computations as the real and virtual contributions have a different number of final states, so they have to be integrated separately over different final state phase spaces.
  To get rid of these difficulties and have an infrared safe prediction, different {\it {subtraction}} methods have been introduced. In the subtraction methods, the analytical calculation is done only for a minimal part which gives rise to singularities.
  The results of partially analytical calculations can act as the local counterterms and once calculated, can be used for any process. We use Catani-Saymour {\text dipole subtraction} method to handle the IR singularities and have IR safe predictions. 
  We follow Ref.~\cite{Catani:1996vz} for dipole subtraction in QCD corrections and Ref.~\cite{Schonherr:2017qcj} for dipole subtraction in EW corrections which also follows Catani-Saymour dipole subtraction method.
  \subsection{The subtraction method}
  \label{subsec_idd_sub_methd}
  
   For a next-to-leading order (NLO) computation, the cross section can be written as 
   \begin{align}
   \sigma=\sigma^{LO}+\sigma^{NLO}\:.
   \label{equ_idd_xsec_lo_nlo}
   \end{align}
   Here $\sigma^{LO}$ is the Born level cross section and $\sigma^{NLO}$ is the NLO correction to $\sigma^{LO}$. Let's say there are $m$ partons in the final state, then we can write 
   \begin{align}
   \sigma^{LO}=\int_m d\sigma^B\:.
   \label{equ_idd_xsec_lo}
   \end{align}
   The Born level computation is done in $4$-dimension.
   For an exclusive NLO cross section the virtual and real parts together will give $\sigma^{NLO}$. The virtual part ($d\sigma^V$) is integrated over $m$ parton phase space and the real part ($d\sigma^R$) is integrated over $m+1$ parton phase space. We write
   \begin{align}
   \sigma^{NLO}=\int d\sigma^{NLO}=\int_{m}d\sigma^V+\int_{m+1} d\sigma^R\:.
   \label{equ_idd_xsec_nlo}
   \end{align}
   The two integrals given in Eq.~\ref{equ_idd_xsec_nlo} are separately divergent in $4$-dimension, but their sum is finite. These pieces have to be regulated separately in $d$-dimension before carrying out the numerical integration.
   In dimensional regularization, the divergences appear as a single pole $\frac{1}{\epsilon}$ for collinear or soft or ultraviolet singularities and as a double pole $\frac{1}{\epsilon^2}$ for collinear and soft singularities. The ultraviolet poles $\frac{1}{\epsilon_{UV}}$ are removed with the one-loop renormalization procedure.
   After removing $\frac{1}{\epsilon_{UV}}$ poles, we write the NLO cross section as
   \begin{align}
   d\sigma^{NLO}=\Big[d\sigma^R-d\sigma^A\Big]+d\sigma^A+d\sigma^V\:.
   \label{equ_idd_dsgm_da}
   \end{align}
   The $d\sigma^A$ has the same pointwise singular behaviour as $d\sigma^R$ in singular regions. Thus, $d\sigma^A$ can act as a local counterterm for $d\sigma^R$. Integrating over phase space, we write
\begin{align}
\sigma^{NLO}=\int_{m+1}\Big[d\sigma^R-d\sigma^A\Big]+\int_{m+1}d\sigma^A+\int_m d\sigma^V\:.
\label{equ_idd_dsgm_da_int}
\end{align}
 Now we can safely perform the numerical integration in $4$-dimension ($\epsilon\rightarrow 0$) for the first integral in Eq.~\ref{equ_idd_dsgm_da_int} as $d\sigma^R$ and $d\sigma^A$ have the same pointwise singular behaviour. The leftover singularities are associated with the last two integrals in Eq.~\ref{equ_idd_dsgm_da_int}.
  One can perform the analytical integration of $d\sigma^A$ over one-parton subspace in $d$-dimension which leads to $\epsilon$ poles. These poles can be combined with the poles in $d\sigma^V$ and  make the whole expression in Eq.~\ref{equ_idd_dsgm_da_int} non-singular.
  Now one can take $\epsilon\rightarrow 0$ limit and carry out the numerical integration over $m$-parton phase space in $4$-dimension. Then we can write the Eq.~\ref{equ_idd_dsgm_da_int} as
\begin{align}
\sigma^{NLO}=\int_{m+1}\Big[\Big(d\sigma^R\Big)_{\epsilon=0}-\Big(d\sigma^A\Big)_{\epsilon=0}\Big]+\int_{m}\Big[d\sigma^V+\int_1 d\sigma^A\Big]_{\epsilon=0}\:. 
 \label{equ_idd_dsgm_da_int_4}
\end{align}   
  The first and second integral in Eq.~\ref{equ_idd_dsgm_da_int_4} are over $(m+1)$ and $m$-partons phase space respectively and the corresponding numerical integration can be carried out in $4$-dimension. Thus the partonic-Monte Carlo integration can be implemented here to extract the finite contributions.
  
  The Next job is to find the expression for $d\sigma^A$ that should satisfy the following properties : i) it should be independent of a particular physical observables for a given process, ii) it should match exactly with the $d\sigma^R$ in the singular regions in $d$-dimension, 
  iii) it should be integrable analytically in $d$-dimension over single parton sub-phase space which leads to collinear and soft divergences,
  iv) its form should be convenient for the Monte-Carlo integration. 
  $d\sigma^A$ should be constructed in such a way that it should be completely process independent. 
  The $(m+1)$ parton matrix element is factorized in the $m$ parton matrix element and a spitting or eikonal function in the collinear and soft regions respectively.
  The same property has to be reflected in the $d\sigma^A$ also. With this property, we write the factorization formula, called dipole formulae as 
  \begin{align}
  d\sigma^A=\sum_{\text{dipoles}}d\sigma^B\otimes dV_{\text{dipole}}\:.
  \label{equ_idd_dp_fac_frml}
  \end{align}
  Here $d\sigma^B$ is the exclusive Born level cross section. The symbol $\otimes$ represents the convolution and sum over colour and spin indices. The dipole factors $dV_{\text{dipole}}$ are universal configurations i.e., they do not depend on the process.
   The $dV_{\text{dipole}}$ can be computed once for all. There are several dipole terms corresponding to different kinematics configuration of $(m+1)$ partons. Each configuration can be obtained effectively from a two-step process;
    first $m$ parton configuration is produced and then on of these parton goes into two partons. This pseudo two-step process leads to the factorization structure given in Eq.~\ref{equ_idd_dp_fac_frml}.
   The $d\sigma^A$ in Eq.~\ref{equ_idd_dp_fac_frml} is well defined in ($m+1$) parton phase space i.e., momentum is conserved in Eq.~\ref{equ_idd_dp_fac_frml} and it does not depend on any additional phase-space cut-off. According to the definition, there is a one-to-one correspondence between configurations in the singular region of $d\sigma^R$ and each term in $d\sigma^A$.
    Therefore, the first integral in Eq.~\ref{equ_idd_dsgm_da_int_4} is integrable via Monte-Carlo techniques. The $dV_{\text{dipole}}$ in Eq.~\ref{equ_idd_dp_fac_frml} can be integrable analytically over one sub-space. This reduces the ($m+1$) parton phase space integral to $m$ parton phase space integral. 
    Then we can write
     \begin{align}
     \int_{m+1}d\sigma^A=\sum_{\text{dipoles}}\int_{m}d\sigma^B\otimes\int_1 dV_{\text{dipole}}=\int_m \Big[d\sigma^B\otimes {\textit{\textbf{I}}}\Big]\:,
     \label{equ_idd_dsgm_da_i}
     \end{align}
     where the universal factor {\textit{\textbf{I}}} is defined as
     \begin{align}
     {\textit{\textbf{I}}}=\sum_{\text{dipoles}}\int_1 dV_{\text{dipole}}\:.
     \label{equ_idd_dsgm_i_int_v}
     \end{align}
   The  {\textit{\textbf{I}}} term contains all singular poles ($\frac{1}{\epsilon}$, $\frac{1}{\epsilon^2}$), which cancel the divergences in $d\sigma^V$.
   Now we can write Eq.~\ref{equ_idd_dsgm_da_int_4} as 
   \begin{align}
\sigma^{NLO}&=\sigma^{NLO}\{m+1\}+\sigma^{NLO}\{m\}\nonumber\\
&=\int_{m+1}\Big[\Big(d\sigma^R\Big)_{\epsilon=0}-\Big(\sum_{\text{dipoles}}d\sigma^B\otimes dV_{\text{dipole}}\Big)_{\epsilon=0}\Big]+\int_{m}\Big[d\sigma^V+d\sigma^B\otimes {\textit{\textbf{I}}}\Big]_{\epsilon=0}\:. 
 \label{equ_idd_dsgm_i_dv}
\end{align} 
  This is the most general subtraction formula for practical implementation. Till now, the subtraction procedure so far discussed  only applies to the process that does not have any initial-state hadrons. For instance, this can be applied to the processes like $e^+e^-$ annihilation, vector boson decay etc.
  The main difficulties come into the picture as the initial state partons (which come from the hadron) carry a well defined momenta, which spoils the cancellation of collinear singularities that appear in perturbative computation.
  As we have discussed, the left-over collinear singularities are factorized and can be absorbed in process independent parton distribution functions. Now adding the dipole contribution from initial state partons, we rewrite Eq.~\ref{equ_idd_dp_fac_frml} as
  \begin{align}
  d\sigma^A=\sum_{\text{dipoles}}d\sigma^B\otimes\Big( dV_{\text{dipole}}+dV^\prime_{\text{dipole}}\Big)\:.
  \label{equ_idd_dsgm_a_vvp}
  \end{align}
  Here the additional dipole term $dV^\prime_{\text{dipole}}$ has been introduced for the collinear singularities of initial state partons and it has the same behaviour as $d\sigma^R$ in the collinear region.
  The dipole term $dV^\prime_{\text{dipole}}$ are also analytically integrable over one parton sub-space even after the momentum of identified parton is fixed. The same integration is followed in Eq.~\ref{equ_idd_dsgm_da_i} with $d\sigma^A$ given in Eq.~\ref{equ_idd_dsgm_a_vvp}. One gets singular term   {\textit{\textbf{I}}} in Eq.~\ref{equ_idd_dsgm_i_int_v} and additional singular terms, which are absorbed in PDF.
  We can write the final expression for NLO cross section as 
  \begin{align}
  \sigma^{NLO}(p)&=\sigma^{NLO\{m+1\}}(p)+\sigma^{NLO\{m\}}(p)+\int_0^1 dx \:\hat{\sigma}^{NLO\{m\}}(x;xp)\nonumber\\
&=\int_{m+1}\Big[\Big(d\sigma^R(p)\Big)_{\epsilon=0}-\Big(\sum_{\text{dipoles}}d\sigma^B(p)\otimes (dV_{\text{dipole}}+dV^\prime_{\text{dipole}})\Big)_{\epsilon=0}\Big]\nonumber\\
&+\int_{m}\Big[d\sigma^V(p)+d\sigma^B\otimes {\textit{\textbf{I}}}\Big]_{\epsilon=0}+\int_0^1dx\int_m\Big[d\sigma^B(xp)\otimes({\textit{\textbf{P}}}+{\textit{\textbf{K}}})(x)\Big]_{\epsilon=0}\:. 
 \label{equ_idd_dsgm_int_ist_prt}
  \end{align}
  The $\sigma^{NLO\{m+1\}}(p)$ and $\sigma^{NLO\{m\}}$ terms are analogous to those in Eq.~\ref{equ_idd_dsgm_da_int_4}. Here the dependency of momentum $p$ of the initial-state parton has been introduced in Eq.~\ref{equ_idd_dsgm_int_ist_prt}. After the factorization of initial-state collinear singularities  into the PDFs, there are leftover finite reminders. 
  The last term of Eq.~\ref{equ_idd_dsgm_int_ist_prt} represents the finite reminders. This term is a cross section with an additional integration over the momentum fraction of a parton.
  It is written as the convolution of Born level cross section with the x-dependent {\textit{\textbf{P}}}, {\textit{\textbf{K}}} terms. The {\textit{\textbf{P}}} and {\textit{\textbf{K}}} terms are universal i.e., they do not depend on physical observables and the corresponding process.
   These terms only depend on the number of identified partons in the initial states. There is a similar term {\textit{\textbf{H}}} associated with the final state identified partons. This term is irrelevant for us as there are no identified partons in the final states in the process discussed in chapter~\ref{chap_wwh_bb_fus},~\ref{chap_h24nu} and ~\ref{chap_h22e2m}.
\subsection{Factorization in the soft and collinear limits}
\label{subsec_idd_fac_col_sft_lmt}
\begin{figure}[h!]
 \begin{center}
\includegraphics [angle=0,width=0.9\linewidth]{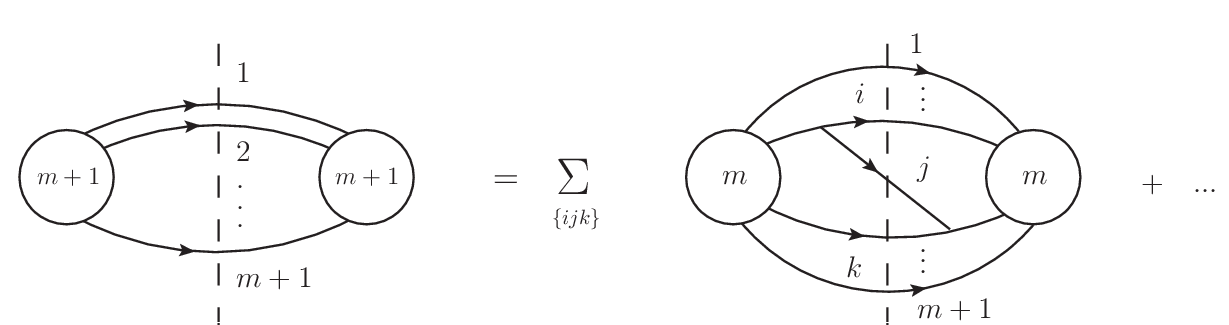}\\
	\caption{Diagrammatic representation for external leg insertion in $\mathcal{M}_{m}$ to get $\mathcal{M}_{m+1}$. The blobs represent Born level matrix elements and their complex conjugates.}
	\label{fig_idd_spct_emt_sum}
	\end{center}
\end{figure}

\begin{figure}[h!]
 \begin{center}
\includegraphics [angle=0,width=0.8\linewidth]{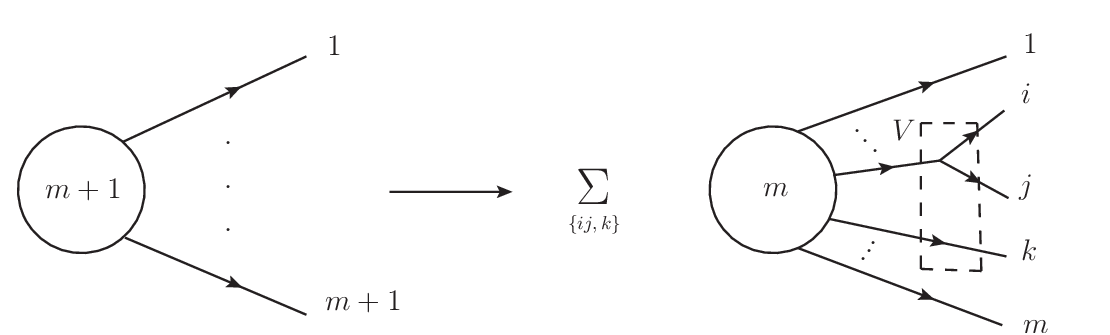}\\
	\caption{Diagrammatic representation of the dipole factorization and formation of $V_{ij,k}$.}
	\label{fig_idd_v_ijk_splt}
	\end{center}	
\end{figure}

  The ($m+1$) parton matrix element square $|\mathcal{M}_{m+1}|^2$ behaves as $\frac{1}{\lambda^2}$ in the soft regime and as $\frac{1}{p_i.p_j}$ in the collinear regime.
  The structural behaviour of $|\mathcal{M}_{m+1}|^2$ in singular regions is universal i.e., it does not depend on the very minute structure of $\mathcal{M}_{m+1}$ itself. As we have mentioned in Sec.~\ref{subsec_idd_sub_methd}, the $\mathcal{M}_{m+1}$ can be constructed from  $\mathcal{M}_{m}$ by inserting an extra parton from all possible external legs of $\mathcal{M}_{m}$.
  Thus, in a singular region, $\mathcal{M}_{m+1}$ should be factorizable with respect to  $\mathcal{M}_{m}$ and a singular entity that depends on the momenta and quantum numbers of the partons in $\mathcal{M}_{m}$.
  In Fig.~\ref{fig_idd_spct_emt_sum}, we have shown diagrammatically the insertion rule of parton $j$ in $\mathcal{M}_{m}$ matrix element  to get $\mathcal{M}_{m+1}$ matrix element. The Fig.~\ref{fig_idd_spct_emt_sum} represents the $|\mathcal{M}|^2$ where the blobs are the matrix element $\mathcal{M}$ and their complex conjugate.
  The ellipse ... represents non-singular terms in both collinear and soft regions. 
  In the singular regime, we can write 
\begin{align}
  |\mathcal{M}_{m+1}|^2\rightarrow  |\mathcal{M}_{m}|^2\otimes V_{ij,k}\:.
  \label{equ_idd_m_v_ijk}
  \end{align}
  Here $V_{ij,k}$ denotes the singular factor which depends only on the momentum and quantum numbers of partons $i$, $j$, and $k$. This factorization is depicted symbolically in Fig.~\ref{fig_idd_v_ijk_splt}. 
  Here the combined ($i$ and $j$) parton is called as emitter and $k$ parton is called as spectator. Because of the `emitter-spectator' structure, this factorization is called {\it{dipole factorization}} whose general expressions are given in Sec.~\ref{subsec_idd_dp_fac_frml}.
 We define the vector $|1,...,m+1\rangle_m$ to represent $\mathcal{M}_{m}$ in helicity and color space. In the soft limit with the parametrization given in Sec~\ref{sec_idd_ir_sng_rlem_dia}, the matrix element square behaves as 
 \begin{align}
   {_{m+1}}\langle 1,...,m+1|&|1,...,m+1\rangle_{m+1}\rightarrow\nonumber\\ &- \frac{1}{\lambda^2}4\pi\mu^{2\epsilon}\alpha_S\:{_{m}}\langle 1,...,m+1|[J^\mu(q)]^\dag J_\mu(q)|1,...,m+1\rangle_{m}\:.
   \label{equ_idd_sft_mat_ele}
 \end{align}
Here $J^\mu(q)$ is the eikonal current for the emission of soft parton $p_j$ . In collinear limit with the parametrization given in Sec~\ref{sec_idd_ir_sng_rlem_dia}, the matrix element square behaves as 
\begin{align}
   {_{m+1}}\langle 1,...,m+1||1,...,m+1\rangle_{m+1}\rightarrow  \frac{1}{p_ip_j}4\pi\mu^{2\epsilon}{_{m}}\langle 1,...,m+1|P_{ij}(z,\epsilon)|1,...,m+1\rangle_{m}\:.
   \label{equ_idd_col_mat_ele}
 \end{align}
 Here $P_{ij}$ is the $d$-dimension Altarelli-Parisi splitting function. The details structure of the eikonal and Altarelli-Parisi splitting function can be found in Ref.~\cite{Catani:1996vz}. The Eq.~\ref{equ_idd_sft_mat_ele} and~\ref{equ_idd_col_mat_ele} are the basis of the factorization given in Eq.~\ref{equ_idd_m_v_ijk}.
\subsection{Dipole factorization formulae}
\label{subsec_idd_dp_fac_frml}
\begin{figure}[h!]
 \begin{center}
\includegraphics [angle=0,width=1\linewidth]{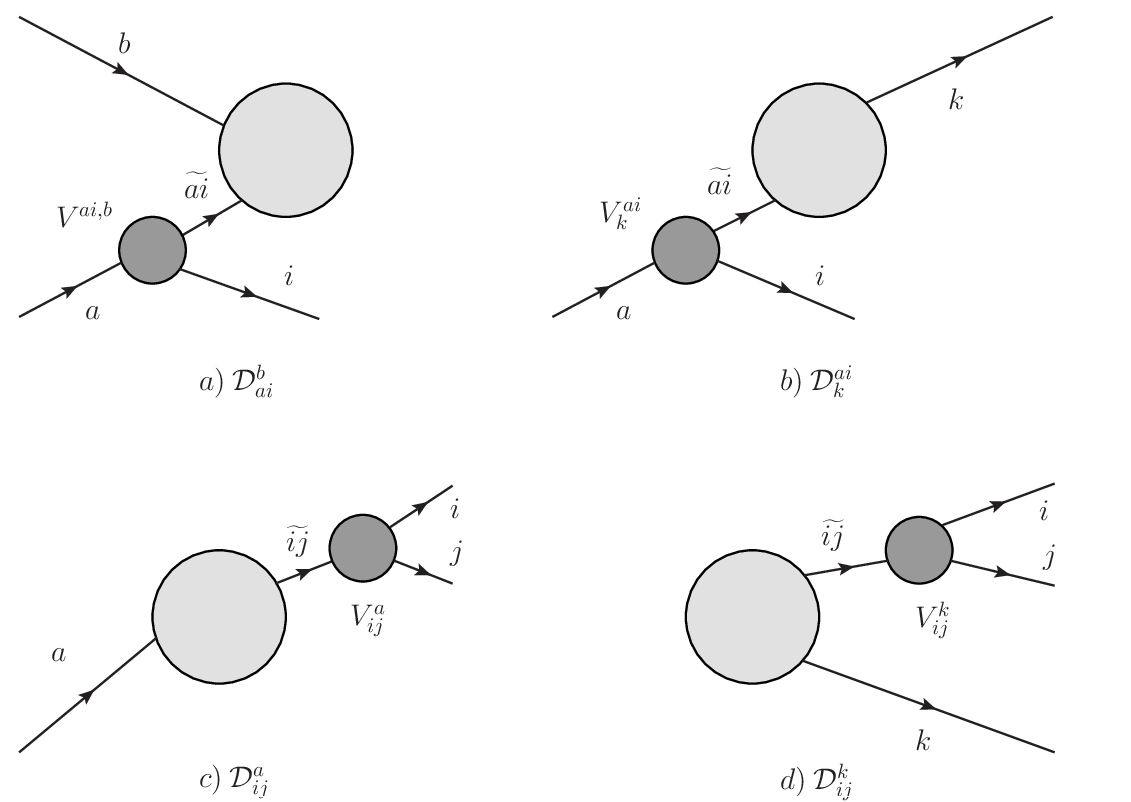}\\
	\caption{Diagrammatically representation of dipoles. The big blobs represent the matrix element of $m$-parton and the small blobs represent splittings.}
	\label{fig_idd_dijk_confg}
	\end{center}	
\end{figure}

 For a given process, there can be several dipole contributions related to each parton. As we have discussed, there can be final state partons and identified initial state partons.
  The dipole contribution configurations depend on whether the emitter and spectator are in the initial and/or in the final state. We only consider the identified partons in the initial state, not in the final state. 
  There will be four configurations depending on the position of emitters and spectators. We denote identified partons in the initial state by $a$ and $b$, whereas we denote the final state partons by $i$, $j$ and $k$. 
  The four possible dipole configuration have been shown in Fig.~\ref{fig_idd_dijk_confg}.
 In dipole contribution ($a$) in Fig.~\ref{fig_idd_dijk_confg}, the emitter is associated with the identified initial state parton $a$ and the final state parton $i$; and the spectator is other identified initial state parton $b$. 
  The emitter configuration in dipole contribution ($b$) in Fig.~\ref{fig_idd_dijk_confg} is same as the dipole contribution ($a$) and the spectator is the final state parton $k$.
  In dipole contribution $(c)$ in Fig.~\ref{fig_idd_dijk_confg}, the emitter is associated with the final state parton $i$ and $j$, and the spectator is the initial state parton $a$. 
  The configuration ($d$) in Fig.~\ref{fig_idd_dijk_confg}, represents the dipole contribution with the emitter configuration same as dipole contribution ($c$) and with the spectator $k$ in the final state parton.
  We will be discussing only the following two dipole contributions as these contributions are relevant for the processes discussed in chapter~\ref{chap_wwh_bb_fus} and~\ref{chap_h22e2m}.
  \subsubsection{Final state singularities with no initial state partons :}
  \label{subsubsec_idd_fss_no_isp}
  In the limit $p_i.p_j\rightarrow 0$ in collinear and soft region, the matrix element square with no initial state parton can be written as 
\begin{align}
   {_{m+1}}\langle 1,...,m+1||1,...,m+1\rangle_{m+1}=\sum_{k\neq i,j}\mathcal{D}_{ij,k}(p_1,...,p_{m+1})+...\:,
   \label{equ_idd_fss_no_isp}
 \end{align}    
  where the ellipse ... denotes the non-singular terms in the limit $p_i.p_j\rightarrow 0$. The dipole contribution $\mathcal{D}_{ij,k}$ is given as
  \begin{align}
   \mathcal{D}_{ij,k}(p_1,...,p_{m+1})=-\frac{1}{2p_i.p_j}\:{_m}\langle 1,...,\widetilde{ij},...,\widetilde{k},...,m+1|\frac{\textit{\textbf{T}}_k.\textit{\textbf{T}}_{ij}}{\textit{\textbf{T}}^{\:2}_{ij}}\:\textit{\textbf{V}}_{ij,k}|1,...,\widetilde{ij},...,\widetilde{k},...,m+1\rangle_{m}\:.
   \label{equ_idd_fss_dijk}
 \end{align}    
  In Eq.~\ref{equ_idd_fss_no_isp} and ~\ref{equ_idd_fss_dijk}, the ($m+1$) parton matrix reduced to $m$ parton matrix element by replacing the partons $i$ and $j$ with parton $\widetilde{ij}$; and the parton $k$ with the parton $\widetilde{k}$.
  The parton $\tilde{ij}$ is called emitter and $\tilde{k}$ is called spectator. The momenta for emitter and spectator are 
  \begin{align}
 \widetilde{p}^\mu_{ij}=p_i^\mu+p_j^\mu-\frac{y_{ij,k}}{1-y_{ij,k}}p^\mu_k\:,\quad {\text{and}}\quad \widetilde{p}^\mu_k=\frac{1}{1-y_{ij,k}}p^\mu_k\:,
  \end{align}
  where $y_{ij,k}$ is given by
  \begin{align}
  y_{ij,k}=\frac{p_ip_j}{p_ip_j+p_jp_k+p_kp_i}\:.
  \end{align}
   In the replacement $\{i,j,k\}\rightarrow \{\widetilde{ij},\widetilde{k}\}$, the momentum conservation $p_i^\mu+p_j^\mu+p_k^\mu=\widetilde{p}_{ij}^\mu+\widetilde{p}_{k}^\mu$ has been implemented. 
   The $\textit{\textbf{T}}_{ij}$ and $\textit{\textbf{T}}_{k}$ are the color charges of the emitter and spectator. The definition and algebra of color charge \textit{\textbf{T}} have been describe in detail in Ref.~\cite{Catani:1996vz}. The $\textit{\textbf{V}}_{ij,k}$ is the singular matrix that depends on $i$, $j$, and $k$. It is related to the $d$-dimension Altarelli-Parisi spitting function. For a fermion split into fermion and gluon, the $\textit{\textbf{V}}_{ij,k}$ is defined as
   \begin{align}
   \langle s|\textit{\textbf{V}}_{qg,k}(\tilde {z}_i;y_{ij,k})|s\prime \rangle=8\pi\mu^{2\epsilon}\alpha_S C_F\Big[\frac{2}{1-\tilde{z}_i(1-y_{ij,k})}-(1+\tilde{z}_i)-\epsilon(1-\tilde{z}_i)\Big]\delta_{ss^\prime}\:,
   \end{align}
   where the kinematic factor $\tilde{z}_i$ is defined as
   \begin{align}
   \tilde{z}_i=\frac{p_ip_k}{p_jp_k+p_ip_k}=\frac{p_i\tilde{p}_k}{\tilde{p}_{ij}\tilde{p}_k}\:.
   \end{align}
  There are other $\textit{\textbf{V}}_{ij,k}$ related to quark-quark and gluon-gluon splitting, which we are not discussing here as we do not need them. They can be found in details in Ref.~\cite{Catani:1996vz}.
  \subsubsection{Initial state singularities with two initial state partons :}
  \label{subsubsec_idd_iss_tisp}
  The dipole factorization formula with two initial state parton $a$ and $b$ in the limit $p_a.p_i\rightarrow 0$ is given by
  \begin{align}
   {_{m+1,ab}}\langle 1,...,&m+1;a,b||1,...,m+1;a,b\rangle_{m+1,ab}\nonumber\\
   &=\sum_{k\neq i}\mathcal{D}_{k}^{ai}(p_1,...,p_{m+1};p_a,p_b)+\mathcal{D}_{b}^{ai}(p_1,...,p_{m+1};p_a,p_b)+...\:.
   \label{equ_idd_iss_tisp}
 \end{align}    
  In Eq.~\ref{equ_idd_iss_tisp}, the first dipole contribution $\mathcal{D}_k^{ai}$ with the initial state parton $\widetilde{ai}$ as the emitter and final state parton $k$ as the spectator is  
  \begin{align}
   \mathcal{D}^{ai}_k(p_1,.&..,p_{m+1};p_a,p_b)=-\frac{1}{2p_a.p_i}\:\frac{1}{x_{ik,a}}\nonumber\\
   &{_{m,ab}}\langle 1,...,\widetilde{k},...,m+1;\widetilde{ai},b|\frac{\textit{\textbf{T}}_k.\textit{\textbf{T}}_{ai}}{\textit{\textbf{T}}^{\:2}_{ai}}\:\textit{\textbf{V}}^{ai}_k|1,...,\widetilde{k},...,m+1;\widetilde{ai},b\rangle_{m,ab}\:.
   \label{equ_idd_iss_daik}
 \end{align} 
  The momenta of the emitter and spectator are 
  \begin{align}
  \widetilde{p}^\mu_{ai}=x_{ik,a}\:p^\mu_a\quad {\text {and}}\quad\widetilde{p}^\mu_{k}=p_k^\mu+p_i^\mu-(1-x_{ik,a})p^\mu_a\:,
  \end{align}
  where 
  \begin{align}
  x_{ik,a}=\frac{p_kp_a+p_ip_q-p_ip_k}{(p_k+p_i)p_a}\:.
  \end{align}
  The singular factor $\textit{\textbf{V}}^{ai}_k$ for quark split into gluon-quark and gluon split in quark-antiquark pair are 
   \begin{align}
   \langle s|\textit{\textbf{V}}^{q_ag_i}_k(x_{ik,a};u_i)|s\prime \rangle=8\pi\mu^{2\epsilon}\alpha_S C_F\Big[\frac{2}{1-x_{ik,a}+u_i}-(1+x_{ik,a})-\epsilon(1-x_{ik,a})\Big]\delta_{ss^\prime}
   \end{align}
  and 
  \begin{align}
   \langle s|\textit{\textbf{V}}^{g_a\bar{q}_i}_k(x_{ik,a})|s\prime \rangle=8\pi\mu^{2\epsilon}\alpha_S T_R\Big[1-\epsilon-2x_{ik,a}(1-x_{ik,a})\Big]\delta_{ss^\prime}
   \end{align}
  respectively, where $u_i=\frac{p_ip_a}{p_ip_a+p_kp_a}$.
  In Eq.~\ref{equ_idd_iss_tisp}, the second dipole contribution $\mathcal{D}_b^{ai}$ with the initial state parton $\widetilde{ai}$ as the emitter and other initial state parton $b$ as the spectator is given by
  \begin{align}
   \mathcal{D}^{ai,b}(p_1,.&..,p_{m+1};p_a,p_b)=-\frac{1}{2p_a.p_i}\:\frac{1}{x_{i,ab}}\nonumber\\
   &{_{m,ab}}\langle \widetilde{1},...,\widetilde{m+1};\widetilde{ai},b|\frac{\textit{\textbf{T}}_b.\textit{\textbf{T}}_{ai}}{\textit{\textbf{T}}^{\:2}_{ai}}\:\textit{\textbf{V}}^{ai,b}|\widetilde{1},...,\widetilde{m+1};\widetilde{ai},b\rangle_{m,ab}\:.
   \label{equ_idd_iss_daib}
 \end{align} 
  The emitter momentum will be parallel to $p_a$ and it can be written as 
  \begin{align}
  \widetilde{p}^\mu_{ai}=x_{i,ab}\:p^\mu_a\:,\quad {\text {where}}\quad x_{i,ab}=\frac{p_ap_b-p_ip_a-p_ip_b}{p_ap_b}\:.
  \label{equ_idd_iss_mom_pai}
  \end{align}
   The spectator momentum $p_b$ will be unchanged as a result all final state momenta $k_\mu$ will be rescaled as 
  \begin{align}
  \widetilde{k}^\mu_{j}=k^\mu_{j}-\frac{2k_j.(K+\widetilde{K})}{(K+\widetilde{K})^2}\:(K+\widetilde{K})^\mu+\frac{2k_j.K}{K^2}\widetilde{K}^\mu\:,
  \label{equ_idd_iss_mom_kj}
  \end{align}
  where the momenta $K^\mu$ and $\widetilde{K}^\mu$ are defined as
  \begin{align}
  K^\mu=p_a^\mu+p_b^\mu-p_i^\mu\:,\quad \widetilde{K}^\mu=\widetilde{p}_{ai}^\mu+p_b^\mu\:.
  \end{align}
  One can check the momentum conservation with the new set of momenta given in Eq.~\ref{equ_idd_iss_mom_pai} and~\ref{equ_idd_iss_mom_kj}. The analogous spitting function $\textit{\textbf{V}}^{ai,b}$ is given as
   \begin{align}
   \langle s|\textit{\textbf{V}}^{q_ag_i,b}(x_{i,ab})|s\prime \rangle&=8\pi\mu^{2\epsilon}\alpha_S C_F\Big[\frac{2}{1-x_{i,ab}}-(1+x_{i,ab})-\epsilon(1-x_{i,ab})\Big]\delta_{ss^\prime}\:,\nonumber\\
   \langle s|\textit{\textbf{V}}^{g_a\bar{q}_i,b}(x_{i,ab})|s\prime \rangle&=8\pi\mu^{2\epsilon}\alpha_S T_R\Big[1-\epsilon-2x_{i,ab}(1-x_{i,ab})\Big]\delta_{ss^\prime}\:.
   \end{align}
  \subsection{Insertion Operators}
  \label{subsec_idd_ins_oprt}
  In Eq.~\ref{equ_idd_dsgm_int_ist_prt}, we have talked about the universal insertion operators $\textit{\textbf{I}}$, $\textit{\textbf{P}}$ and $\textit{\textbf{K}}$. The operator $\textit{\textbf{I}}$ removes all singular poles ($\frac{1}{\epsilon},\:\frac{1}{\epsilon^2}$) from $d\sigma^V$.
  We have discussed splitting functions $\textit{\textbf{V}}_{ij,k}$ ($\textit{\textbf{V}}_j^{\:ai},...$) in the last section. The one body phase space integration in $d$-dimension over these spitting functions will give rise to the universal operator $\textit{\textbf{I}}$.
  By doing so, one can find the below expression for the insertion operator $\textit{\textbf{I}}$,
  \begin{align}
  \textit{\textbf{I}}(\{p\};\epsilon)=-\frac{\alpha_S}{2\pi}\frac{1}{\Gamma(1-\epsilon)}\sum_I\frac{1}{\textit{\textbf{T}}_I^{\:2}}\mathcal{V}(\epsilon)\sum_{J\neq I}\textit{\textbf{T}}_I.\textit{\textbf{T}}_J\:\Big(\frac{4\pi\mu^2}{2p_I.p_J}\Big)^\epsilon\:.
  \label{equ_idd_ip_exp}
  \end{align}
  Here the $\{p\}$ denote the set of parton momenta where the information of partons (whether incoming or outgoing or identified) are not specified. The $\textit{\textbf{T}}_I$ is the colour charge of parton $I$ and $\mu$ is the factorization scale. The singular factor $\mathcal{V}_I(\epsilon)$ in Eq.~\ref{equ_idd_ip_exp} contains all poles and depends on parton flavour. The $\mathcal{V}_I(\epsilon)$ is given as
  \begin{align}
  \mathcal{V}_I(\epsilon)=\textit{\textbf{T}}_I^{\:2}\Big(\frac{1}{\epsilon^2}-\frac{\pi^2}{3}\Big)+\gamma_I\frac{1}{\epsilon}+\gamma_I+K_I+\mathcal{O}(\epsilon)\:.
  \end{align}
  The values for $\gamma_I$ and $K_I$ for different parton flavour are 
  \begin{align}
  \gamma_q=\gamma_{\bar{q}}=\frac{3}{2}C_F\:,\quad& \gamma_g=\frac{11}{6}C_A-\frac{2}{3}T_RN_f\:,\nonumber\\
 K_q=K_{\bar{q}}=\Big(\frac{7}{2}-\frac{\pi^2}{6}\Big)C_F\:,\quad& K_g=\Big(\frac{67}{18}-\frac{\pi^2}{6}\Big)C_A-\frac{10}{9}T_RN_f\:.
  \end{align}
  The insertion operator \textit{\textbf{I}} makes $d\sigma^V$  divergenceless and add finite contributions to m parton $\sigma^{NLO}$.
  
  As we have mentioned, the last integral in Eq.~\ref{equ_idd_dsgm_int_ist_prt} is the finite reminder of factorization of initial state collinear singularity into PDF. The integral can be written for one identified parton $a$ with two initial-state partons (a,b) as
  \begin{align}
  \int_0^1& dx\hat{\sigma}^{NLO\{m\}}_{ab}(x;xp_a,p_b,\mu_F^2)=\sum_{a^\prime}\int_0^1 dx\int_m\Big[d\sigma^B_{a^\prime b}(xp_a,p_b)\otimes(\textit{\textbf{K}}+\textit{\textbf{P}})^{a,a^\prime}(x)\Big]_{\epsilon=0}\nonumber\\
  &=\sum_{a^\prime}\int_0^1 dx\int d\phi^{(m)}(xp_a,p_b)\nonumber\\
  &\quad\quad\quad{_{m,a^\prime b}}\langle 1,...,m;xp_a,b|\Big(\textit{\textbf{K}}^{\:a,a^\prime}+\textit{\textbf{P}}^{\:a,a^\prime}(xp_a,x;\mu_F^2)\Big)|1,...,m;xp_a,b\rangle_{m,a^\prime b}\:.
  \label{equ_idd_kp_exp}
  \end{align}
  The same integral for the parton $b$ is completely analogous to Eq.~\ref{equ_idd_kp_exp} with the replacement $xp_a\rightarrow p_a$, $p_b\rightarrow xp_b$ and $\sum_{a^\prime} \rightarrow \sum_{b^\prime}$.
  The colour-charge operators \textit{\textbf{K}} and  \textit{\textbf{P}} are given as
  \begin{align}
  &\textit{\textbf{P}}^{\:a,a^\prime}(\{p\};xp_a,x;\mu_F^2)=\frac{\alpha_S}{2\pi}P^{aa^\prime}(x)\frac{1}{\textit{\textbf{T}}^{\:2}_{a^\prime}}\sum_{I\neq a^\prime}\textit{\textbf{T}}_I.\textit{\textbf{T}}_{a^\prime}\:\text{ln}\frac{\mu_F^2}{2xp_a.p_I}\:,\nonumber\\
  &\textit{\textbf{K}}^{\:a,a^\prime}(x)=\frac{\alpha_S}{2\pi}\Big\{\overline{K}^{aa^\prime}(x)-\overline{K}^{aa^\prime}_{F.S.}(x)\nonumber\\
  &\quad\quad\quad+\delta^{aa^\prime}\sum_{i}\textit{\textbf{T}}_i.\textit{\textbf{T}}_a\:\frac{\gamma_i}{\textit{\textbf{T}}^{\:2}_i}\Big[\Big(\frac{1}{1-x}\Big)_++\delta(1-x)\Big]\Big\}-\frac{\alpha_S}{2\pi}\textit{\textbf{T}}_b.\textit{\textbf{T}}_{a^\prime}\frac{1}{\textit{\textbf{T}}^{\:2}_{a^\prime}}\widetilde{K}^{aa^\prime}(x)\:.
  \end{align}
  The expressions for $P^{aa^\prime}$ and $K^{aa^\prime}$ are a bit complicated. They are related to Atlarelli-Parisi splitting functions and are written in terms of plus functions. Their explicit expressions have been given in Ref~\cite{Catani:1996vz}. We use these expressions to calculate the collinear reminder for our process in the chapter~\ref{chap_wwh_bb_fus}.

\chapter{$W^+W^-H$ production through bottom quarks fusion
at hadron colliders}
\label{chap_wwh_bb_fus}
\section{Introduction}
\label{sec:intro}


In this chapter, we are interested in the coupling of the Higgs boson with the $W$ and $Z$  bosons. In particular, we are interested in the quartic $VVHH$ couplings.
%
  We recall the discussion in the introduction chapter about the bounds on $HHH$ and $VVHH$ couplings and also about the processes where we can probe these couplings.
     We consider the process  $pp \to HWW$ at hadron colliders. This process can help us in measuring
    $HHWW$ coupling, independent of $HHZZ$ coupling.
   These processes can take place by both quark-quark \cite{Baglio:2015eon} and gluon-gluon
   scattering. At a 100 TeV collider, gluon-gluon scattering and bottom-bottom quark
    scattering give important contributions. These contributions depend
    on $HHWW$ coupling. The gluon-gluon contribution is discussed in \cite{Agrawal:2019ffb}. This contribution is smaller than the contribution of bottom-bottom scattering. The contribution of  bottom-bottom scattering is only about $15-20\%$ of the light quarks
    scattering contribution at the 100 TeV center-of-mass energy (CME) and at the
    leading order (LO), light quarks contribution does not depend
    on $WWHH$ coupling.
    The dependence on this quartic coupling, $WWHH$, can be enhanced if we measure
    the polarization of the final state $W$ bosons.  
    There is significant enhancement of the fraction
    of the bottom-bottom scattering events, when both $W$ bosons in the 
    final states are longitudinally  polarized. The ATLAS and CMS collaborations have measured
    the $W$ polarization at the LHC~\cite{Chatrchyan:2011ig,Aad:2012ky,Aaboud:2019gxl}. We compute the one-loop QCD corrections
    to various combinations of final state $W$ bosons polarization. The longitudinally polarized
    $W$ boson final states receive largest corrections, leading
    to even larger fraction of events with bottom-bottom scattering.  We also scale the $WWHH$ coupling and examine the effect of the NLO QCD corrections and the measurement of the polarization of W bosons. It appears that an analysis of $WWH$ events, when both the $W$ bosons
    are longitudinally polarized, can help in determining the
    $WWHH$ coupling.
    
       The chapter is organized as follows. The second and third sections describe the process and the details of the calculations. In the fourth section, we present the
     numerical results, and the last section has the conclusions.

\section{The Process}
\label{sec:prcs}
We are interested in quark-quark scattering  for the production of $WWH$. To study $WWHH$ coupling,
we consider this process in five-flavour scheme.
We study the process $b\:\bar{b}\rightarrow W^+W^-H$ at hadron colliders. 
We take bottom quarks as massless but at the same time, we consider $b\:\bar{b}\:H$ 
Yukawa coupling, which is proportional to the mass of the bottom quark. With this consideration, 
the diagrams with $WWHH$ coupling would appear, with the Higgs boson coupling to the bottom 
quark. This coupling would not appear at the leading order (LO) for the other quarks in the initial state.  
This channel has been discussed only with $t\:\bar{t}\:H$ Yukawa couplings \cite{Baglio:2016ofi} but not with $b\:\bar{b}\:H$ Yukawa couplings.

\begin{figure}[!]
\includegraphics [angle=0,width=1\linewidth]{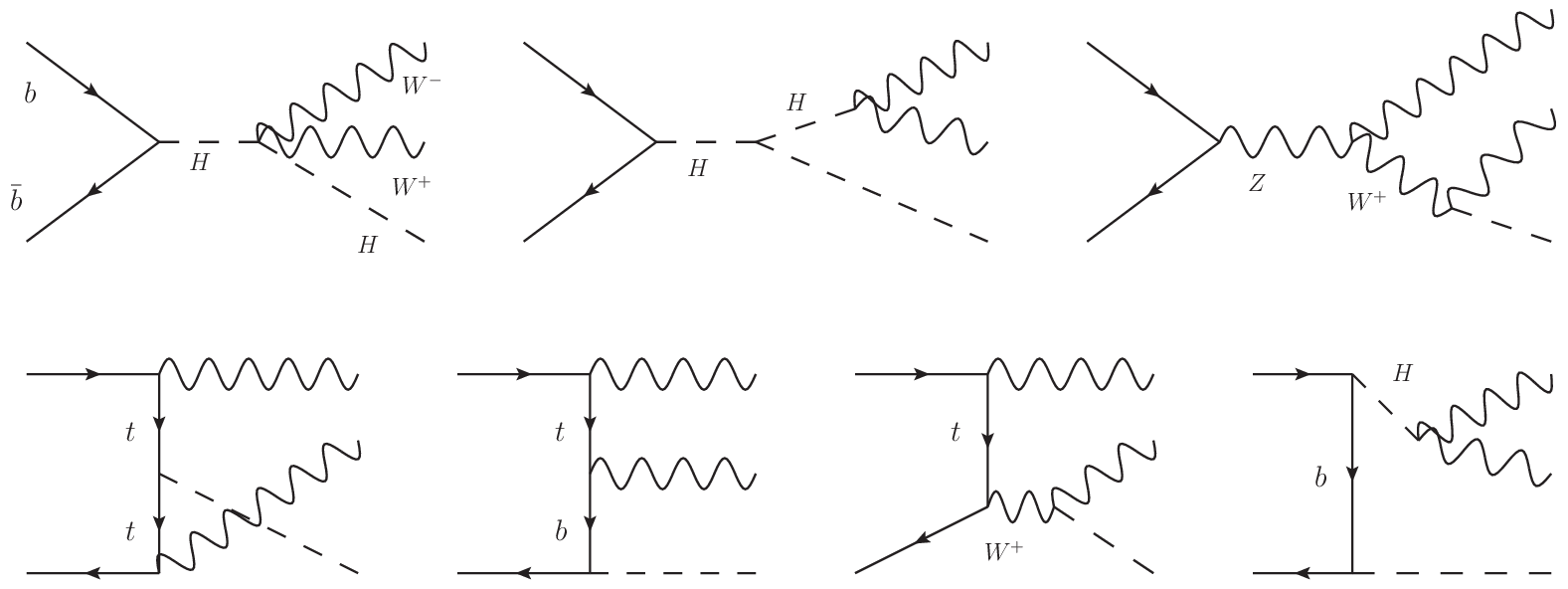}\\
	\caption{A few LO Feynman diagrams for $W^+W^-H$ production in $b\bar{b}$ channel. Diagrams in the upper row are $s$-channel diagrams and in the lower row are $t$-channel diagrams.  }
	\label{fig:bb2wwh_tree_dia}
\end{figure}

At the LO there are 20 diagrams -- 9 s-channel and 11 t-channel. A representative
set of diagrams are displayed in Fig.~\ref{fig:bb2wwh_tree_dia}. Only one of the diagrams has 
$WWHH$ coupling, which is one of our main points of interest.
We vary $WWHH$ coupling in order to see its impact on the cross section for the different center of mass energies. There is no strong coupling dependency in the LO diagrams; they solely depend on electroweak couplings. Some of
 the $t$-channel diagrams depend on $t\:\bar{t}\:H$ Yukawa couplings and give large contributions to the LO cross section, due to the top-quark mass dependency of $t\:\bar{t}\:H$ Yukawa coupling.

\begin{figure}[!]
\includegraphics [angle=0,width=1\linewidth]{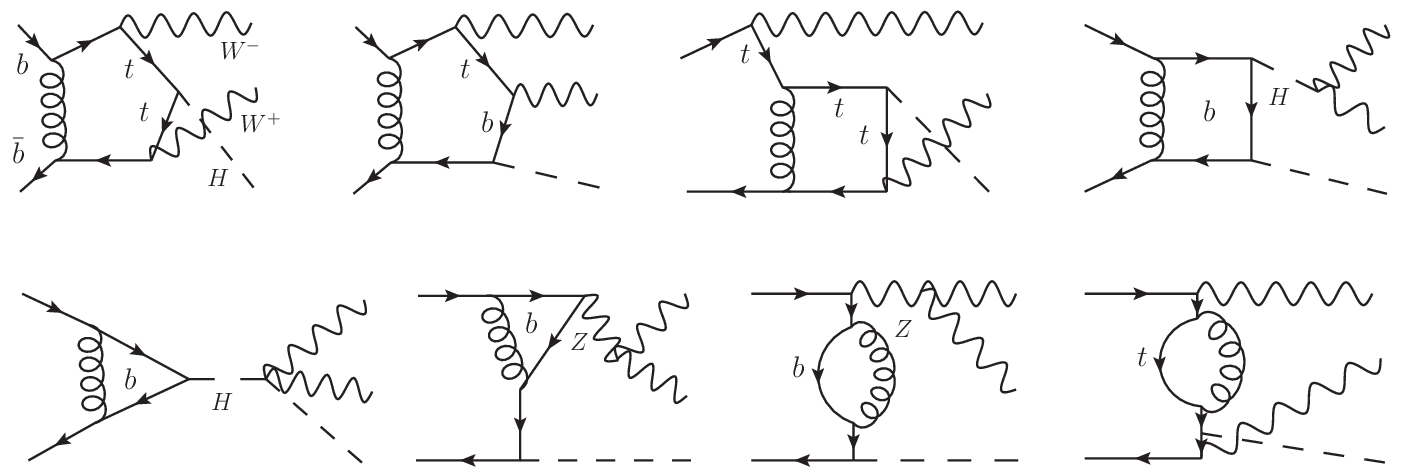}\\
	\caption{Few sample one-loop Feynman diagrams for $W^+W^-H$ production in $b\bar{b}$ channel. }
	\label{fig:bb2wwh_loop_dia}
\end{figure}

\begin{figure}[!]
\includegraphics [angle=0,width=1\linewidth]{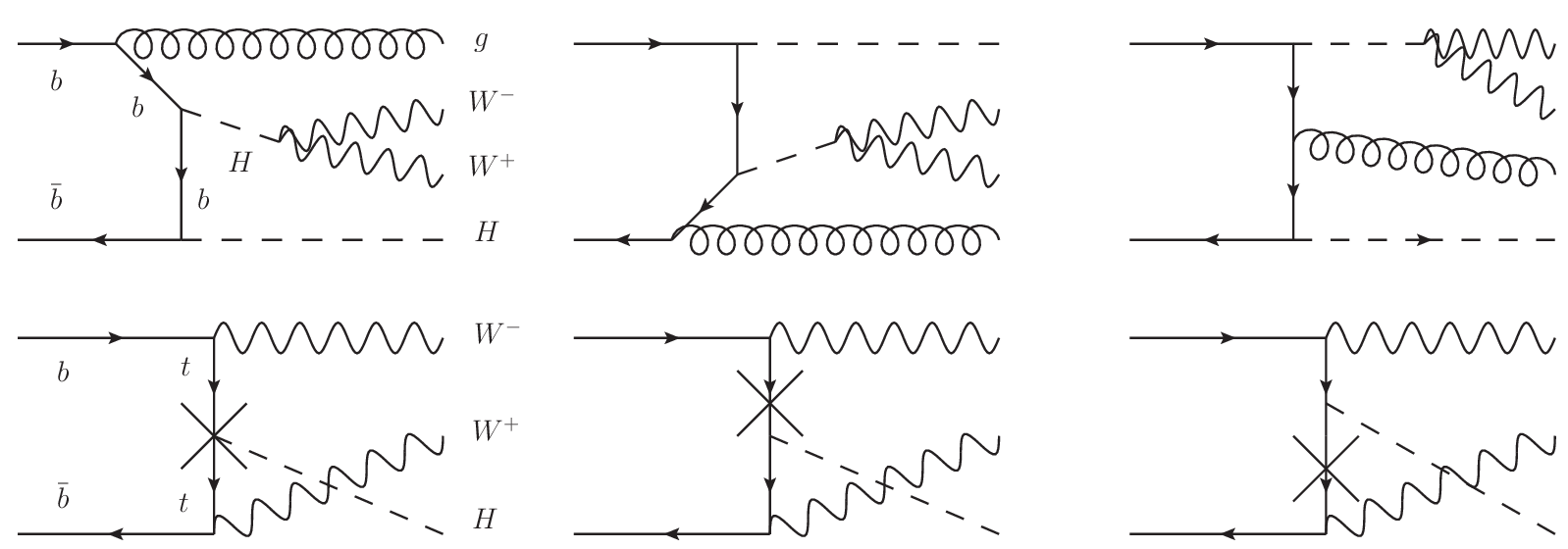}\\
	\caption{Few sample real emission Feynman diagrams for the sub-process $b\:\bar{b}\rightarrow W^+W^-H\:g$ and vertex CT diagrams and self-energy CT diagrams for the process $b\:\bar{b}\rightarrow W^+W^-H$. }
	\label{fig:bb2wwh_real_ct_dia}
\end{figure}

To compute the one-loop QCD corrections to this process, we need to include one-loop diagrams
and next-to-leading order (NLO) tree-level diagrams. The one-loop diagrams can be categorized as pentagon, box, triangle as well as bubble diagrams. There are $3$ pentagon diagrams, $14$ box diagrams, $34$ triangle diagrams, and $14$ bubble diagrams.  A few representative  NLO diagrams are displayed in Fig.~\ref{fig:bb2wwh_loop_dia}. There is only one one-loop diagram (triangle) which has $WWHH$ coupling. Bubble diagrams are UV divergent and a few triangle diagrams are also UV divergent. To remove UV divergence from the amplitude, counterterm (CT) diagrams need to be added to the virtual amplitudes. There are $15$ vertex CT diagrams and $14$ self-energy CT diagrams.  A set of CT diagrams are shown in Fig.~\ref{fig:bb2wwh_real_ct_dia}. Also, most of the virtual diagrams are infrared (IR) singular. In order to remove IR singularities from the virtual diagrams, one needs to include real emission diagrams. There are three such processes. These processes are a) $b\bar{b}\rightarrow W^+W^-Hg$, b) $g\bar{b}\rightarrow W^+W^-H\bar{b}$ and c) $bg\rightarrow W^+W^-Hb$. There are $54$ Feynman diagrams for each of these processes. We have shown a few diagrams for the first sub-process in Fig.~\ref{fig:bb2wwh_real_ct_dia}. All these diagrams have been generated using a {\tt Mathematica} package, {\tt FeynArts} \cite{Hahn:2000kx}.

\section{Calculations and Checks}
\label{sec:calc_check}

 We have to perform $2 \to 3$ and $2 \to 4$ tree level and $2 \to 3$
 one-loop level calculations. For the calculation, we use helicity methods.
 As a starting point, we consider a few prototype diagrams in each case. With suitable
 crossing, and coupling choices, we can compute rest of the diagrams.
  We compute helicity amplitudes at the matrix element level for the
  prototype diagrams. These helicity amplitudes can be used to probe the physical observables dependent on the polarization of external particles. As mentioned before, the $b$-quarks are treated as massless quarks because of their small mass and we use massless spinors for $b$-quarks.
  As discussed in chapter~\ref{chap_shf}, 
  the tree-level helicity amplitudes can be written in terms of the spinor products $\langle pq\rangle$ and $[pq]$. For the one-loop amplitudes, the vector  current $\langle p \gamma^\mu q]$ has been used.
   We adopt four-dimensional-helicity (FDH) scheme \cite{BERN1992451,Gnendiger:2017pys} to compute the amplitudes. In this scheme, all spinors, $\gamma$-matrices algebra are computed in $4$-dimensions. We use package {\tt FORM} \cite{Vermaseren:2000nd} to implement the helicity formalism. 
In this calculation, we don't have fermion loops; so there are no traces
of matrices involving $\gamma_5$. Therefore in our scheme, we have treated $\gamma_5$ with properties same as that in four-dimension \cite{Shao:2011zza,Gnendiger:2017rfh}.

Using {\tt FORM}, we write helicity amplitude in terms of spinorial objects, scalar products of momenta and polarizations. For the one-loop calculations,
we also have tensor and scalar integrals. The one-loop scalar integrals are computed using the package OneLoop \cite{vanHameren:2010cp}. We use an in-house reduction code, OVReduce \cite{Agrawal:2012df,Agrawal:1998ch}, to compute tensor integrals in dimensional regularization. Finally, the phase space integrals have been done with the advanced Monte-Carlo integration ({\tt AMCI}) package \cite{Veseli:1997hr}. In {\tt AMCI}, the {\tt VEGAS}\cite{Lepage:1977sw} algorithm is implemented using parallel virtual machine ({\tt PVM}) package \cite{10.7551/mitpress/5712.001.0001}.

Few checks have been performed to validate the amplitudes. The one-loop
amplitudes have both ultraviolet (UV) and infrared (IR) singularities. 
UV singularities are removed by using counter-term (CT) diagrams, and
the IR divergences are removed using Catani-Seymour (CS) dipole substraction
methods. Cancellation of these divergences are powerful checks on the
calculation.
All UV singularities are removed by fermionic mass and wave function renormalization. There are no UV singularities coming from pentagon and box diagrams as there are no 4-point box tensors in those amplitudes. UV singularities are coming from the triangle as well as bubble diagrams. The appropriate vertex and self-energy counterterms (CT) diagrams have been added in total amplitude, which gives renormalized amplitude. A few sample CT diagrams are depicted in Fig. 4. We use the $\rm \overline{MS}$ scheme for massless fermions and the on-shell subtraction scheme for massive fermions.

The next check is infrared (IR) singularity cancellation. We implement the Catani-Seymour dipole subtraction method \cite{Catani:1996vz} for the cancellation of IR singularities. Except bubble diagrams, all other virtual diagrams are IR singular. 

 As discussed before, we use the FDH scheme, so we take $\textit{\textbf{I}}$ term in the FDH scheme. The term $\textit{\textbf{I}}$ discussed in Sec.~\ref{subsec_idd_ins_oprt} is in conventional dimensional regularization (CDR) scheme and in any other regularization scheme (RS) it is given as \cite{Catani:1996pk}
\begin{eqnarray}
\textit{\textbf{I}}^{\:\rm RS}( \{p\},\epsilon) = \textit{\textbf{I}}^{\:\rm CDR}(\{p\},\epsilon) -\frac{\alpha_s}{2\pi}\sum_I\widetilde{\gamma}_I^{\:\rm RS} + \mathcal{O}(\epsilon)\:.
\end{eqnarray}
In the FDH scheme, $\widetilde{\gamma}_I^{\:\rm RS}$ are 
\begin{eqnarray}
\widetilde{\gamma}_q^{\:\rm FDH} = \widetilde{\gamma}_{\bar{q}}^{\:\rm FDH}=\frac{1}{2}C_F\:,\quad \widetilde{\gamma}_g^{\:\rm FDH}=\frac{1}{6}C_F\:.
\end{eqnarray} 
This will only effect the finite contributions. We compute the $\textit{\textbf{I}}$ term for this process and we have checked that the second integral of Eq.~\ref{equ_idd_dsgm_int_ist_prt} is IR safe. 
We also calculate the $\textit{\textbf{P}}$ and $\textit{\textbf{K}}$ terms as in this process there are two identified initial state partons.

There are three real emission sub-processes that can contribute to $\sigma^{NLO}$. These processes are 
\begin{equation}
a)\quad b\:\bar{b}\rightarrow W^+W^-H\:g\quad
b)\quad g\:\bar{b}\rightarrow W^+W^-H\:\bar{b}\quad
 c) \quad b\:g\rightarrow W^+W^-H\:b\:,
\end{equation}
as these processes mimic the Born level process in the soft and collinear regions. Due to large contributions, top resonance in the last two processes jeopardizes the perturbative calculation. The cross sections for these two processes are five to six times higher than the Born level cross section. One can not remove those top resonant diagrams as it will affect the gauge invariance and we have checked that the interference between resonant and non-resonant diagrams coming from the off-shell region is large, which will again ruin the perturbative computations. There are several techniques to remove these on-shell contributions safely \cite{Grazzini:2016ctr,Denner:2012yc,Cascioli:2013wga,Gehrmann:2014fva}. One can also restrict resonant top momenta out of the on-shell region and can have contributions only from the off-shell region. To implement the last technique with a standard jet veto, one needs a very large number of phase space points to get a stable cross section. The implementation of these techniques is beyond the scope of this paper. Instead of these techniques, we exclude the last two channels by assuming $b$-quark tagging with $100\%$ efficiency \cite{Bierweiler:2012kw,Baglio:2016ofi}.
\section{Numerical Results}
\label{sec:numr_res}
The sub-process $b\:\bar{b}\rightarrow W^+W^-H$ gives a significant contribution to the main process $p\:p\rightarrow W^+W^-H$. We calculate the NLO QCD contribution to this process. In particular we focus on 
the corrections to  cross sections and distributions for various polarization configurations of the final state particles. We also probe the variation of cross sections with $WWHH$ anomalous coupling.
Some of the Feynman diagrams, tree-level diagrams, as well as one-loop diagrams are heavy vector bosons, Higgs boson and top quark mediated. We use complex-mass scheme (CMS) \cite{DENNER200622} throughout our calculation to handel the resonance instabilities coming from these massive unstable particles. We take Weinberg angle as $\cos^2\theta=(m_W^2 - i\Gamma_Wm_W)/(m_Z^2 - i\Gamma_Zm_Z)$. The input SM parameters are \cite{GRAZZINI2020135399}: $m_W = 80.385$ GeV, $\Gamma_W=2.0854$ GeV, $m_Z = 91.1876$ GeV, $\Gamma_Z=2.4952$ GeV, $m_H=125$ GeV, $\Gamma_H=0.00407$ GeV, $m_t=173.2$ GeV, $\Gamma_t = 1.44262$ GeV. For the bottom-quark mass, we have used the running mass, as we have renormalized bottom quark mass in  $\rm \overline{MS}$ scheme. We have used $m_b=2.8 $ GeV which can be obtained by running the mass $m_b=4.92 $ GeV at bottom mass scale
to the Higgs boson mass scale \cite{Harlander:2003ai}. We have used this value for both LO and the NLO calculations. For the top quark, we have renormalized in on-shell scheme. There are several pieces in the one-loop calculation that contribute to total $\sigma^{NLO}$. As we have discussed above, virtual amplitudes, CT amplitudes, dipole $\textit{\textbf{I}}$, $\textit{\textbf{P}}$ and $\textit{\textbf{K}}$ terms, dipole subtracted real emission amplitudes contribute to the finite part. We find that there are significant contributions from all these pieces except dipole subtracted real emission amplitudes, which give an almost vanishing contribution.

We use {\tt CT14lo}  and {\tt CT14nlo} PDF sets \cite{Dulat:2015mca} for LO ($\sigma^{LO}$) and NLO ($\sigma^{NLO}$) cross sections calculation. We use these PDF sets through {\tt LHAPDF} \cite{Whalley:2005nh} libraries. As mentioned before, we calculate the cross sections in three different CMEs corresponding to current and proposed future colliders. We choose renormalization ($\mu_R$) and factorization ($\mu_F$) scales dynamically as 
\begin{equation}
\mu_R=\mu_F=\mu_0=\frac{1}{3}\Big(\sqrt{p_{T,W^+}^2+M^2_{W}}+\sqrt{p_{T,W^-}^2+M^2_{W}}+\sqrt{p_{T,H}^2+M^2_{H}}\Big),\label{eq:scale_mu}
\end{equation} 
where $p_{T,W}$, $p_{T,H}$ are the transverse momenta and $M_W$, $M_H$ are the masses of $W$ and Higgs bosons. We measure the scale uncertainties by varying both $\mu_R$ and $\mu_F$ independently by a factor of two around the $\mu_0$ given in Eq. \ref{eq:scale_mu}.
\subsection{Results for the SM}
\label{subsec:numr_res_sm}
We have listed the cross sections for different CMEs with their respective scale uncertainties in Table~\ref{table:bb2WWH_qcd}. As we see in Table~\ref{table:bb2WWH_qcd} the LO cross sections are $217$, $1086$ and $15258$ {\it ab}, whereas NLO cross sections are $289$, $1559$ and $23097$ {\it ab} at $14$, $27$ and $100$ TeV CMEs respectively. The cross section rapidly increases with CME as PDFs for $b$-quarks are small for lower energies. The relative enhancements \big(RE $=\frac{\sigma^{\rm NLO}_{\rm QCD}-\sigma^{\rm LO}}{\sigma^{\rm LO}}$\big) due to NLO QCD correction are also presented in that table. The RE also increases with CME and it is $33.2\%$, $43.6\%$ and $51.4\%$ for $14$, $27$ and $100$ TeV CMEs respectively. We have calculated scale uncertainty as the relative change in the cross sections for the different choices of scales within bound $0.5 \mu_0\leq\mu_R/\mu_F\leq 2\mu_0$. We see that the NLO uncertainties are a little bit higher than the LO. As there is no strong coupling ($\alpha_s$) at the Born level, the LO uncertainties come from 
the factorization scale, whereas at the NLO the uncertainties come from both, factorization as well as renormalization scales. To see the different scale uncertainties separately, we vary $\mu_R$ and $\mu_F$ independently. We see the renormalization scale uncertainty varies from $\sim -11\%$ to $\sim 0.7 \%$ and the factorization scale varies from $\sim-15.7\%$ to $\sim 17.3 \%$ at NLO depending on CMEs from $14$ to $100$ TeV.

  To get a better understanding, let us consider the diagrams in Fig.~\ref{fig:bb2wwh_tree_dia}, 
  which make contributions at the LO. These diagrams can be classified into four categories -- 1) The diagrams with
  one bottom-Yukawa coupling, 2) the diagrams with one top-Yukawa coupling, 3)
  the diagrams without these Yukawa couplings, 4) The diagrams with
  two bottom-Yukawa couplings. At the tree level, because of a change of helicity at 
  the Yukawa vertex, the diagrams of the first category do not interfere with 
  the other three categories. The second and fourth categories of 
  diagrams are t-channel diagrams. They have same helicity structure
  as the third category of diagrams. The diagrams with two bottom-Yukawa
  couplings make a very small contribution. The main contribution comes from 
  the individual square of the matrix elements of first three categories. In particular,
  the square of matrix elements of the second and third categories are individually quite 
  large, but there is also a sizable destructive interference between these two 
  categories of diagrams.  At the NLO level, in addition, we need to include contributions
  of the interference between the LO order diagrams and one-loop 
  diagrams. The relative contributions of these terms is discussed below.

\begin{table}[H]
\begin{center}
\begin{tabular}{|c|c|c|c|} 
\hline
 CME(TeV)& $\sigma^{LO}$[ab] & $\sigma^{NLO}_{QCD}$[ab]&RE\\ 
\hline
 $14$ & $217^{+16.1\%}_{-18.9\%}$&$289^{+17.6\%}_{-20.8\%}$ &$33.2\%$\\ 
$27$& $1086^{+19.2\%}_{-20.5\%}$ & $1559^{+18.0\%}_{-20.8\%}$&$43.6\%$\\ 
$100$&  $15258^{+22.0\%}_{-20.9\%}$& $23097^{+20.6\%}_{-21.0\%}$&$51.4\%$ \\ 
\hline
\end{tabular}
\caption{The LO and NLO cross sections for different collider CMEs with their respective scale uncertainties. RE is the relative enhancement of the total cross section from the Born level cross section.}
\label{table:bb2WWH_qcd}
\end{center}
\end{table}

As discussed before, we probe the contributions from different polarization configurations of the final state $W$ bosons to the LO and NLO cross sections. The right-handed, left-handed and longitudinal polarization of a $W$ boson are denoted as `+', `-', and `0'.
 The contributions of different nine polarization combinations of final state $W$ bosons are given in Table~\ref{table:bb2WWH_pol} for $14$, $27$ and $100$ TeV CMEs.  We see that the large contributions are coming from the longitudinal polarization states and among them, the `00' combination gives the largest contribution to the total cross sections.  Relative enhancement (RE) for the `00' combination increases with the CME and it becomes $\sim 117\%$ at $100$ TeV. In the $R_{\xi}$ gauge, the pseudo Goldstone bosons couple to massive fermions with a coupling proportional to the 
 mass of the fermion. These pseudo Goldstone bosons 
 represents the longitudinal polarization state of
 a $W$ boson. This leads to larger values of the cross section  in longitudinal polarization combinations due to heavy fermion mediated diagrams. These longitudinal polarization modes are useful for background suppression to this process. The background may come from the processes with gauge bosons or gluons or photons couplings with light quarks. The negligible masses of the light quarks ($u,d,s$ and $c$) lead to the suppression of backgrounds in polarization combinations that include longitudinal polarization. 
\begin{table}
\begin{center}

\begin{tabular}{|c|c|c|c|c|c|c|c|c|c|}
\hline
\multirow{1}{*}{Pol.} & \multicolumn{3}{c|}{$14$\:TeV\:} & %
    \multicolumn{3}{c|}{$27$\:TeV\:}& %
    \multicolumn{3}{c|}{$100$\:TeV\:}\\
\cline{2-10}
\multirow{1}{*}{com.} & $\sigma^{ LO}$ & $\sigma^{NLO}_{QCD}$ & RE(\%) &$\sigma^{LO}$ & $\sigma^{NLO}_{QCD}$& RE(\%)&$\sigma^{ LO}$ & $\sigma^{NLO}_{QCD}$ & RE(\%)\\
\hline
 $++$ &$13$&$18$&$38.5$&$60$&$88$&$46.7$&$702$&$1056$&$50.4$ \\
$+-$&$18$& $25$ &$38.9$&$82$&$127$&$54.9$ &$965$&$1499$&$55.3$\\
$+0$&$37$ & $49$&$32.4$&$187$& $266$& $42.2$&$2568$&$3336$&$29.9$\\
$-+$& $4$& $6$&$50.0$&$19$& $28$ &$47.4$&$229$&$334$&$45.9$\\
$--$&$13$ & $18$&$38.5$&$61$& $89$ &$45.9$&$707$&$1044$&$47.7$\\
$-0$&$22$&  $28$&$27.3$&$108$& $144$ &$33.3$&$1454$&$1346$&$-7.4\quad$\\
$0+$& $22$& $28$&$27.3$&$109$& $145$&$33.0$ &$1470$&$1216$&$-17.3\quad$\\
$0-$&$37$& $49$&$32.4$&$186$& $268$&$44.1$ &$2583$&$3151$&$22.0$\\
$00$&$51$&  $67$&$31.4$&$274$& $404$ &$47.4$&$4490$&$9748$&$117.1$\\
\hline
$\sum$&$217$&  $289$&$32.2$&$1086$& $1559$ &$43.6$&$15258$&$23097$&$51.4$\\
\hline

\end{tabular}
\caption{The LO and NLO cross sections and their relative enhancements (RE) for different polarization combinations of final state $W$ bosons and their sum at $14$, $27$ and $100$ TeV CMEs. The results are in {\it ab} unit.}
\label{table:bb2WWH_pol}
\end{center}
\end{table}

At the 100 TeV CME, the NLO corrections are largest for `00' combination
of $W$ boson polarization. In this case, the largest positive contribution comes
from the interference of the second category LO diagrams and one-loop diagrams
corresponding to the third category and vice-versa. But the interference
of second category LO diagrams and corresponding one-loop diagrams
is negative; the same is true for the interference
of the third category LO diagrams and corresponding one-loop diagrams. 
A small positive contribution is also obtained from the interference of the 
first category LO diagrams and corresponding one-loop diagrams. These 
diagrams are responsible for the $WWHH$ coupling dependence of the
process.

To find the relative contribution of the bottom-bottom scattering to the $pp\rightarrow W^+W^-H$ process, we compute the cross sections in other $q\bar{q}$ channels along with the $b\bar{b}$ channel. The results are presented in Table~\ref{table:qq2WWH_qcd}. The cross sections in $q\bar{q}$ channels ($4$FNS) have been calculated using {\tt MagGraph5\_aMC5@NLO} \cite{Alwall:2014hca}. {\tt MagGraph5\_aMC5@NLO} cannot compute the one-loop QCD corrections to the $b\bar{b}$ channel due to the presence of the
resonances in the diagrams. As we see in Table~\ref{table:qq2WWH_qcd}, the $b\bar{b}$ channel gives significant contributions to the full process $pp\rightarrow W^+W^-H$. The $b\bar{b}$ channel contributes $\sim2.3\%$ to the LO and $\sim2.1\%$ to the NLO cross sections  at $14$ TeV and  $\sim14.1\%$ to the LO and $\sim12.5\%$ to the NLO  cross sections at $100$ TeV of process $pp\rightarrow W^+W^-H$. These numbers are calculated without the channels $gg\rightarrow W^+W^-H$, which can also add a significant contribution to the process $pp\rightarrow W^+W^-H$~\cite{Agrawal:2019ffb,Baglio:2016ofi}. If one adds $gg$ channel, these numbers will be changed accordingly. As we see in Table~\ref{table:qq2WWH_qcd}, the corrections are pretty high in $q\bar{q}$ channels ($4$FNS). In those channels, {\tt MadGraph5\_aMC@NLO} includes all real emission diagrams and the results are complete, but we impose jet veto on $b$-quarks with $100\%$ efficiency for real emission diagrams to overcome certain difficulties discussed in Sec. \ref{sec:calc_check}. The proper inclusion of all real emission diagrams may increase the QCD correction significantly in $b\bar{b}$ channel.  

\begin{table}[H]
\begin{center}

\begin{tabular}{|c|c|c|c|c|}
\cline{1-5}
{ \multirow{2}{*}{channel}}& \multicolumn{2}{c|}{$14$\:TeV\:} & %
    \multicolumn{2}{c|}{$100$\:TeV\:}\\
\cline{2-5}
 & $\sigma^{LO}$[ab] & $\sigma^{NLO}_{QCD}$[ab] &$\sigma^{LO}$[ab] & $\sigma^{NLO}_{QCD}$[ab] \\
\hline 
$4$FNS& $9460$&$13250$&$108100$&$185100$ \\
\hline
$b\bar{b}$& $217$&$289$&$15258$&$23097$ \\
\hline
\end{tabular}
\end{center}
\caption{The LO and NLO cross sections in $4$FNS and the $b {\bar b}$ channel at $14$ and $100$ TeV CMEs. The results for $4$FNS have been obtained using {\tt MagGraph5\_aMC5@NLO}. The $b\bar{b}$ channel results are from our code.  }
\label{table:qq2WWH_qcd}
\end{table}

\begin{figure}[!hbt]
\includegraphics [angle=0,width=0.5\linewidth]{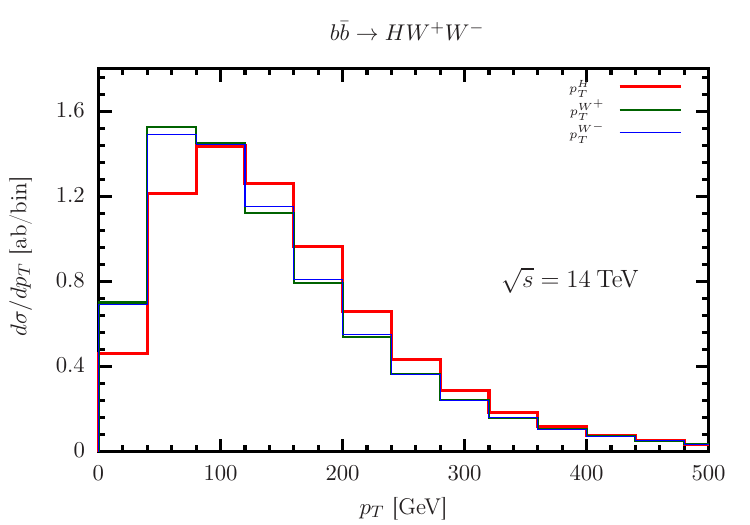}
\includegraphics [angle=0,width=0.5\linewidth]{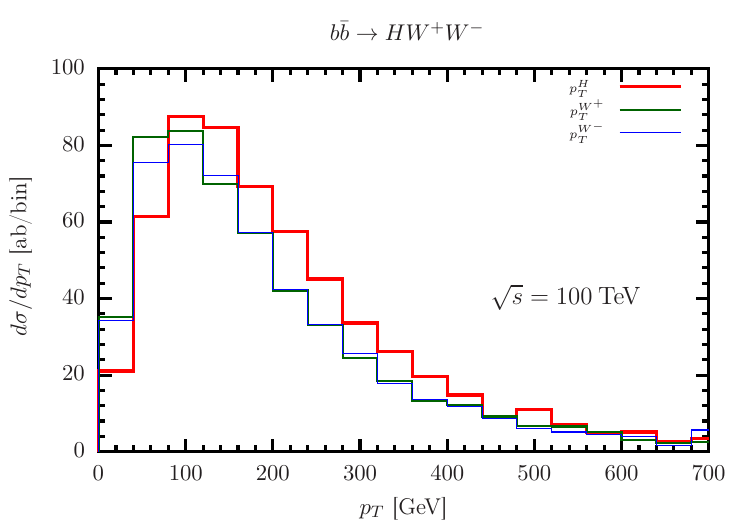}
\includegraphics [angle=0,width=0.5\linewidth]{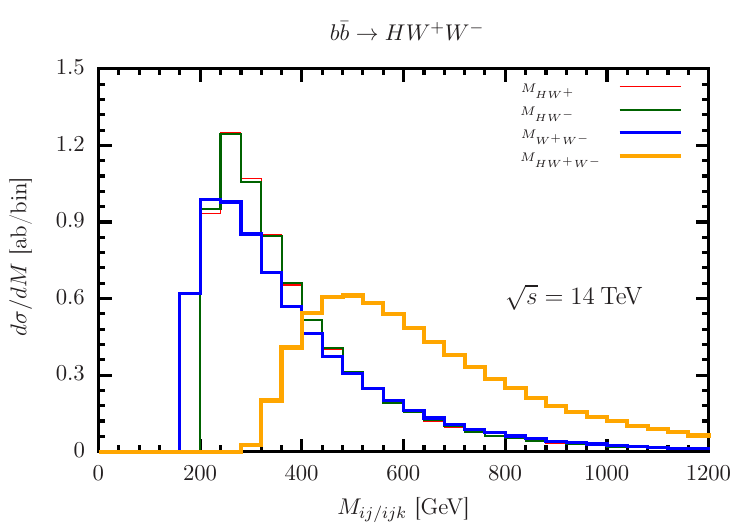}
\includegraphics [angle=0,width=0.5\linewidth]{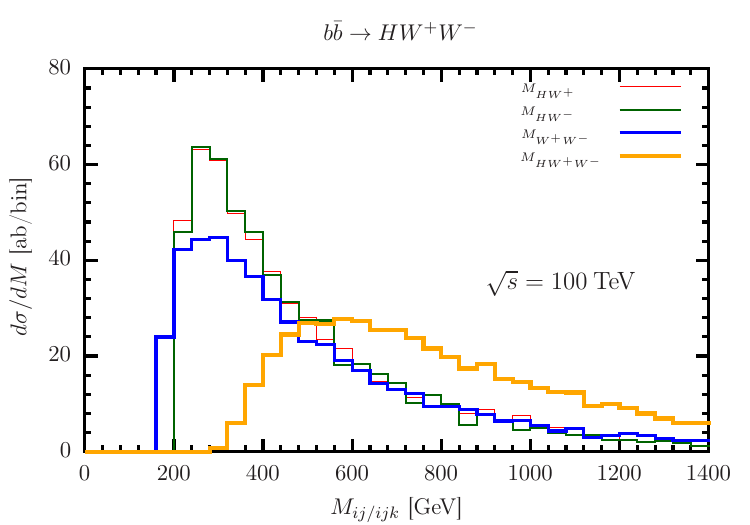}
\caption{ The NLO differential cross section distribution with respect to transverse momentum ($p_T$) and invariant masses ($M_{ij/ijk}$) for $14$ and $100$ TeV CMEs.}
\label{fig:bb2wwh_NLO_pt_inv_mas}
\end{figure}

\begin{figure}[!hbt]
\includegraphics [angle=0,width=0.5\linewidth]{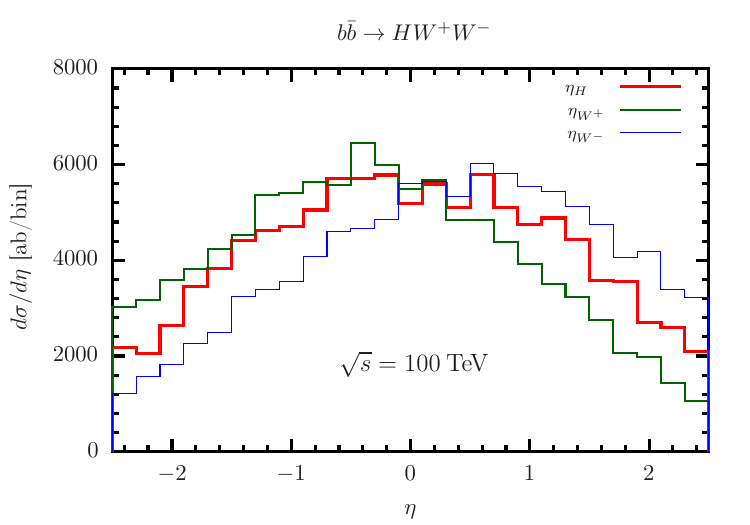}
\includegraphics [angle=0,width=0.5\linewidth]{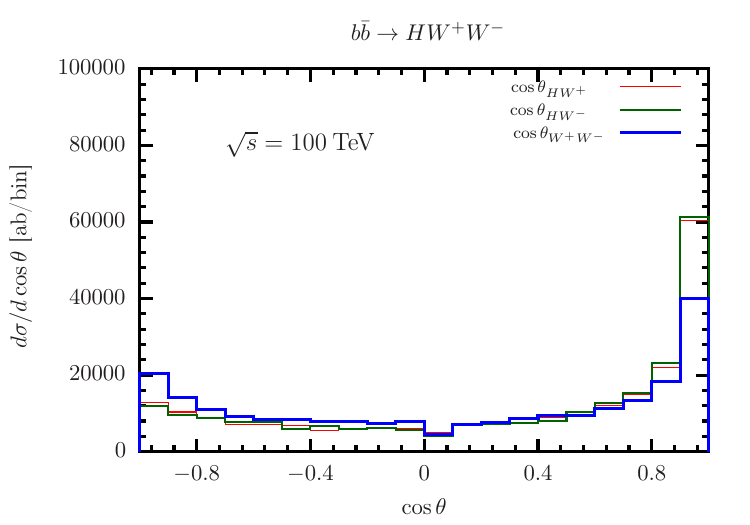}
\caption{ The NLO differential cross section distribution with respect to rapidity ($\eta$) and cosine angle ($\cos\theta$) between the two final state particles at $100$ TeV CME. Plots for 14 TeV are similar.} 
\label{fig:bb2wwh_NLO_eta_cos0}
\end{figure}

\begin{figure}[!hbt]
\includegraphics [angle=0,width=0.5\linewidth]{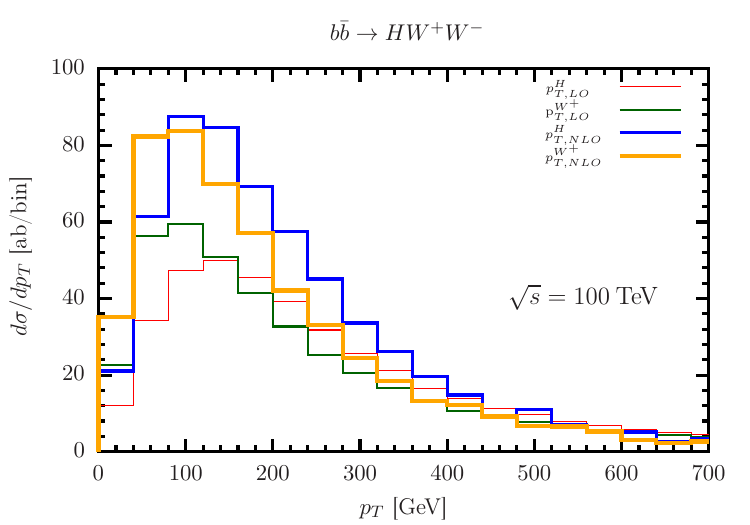}
\includegraphics [angle=0,width=0.5\linewidth]{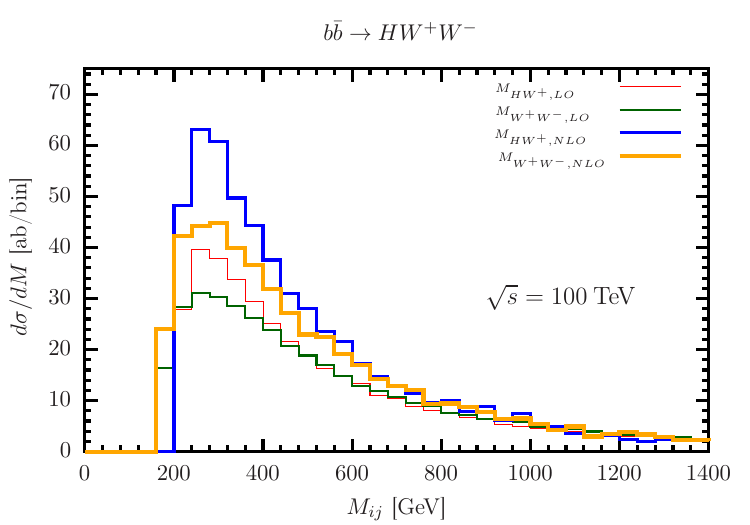}
	\hspace*{4cm} \includegraphics [angle=0,width=0.5\linewidth]{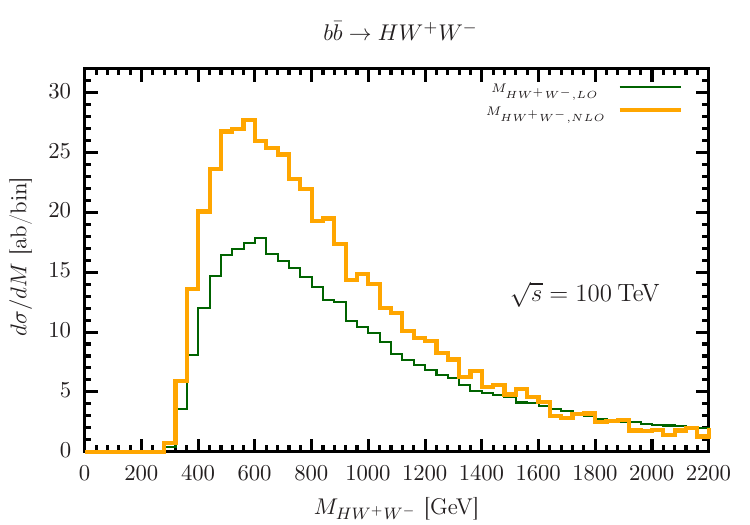}
\caption{The LO and NLO differential cross section distribution with respect to transverse momentums ($p_T$) and invariant masses ($M_{ij/ijk}$) for $100$ TeV CME.} 
\label{fig:bb2wwh_LO_NLO_pt_inv_mas}
\end{figure}

We have plotted a few different kinematical distributions at the NLO level in Fig.~\ref{fig:bb2wwh_NLO_pt_inv_mas} and Fig.~\ref{fig:bb2wwh_NLO_eta_cos0}. In Fig.~\ref{fig:bb2wwh_NLO_pt_inv_mas}, the upper-panel histograms are for the transverse momentum($p_T$) of final state particles at $14$ and $100$ TeV CMEs. As expected $W$ bosons $p_T$ distributions almost coincide with each other. The $p_T$ distributions of the Higgs boson is a bit harder. The differential cross sections are maximum around $p_T=100$ TeV for the Higgs boson and near $p_T=80$ TeV for the $W$ bosons. In the lower panel of Fig.~\ref{fig:bb2wwh_NLO_pt_inv_mas}, we have plotted the histograms for the different invariant masses ($M_{ij,ijk}$) at $14$ and $100$ TeV CMEs. Invariant mass thresholds are around $\sim210$, $\sim170$, $\sim290$ TeV and distributions are peaked around $250$, $230$, $490$ TeV for $M_{HW}$, $M_{WW}$ and $M_{HWW}$ respectively. In Fig.~\ref{fig:bb2wwh_NLO_eta_cos0}, we have plotted differential cross sections with respect to rapidity ($\eta$) of final state particles and cosine angle ($\cos\theta$) between the two final state particles for $100$ TeV CME. The distributions have maxima around $\eta=$ $0$, $-0.4$ and $0.4$ for the Higgs boson, $W^+$ and $W^-$ boson respectively. From the $\cos\theta$ plot in Fig.~\ref{fig:bb2wwh_NLO_eta_cos0}, it is clear that maximum contributions come when two final state particles are near to collinear region i.e, $\theta\sim0\:\rm{or}\:\pi$. In Fig.~\ref{fig:bb2wwh_LO_NLO_pt_inv_mas}, we have plotted the LO and the NLO  distributions to show the effect of the one-loop QCD corrections. The distributions are for only 100 TeV CME. The behavior for the 14 TeV CME is similar.
In the upper half of Fig.~\ref{fig:bb2wwh_LO_NLO_pt_inv_mas}, $p_T$ distributions are plotted and in the lower half of Fig.~\ref{fig:bb2wwh_LO_NLO_pt_inv_mas}, invariant masses have been plotted at $100$ TeV CME. We see an increase for the smaller values of the kinematic variables in all the plotted distributions.
\subsection{Anomalous coupling effect}
\label{subsec:numr_res_bsm}
As we discussed in the introduction, $VVHH$ coupling in SM is only loosely bound so far. We allow $WWHH$ coupling to deviate from the SM value in the search for new physics in the context of $\kappa$-framework~\cite{LHCHiggsCrossSectionWorkingGroup:2012nn,Ghezzi:2015vva}. In the $\kappa$ framework, only SM couplings deviate by a scale factor. $\kappa$ is defined as the deviation from the SM coupling. It is a scale factor. Although $WWH$ and $WWHH$ couplings in the SM are related but in many Effective Field Theory frameworks, these couplings can vary independently~\cite{Bishara_2017}. As there is no QCD correction to $WWHH$-vertex, the anomalous coupling will not affect the renormalization.  We have checked that the UV and IR poles cancel with the same CTs and dipole terms as in the SM. We denote the deviation of $VVHH$ coupling from the SM as $\kappa_{V_2H_2}$ and $\kappa_{V_2H_2}=1$ in the SM. In this framework, we vary $\kappa_{V_2H_2}$ from $-2.0$ to $2.0$ and calculate the relative increment \big($\rm {RI}=\frac{\sigma_{\kappa}-\sigma_{SM}}{\sigma_{SM}}$\big) in the total cross section, whereas the $\kappa$ for other SM couplings are set to $1$. We choose $\kappa_{V_2H_2}=-2.0,\:-1.0,\:1.5,\:2.0$ and tabulate the results for the LO and NLO cross sections at $14$ and $100$ TeV CMEs in Table~\ref{table:bb2WWH_BSM}. It is clear from Table~\ref{table:bb2WWH_BSM} that cross sections are lower than SM predictions when $\kappa_{V_2H_2}$ is positive and higher than the SM predictions when $\kappa_{V_2H_2}$ is negative.
(We note that this is due to the interference pattern within
the diagrams of the first category.)  There is not a significant relative increment ($-0.3\rm {\:to\:}+2.1\%$) at $14$ TeV. At $100$ TeV, relative increments vary from $-2.2\%$ to $+11.4\%$ for the LO cross section and from $-2.1\%$ to $+10.3\%$ for the NLO cross section. There is also $HHH$ coupling involved in this process. We also observe the $HHH$ anomalous coupling effect on the total cross sections. We vary the corresponding $\kappa_{H_3}$ from $0.5$ to $2.0$. We see that there is no significant change in the LO as well as the NLO cross sections and the relative increment are smaller than $1\%$ for $14$ and $100$ TeV CMEs. We see something very interesting in 
Table~\ref{table:bb2WWH_BSM_00}. The cross sections for the two longitudinally polarized $W$ bosons configuration have a stronger dependence on the $\kappa_{V_2H_2}$. For the NLO cross sections the dependence is almost twice as strong as in the total cross sections. This again demonstrates the importance of measuring the polarization of the W bosons. However this dependence is weaker as compared to the LO cross sections. The difference in this dependence
underlines the importance of considering the NLO corrections.
\begin{table}
\begin{center}
\begin{tabular}{|c|c|c|c|}
\hline
CME(TeV)&$\kappa_{V_2H_2}$&$\sigma^{LO}$[ab]\quad\quad RI\quad\quad\quad&$\sigma^{NLO}$[ab]\quad\quad RI\quad\quad\quad\\
\hline
\multirow{5}{*}{$14$}
&$1.0$ (SM)&$217\quad\quad\quad\quad\quad\quad$&$289\quad\quad\quad\quad\quad\quad$\\
\cline{2-4}
&$\phantom{-}2.0$&$216$\:\:\quad\quad$[-0.5\%]$&$288$\:\:\quad\quad$[-0.3\%]$\\
&$\phantom{-}1.5$&$216$\:\:\quad\quad$[-0.5\%]$&$289$\:\:\quad\quad\quad$[0.0\%]$\\
&$-1$&$220$\:\:\quad\quad$[+1.4\%]$&$293$\quad\quad\:\:$[+1.4\%]$\\
&$-2.0$&$222$\:\:\quad\quad$[+2.3\%]$&$295$\:\:\quad\quad$[+2.1\%]$\\
\hline
\multirow{5}{*}{$100$}
&$1.0$(SM)&$15258\quad\quad\quad\quad\quad\quad\:\:\:$&$23097\quad\quad\quad\quad\quad\quad$\\
\cline{2-4}
&$\phantom{-}2.0$&$14925$\:\:\quad\quad$[-2.2\%]\quad$&$22607$\:\:\quad\quad$[-2.1\%]$\\
&$\phantom{-}1.5$&$15048$\:\:\quad\quad$[-1.4\%]\quad$&$22810$\:\:\quad\quad$[-1.2\%]$\\
&$-1.0$&$16296$\:\quad\quad$[+6.8\%]\quad$&$24760$\quad\:\quad$[+7.2\%]$\\
&$-2.0$&$16997$\:\quad\quad$[+11.4\%]\quad$&$25465$\quad\quad$[+10.3\%]$\\
\hline
\end{tabular}
\caption{Effect of anomalous $WWHH$ coupling on the LO and NLO cross sections at the $14$ and $100$ TeV CMEs.}
\label{table:bb2WWH_BSM}
\end{center}
\end{table}

\begin{table}
\begin{center}
\begin{tabular}{|c|c|c|c|}
\hline

$\kappa_{V_2H_2}$&$\sigma^{LO}$[ab]\quad\quad RI\quad\quad\quad&$\sigma^{NLO}$[ab]\quad\quad RI\quad\quad\quad\\
\hline
$1.0$ (SM)&$4490\quad\:\:\quad\quad\quad\quad\quad$&$9748\:\:\quad\quad\quad\quad\quad\quad$ \\
$\phantom{-}2.0$&$4159$\:\:\quad\quad$[-7.4\%]$&$9544$\:\:\quad\quad$[-2.1\%]$\\
$\phantom{-}1.5$&$4333$\:\:\:\quad\quad$[-3.5\%]$\:&$9654$\quad\quad\:\:$[-1.0\%]$\:\:\\
$-1.0$&$5493$\:\quad\quad$[+22.3\%]$&$11117$\quad\quad\:\:$[+14.0\%]$\\
$-2.0$&$6164$\:\quad\quad$[+37.2\%]$&$11993$\:\:\quad\quad$[+23.0\%]$\\
\hline
\end{tabular}
\caption{Effect of anomalous $VVHH$ coupling on `00' mode at $100$ TeV CME.}
\label{table:bb2WWH_BSM_00}
\end{center}
\end{table}

\begin{figure}[!hbt]
\includegraphics [angle=0,width=0.5\linewidth]{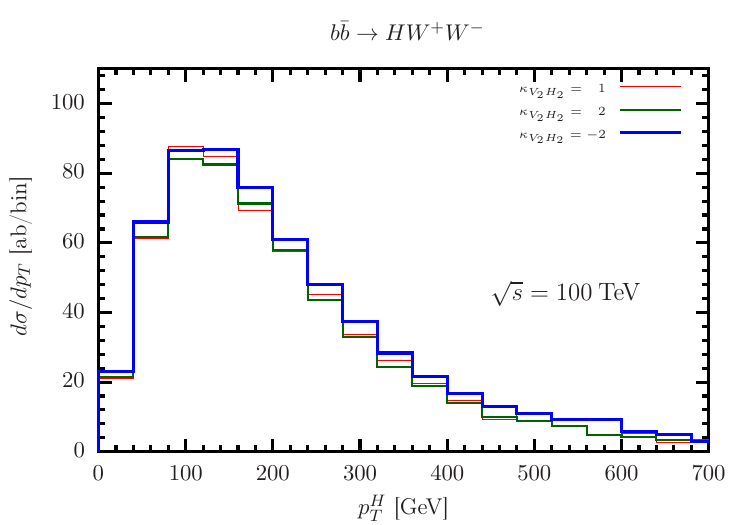}
\includegraphics [angle=0,width=0.5\linewidth]{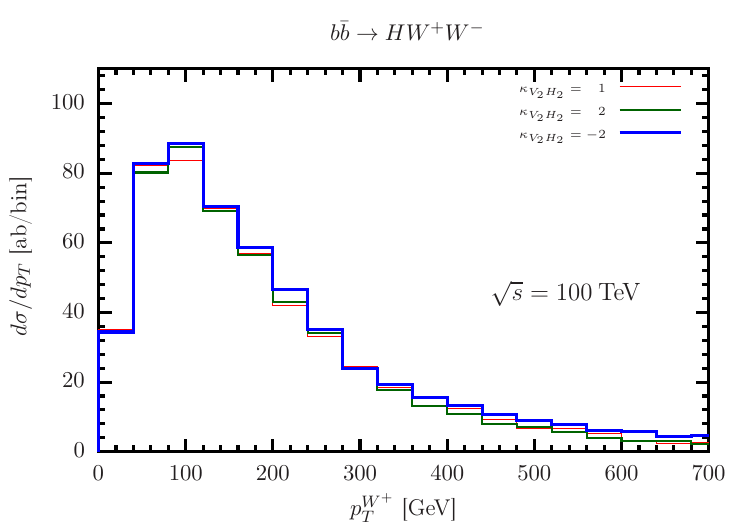}
\includegraphics [angle=0,width=0.5\linewidth]{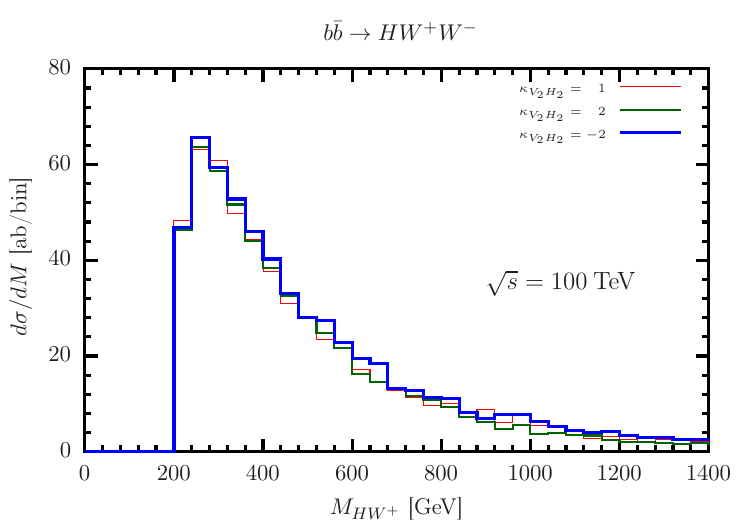}
\includegraphics [angle=0,width=0.5\linewidth]{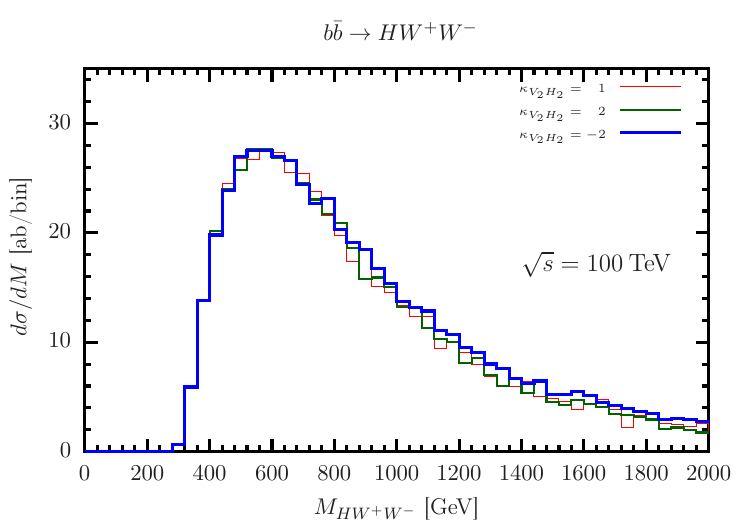}
\caption{Effect of anomalous $VVHH$ coupling on differential cross section distribution at $100$ TeV CME. Upper panel plots are for the transverse momentum of Higgs boson($p^H_T$) and $W^+$ boson ($p_T^{W^+}$). Lower panel plots are for the  $H\textendash W^+$      ($M_{HW^+}$) and $H\textendash W^+\textendash W^-$ ($M_{HW^+W^-}$) invariant masses.}  
\label{fig:bb2wwh_BSM_NLO_pt_inv_mas}
\end{figure}

In Fig.~\ref{fig:bb2wwh_BSM_NLO_pt_inv_mas}, we have plotted the NLO differential cross section distributions for the Higgs boson and $W^+$ boson transverse momenta,
and different invariant masses. The maxima of the differential cross sections are 
about at the same value  as for the SM. As there is not that much increase for $\kappa_{V_2H_2}=2$, the corresponding distributions nearly overlap with the SM. On the other hand, we see a sharp deviation in distributions from the SM for $\kappa_{V_2H_2}=-2$. Interesting fact about the negative $\kappa_{V_2H_2}$ is that
the distribution are harder. This difference in the shape can be used in putting a strong bound on the coupling. One could put a cut on $p_T^{W}$, or one of the plotted
invariant masses to select events with a larger component of anomalous events.
\section{Conclusion}
\label{sec:conclusion}
    In this chapter, we have focused on the NLO QCD corrections to $b {\bar b} \to WWH$. This process has 
	significant dependence on $WWHH$ coupling. But, the contribution of this process to $pp \to WWH$
	is only about $15-20\%$ of that of light quark scattering. 
This is where the consideration of the polarization of  the W bosons helps. 
When  both the $W$ bosons are longitudinally polarized, then this fraction can increase to about $50\%$.
 It turns out that the NLO QCD corrections are
also the largest for this polarization configuration, making the
dependence on the $WWHH$ coupling even stronger.
For example, at the 100 TeV CME, the NLO corrections are about $51\%$, but the corrections are about $117\%$, when both final state $W$ bosons are longitudinally polarized.
Our study suggests that the measurement of the polarization of the final state
$W/Z$ bosons can be a useful tool to measure the
couplings of the vector bosons and Higgs boson. We have also examined the
effect of the variation of $\kappa_{V_2H_2}$. The variation in the cross section
can be twice as large when we consider longitudinally polarized $W$ bosons. In addition, we find that the invariant mass and the $p_T^{W}$ distributions are considerably harder for the negative values of $\kappa_{V_2H_2}$. This can also be useful to put a stronger bound on the coupling. However, to find the bound, one would need to do a detailed background analysis which we leave for the future.

\chapter[Effect of anomalous HHH and ZZHH couplings on
the decay width\\ of $H\rightarrow \nu_e\bar{\nu}_e\nu_\mu\bar{\nu}_\mu$]{Effect of anomalous HHH and ZZHH couplings on
the decay width of $H\rightarrow \nu_e\bar{\nu}_e\nu_\mu\bar{\nu}_\mu$}
\label{chap_h24nu}
\section{Introduction}

\label{sec_h24nu_intro}
%
 The couplings of the Higgs boson can be determined either
 through a production process or the decays. Some of the
 couplings of the Higgs boson, like its self couplings and
 quartic couplings with gauge bosons are hard to determine.
 Even at the high-luminosity LHC (HL-LHC), it is not clear
 if these couplings can be measured with good enough precision.
 This is because of the requirement of multiple Higgs bosons
 and/or gauge bosons
 in the production process.
 Another avenue to determine these couplings is electroweak
 radiative corrections to either a single Higgs boson production
 process (like vector-boson fusion or associated production
 with a vector boson) or decays like $H \to V V^*$ (V is either
 W or Z boson.). One can also study two loop corrections to
 a few processes to explore the possibilities of measuring
 these couplings.

  In this chapter, we consider the decay process $H\rightarrow \nu_e\bar{\nu}_e\nu_\mu\bar{\nu}_\mu$. We compute one-loop electroweak corrections to this process. These corrections depend on the trilinear Higgs boson
  coupling $HHH$ and the quartic coupling $ZZHH$. We explore the possibility of measuring these couplings in this decay process. 
  We use $\kappa$-framework to determine this dependence. We have focused initially
  at this process to avoid complications due to final state radiation
  when there is a charged lepton in the final state. That process has been discussed in the next chapter.
   There are a few groups who have been studied electroweak corrections to $H\rightarrow 4l$ channel \cite{BREDENSTEIN2006131,Boselli:2015aha}.  There are loose experimental bounds on these couplings.
   We recall the discussion in the introduction chapter about the present bounds on $HHH$ and $VVHH$ couplings.   
    The ATLAS collaboration has put a bound on $VVHH$ coupling using the vector boson fusion (VBF) mechanism of
 a pair of Higgs boson, and using 126 fb$^{-1}$ of data at 13
    TeV. They obtained a  bound of $ -0.43 < \kappa_{V_2H_2} < 2.56$
 at 95$\%$ confidence level \cite{Aad_2021}. Here $ \kappa_{V_2H_2}$ is the scaling factor for the $VVHH$ coupling. However, in this process, both couplings $WWHH$ and $ZZHH$ are present. The
    process $p p \to HHV$, with a $W$ or a $Z$ boson, allows us to separately measure $WWHH$ and $ZZHH$ couplings. The expected bound
    from the $WHH$ production
  at the HL-LHC is $ -9.4 < \kappa_{V_2H_2} < 7.9$ \cite{Nordstrom:2018ceg}, which is quite weak. The decay process under consideration
  depends on $ZZHH$, not on $WWHH$.
The $\kappa_{HHH}$, the scaling factor for the $HHH$ coupling
is largely unconstrained by the experimental data.
According to the future projections for HL-LHC, $-2 < \kappa_{HHH} < 8$ \cite{TheATLAScollaboration:2014scd}.
   As this process is sensitive to $HHH$ and $VVHH$ coupling at one-loop level, one can probe the the effect of the anomalous $HHH$ and $VVHH$ couplings in this process within the above mentioned experimental bounds.
 In the next section, we discuss the process and the diagrams that contribute to it. In the section 3, we discuss how we did the calculation. In the section 4, we discuss our results. In the last section, we have some conclusions.

\section{The Process}
\label{sec_h24nu_prcs}
We are interested in calculating the decay width of the decay channel $H\rightarrow \nu_e \bar{\nu}_e \nu_\mu \bar{\nu}_\mu$. As our aim is to see the effect of anomalous $HHH$ and $ZZHH$ couplings in this channel, we calculate NLO electroweak (EW) correction. There are a few one-loop diagrams where we can vary $HHH$ and $ZZHH$ couplings to study the effects on the decay width.

\begin{figure}[!h]
  \begin{center}
\includegraphics [angle=0,width=0.5\linewidth]{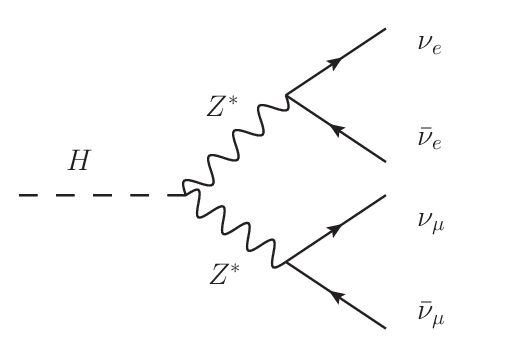}\\
	\caption{LO Feynman diagram for decay $H\rightarrow \nu_e \bar{\nu}_e \nu_\mu \bar{\nu}_\mu$.  }
	\label{fig_h24nu_tree_dia}
	\end{center}
\end{figure}

\begin{figure}[!h]
  \begin{center}
\includegraphics [angle=0,width=1\linewidth]{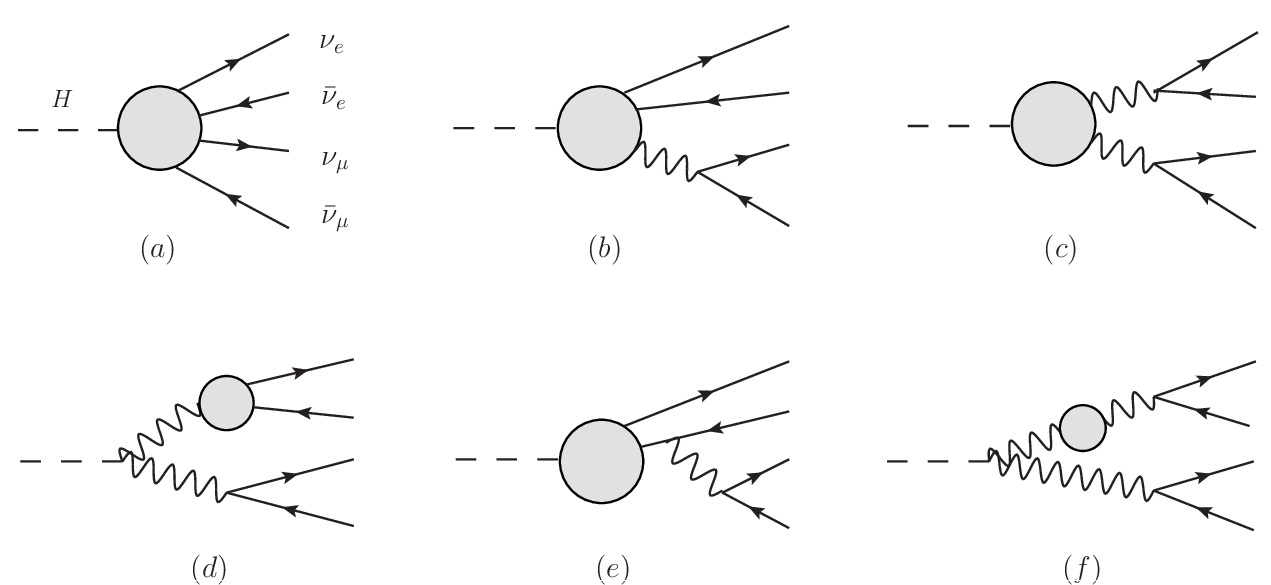}\\
	\caption{Generic NLO EW virtual Feynman diagrams for $H\rightarrow \nu_e \bar{\nu}_e \nu_\mu \bar{\nu}_\mu$.  }
	\label{fig_h24nu_loop_dia}
	\end{center}
\end{figure}

At the leading order (LO), there is only one Feynman diagram as shown in Fig.~\ref{fig_h24nu_tree_dia}, as we allow one $Z$ to decay into muon  neutrinos and another into electron neutrinos. In the calculation of
the one-loop level amplitudes, there are a total of $118$ Feynman diagrams. As the process is $1\rightarrow 4$, the virtual diagrams are of pentagon, box, triangle and bubble-types. The generic one-loop diagrams are shown in Fig.~\ref{fig_h24nu_loop_dia}.

\begin{figure}[!h]
  \begin{center}
\includegraphics [angle=0,width=1\linewidth]{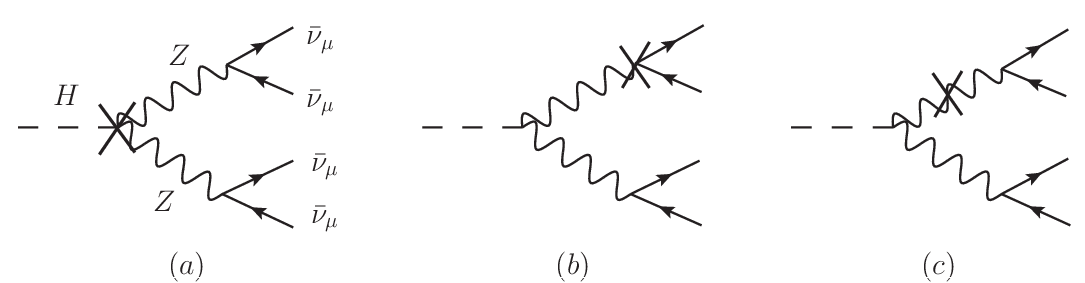}\\
	\caption{Counterterm diagrams for NLO EW correction to $H\rightarrow \nu_e \bar{\nu}_e \nu_\mu \bar{\nu}_\mu$.  }
	\label{fig_h24nu_ct_dia}
	\end{center}
\end{figure}
\begin{figure}[!h]
  \begin{center}
\includegraphics [angle=0,width=1\linewidth]{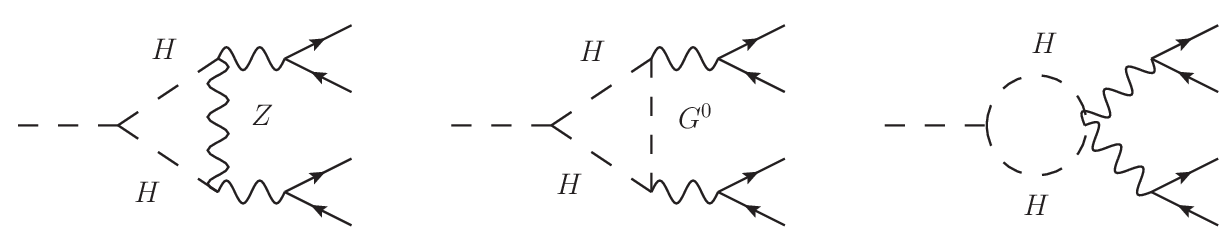}\\
	\caption{ NLO EW virtual diagrams with $HHH$ and $ZZHH$ couplings.  }
	\label{fig_h24nu_hhh_dia}
	\end{center}
\end{figure}

\begin{figure}[!h]
  \begin{center}
\includegraphics [angle=0,width=1\linewidth]{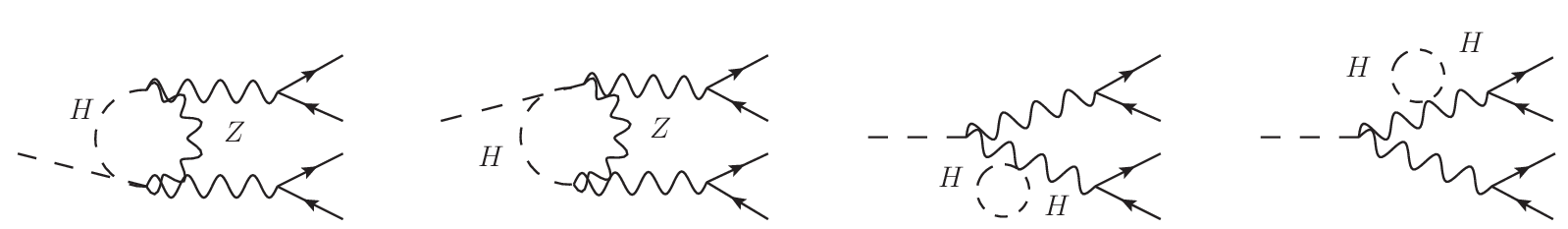}\\
	\caption{ NLO EW virtual diagrams with $ZZHH$ couplings.  }
	\label{fig_h24nu_zzhh_dia}
	\end{center}
\end{figure}
\begin{figure}[!h]
  \begin{center}
\includegraphics [angle=0,width=1\linewidth]{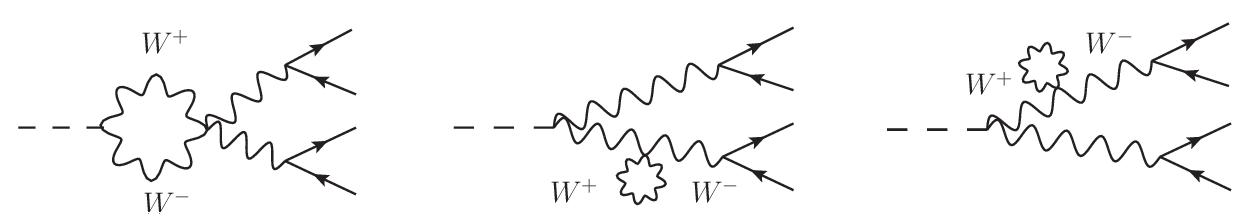}\\
	\caption{ NLO EW virtual diagrams with $ZZWW$ couplings.  }
	\label{fig_h24nu_wwzz_dia}
	\end{center}
\end{figure}

The generic diagram $(a)$ displayed in Fig.~\ref{fig_h24nu_loop_dia}, is a pentagon-type diagram. There are $6$ pentagon-type diagrams in this process.  The generic diagram $(b)$ displayed in Fig.~\ref{fig_h24nu_loop_dia}, is a box-type diagram. There is a another similar generic diagram with other neutrinos that is not shown in Fig.~\ref{fig_h24nu_loop_dia}. There are $12$ box-type diagrams in this process. The generic $(c)$ diagram is the correction to $HZZ$-vertex. This generic diagram has both triangle and bubble-type diagrams. We have a total of $40$ virtual diagrams related to this generic diagram, out of which $30$ are triangle and $10$ are bubble-type diagrams. $HZZ$-vertex correction diagrams are relevant in this study, as there are a few diagrams where we can introduce the effect of anomalous $HHH$, $ZZHH$ and $ZZWW$ couplings. The related diagrams are shown in Fig.~\ref{fig_h24nu_hhh_dia}, Fig.~\ref{fig_h24nu_zzhh_dia} and Fig.~\ref{fig_h24nu_wwzz_dia}. The generic diagram $(d)$ in the second row of Fig.~\ref{fig_h24nu_loop_dia}, is the virtual correction to $Z\nu\bar{\nu}$ vertex. We also have one more similar generic diagram for other $Z\nu\bar{\nu}$ vertex. There are a total of $6$ triangle-type virtual diagrams related to these generic diagrams. The generic diagram $(e)$ in the second row of Fig.~\ref{fig_h24nu_loop_dia}, is the triangle-type diagram and there is a another similar generic diagram with Higgs boson and other neutrinos. There are $8$ triangle-type diagrams related to these generic diagrams. The last generic diagram $(f)$ in Fig.~\ref{fig_h24nu_loop_dia}, represents the $Z$ boson self-energy diagrams. There is another similar generic diagram with another $Z$ boson that has not been shown in Fig.~\ref{fig_h24nu_loop_dia}. All such diagrams are bubble-type diagrams. There is also dependence on $ZZHH$ and $ZZWW$ couplings from the $Z$ boson self-energy diagrams. The related diagrams are shown in Fig.~\ref{fig_h24nu_zzhh_dia} and Fig.~\ref{fig_h24nu_wwzz_dia}. We have also listed the counterterm (CT) diagrams in Fig.~\ref{fig_h24nu_ct_dia}. There are another two CT diagrams similar to diagrams $(b)$ and $(c)$ related to other $Z\nu\bar{\nu}$-vertex and other $Z$ boson self-energy diagrams. These five CT diagrams cancel all UV divergence from the virtual amplitudes. As shown in the Fig.~\ref{fig_h24nu_ct_dia}, diagram $(a)$ is the CT diagram for $HZZ$-vertex, diagram $(b)$ is the CT diagram for $Z\nu\bar{\nu}$-vertex and, diagram $(c)$ is the CT diagram for $Z$ boson self-energy diagrams. The computation of CT diagrams also involve the self-energy 
diagrams corresponding to $Z$ boson, $W$ boson and Higgs boson where we can introduce anomalous  $HHH$, $ZZHH$ and $ZZWW$ couplings. A detailed study about the effect of anomalous coupling has been discussed in Sec.~\ref{subsec_h24nu_renorm_cms}. There are no real emission diagrams as the tree-level diagram (Fig.~\ref{fig_h24nu_tree_dia}) do not have any charged gauge bosons or charged leptons to emit photons. All diagrams have been generated using {\tt FeynArts} \cite{Hahn:2000kx}, a {\tt Mathematica} package.

\section{Calculations}
\label{sec_h24nu_calc_check}
There are a few hundred diagrams at one-loop level. As we treat leptons and light quarks to be massless, a large set of diagrams become zero because of vanishing coupling with the scalars. We also ignore the tadpole diagrams as the renormalization condition will set them to zero. With these considerations, we are left with $118$ diagrams to compute as mentioned in Sec.~\ref{sec_h24nu_prcs}. There are another type of diagrams similar to generic diagram $(f)$ in Fig.~\ref{fig_h24nu_loop_dia}, which are the bubble-type diagrams for the
mixed propagators between a Goldstone boson and $Z$ boson. These diagrams do not contribute.  We use helicity formalism to compute the process amplitudes at one-loop level as well as at tree level. First we classify the one-loop level diagrams in a few set of prototype amplitudes. Then we compute all virtual diagrams with the help of these prototype amplitudes by suitable crossing, mass and coupling choices.
 Following the spinor helicity formalism given in chapter~\ref{chap_shf}, we compute the helicity amplitudes. The tree-level helicity amplitudes can be computed easily with the spinor products $[pq]$ and $\langle pq \rangle$. To calculate one-loop level helicity amplitude, we also use the vector current $\langle p \gamma^\mu q ]$. The functional form  of $[pq]$ and $\langle p \gamma^\mu q ]$ are given in Sec.~\ref{sec_shf_ff_sp_vec}.
  We calculate one-loop amplitudes in 't Hooft-Veltman (HV) scheme \cite{THOOFT1972189}. In this scheme, the loop part (`unobserved') is computed in $d$-dimension and rest of the amplitude (`observed') is computed in $4$-dimension \cite{PhysRevD.84.094021}. The gamma matrix, loop momentum algebra in `unobserved' part has been done in $d$-dimension. In this process, we have two fermion-loop ($t$-quark) diagrams where we face $\gamma^5$-anomaly issues. The detailed discussion on $\gamma^5$-anomaly is in the next sub-section~(\ref{subsec_h24nu_gamma_five_anomaly}).

We use the symbolic manipulation program {\tt FORM} \cite{Vermaseren:2000nd}, to calculate the helicity amplitudes. Using {\tt FORM}, the amplitudes are written in terms of spinor products, scalar product of momenta and different vector objects. One needs to also calculate one-loop scalar and tensor integral for NLO amplitude computation. The one-loop scalar integrals have been calculated using a package {\tt OneLoop} \cite{vanHameren:2010cp}. To calculate the tensor integral, we use an in-house reduction code, OVReduce \cite{Agrawal:2012df,Agrawal:1998ch}. At last, the phase space integral has been computed with the Monte Carlo integration {\tt AMCI}~\cite{Veseli:1997hr} and {\tt VEGAS} algorithm \cite{Lepage:1977sw}. 
The {\tt AMCI} is implemented using parallel virtual machine ({\tt PVM}) package \cite{10.7551/mitpress/5712.001.0001}.
\subsection{$\gamma^5$-anomaly}
\label{subsec_h24nu_gamma_five_anomaly}
\begin{figure}[!h]
  \begin{center}
\includegraphics [angle=0,width=0.8\linewidth]{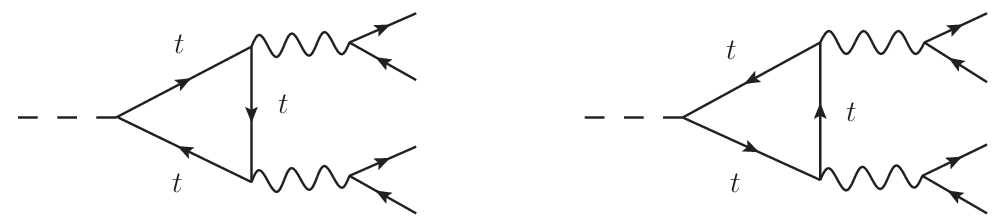}\\
	\caption{ The triangle virtual diagrams with $t$-quark fermion loop.}
	\label{fig_h24nu_top_dia}
	\end{center}
\end{figure}

At one-loop level, we have two triangle diagrams involving $t$-quark fermion loop. The corresponding diagrams are shown in Fig.~\ref{fig_h24nu_top_dia}. To find these amplitudes, one needs to compute trace involving $\gamma^5$-matrices that come from two $Zt\bar{t}$ vertices. This trace is inconsistent, as one can get different results depending on the different starting points of the trace. The formal definition of $\gamma^5=\frac{i}{4!}\epsilon_{\mu\nu\rho\sigma}\gamma^\mu\gamma^\nu\gamma^\rho\gamma^\sigma$ is not consistent in $d$-dimension as the anti-symmetric tensor $\epsilon_{\mu\nu\rho\sigma}$ lives in $4$-dimension, so, $\gamma^5$ do not anti-commute with other $\gamma$-matrices in $d$-dimension. The problem arises because of simultaneous use of cyclic properties of the trace and anti-commutation relation between $\gamma^5$ and other $\gamma$-matrices. Therefore, one of the properties needs to be dropped to get the right result
\cite{PhysRevD.84.094021}. 

To address this issue, two elegant treatments have been introduced so far. One is known as BMHV (Breitenlohner-Maison-'t Hooft-Veltman) scheme and other one is known as KKS (Korner-Kreimer-Schilcher) scheme \cite{PhysRevD.84.094021,Korner:1991sx}.
 In BMHV scheme, the $d$-dimensional objects splits in to $4$-dimensional and $2\epsilon$ dimensional parts.
     In this scheme, $\gamma^5$ matrix anti-commute with the $4$-dimensional $\gamma$-matrices and commute with $2\epsilon$-dimensional part of $\gamma$-matrices.
     This treatment is very tedious for performing the algebra in computing the amplitudes.
 In our calculation, we use KKS scheme to calculate these traces which appear in the loop diagrams shown in Fig.~\ref{fig_h24nu_top_dia}. Following KKS prescription, we take all $\gamma^5$-matrices to a particular vertex (`reading point') by anti-commuting with other $\gamma$-matrices.
  Then we do $d$-dimensional algebra and compute trace. We also follow the same prescription to calculate $Z\nu\bar{\nu}$-vertex correction where we have a current with $d$-dimensional $\gamma$-matrices and  $\gamma^5$-matrices \cite{Garzelli:2009is}. This removes $\gamma^5$-anomaly from our calculation.
\subsection{Renormalization and CMS at one-loop}
\label{subsec_h24nu_renorm_cms}
We use on-shell renormalization scheme to calculate the required CTs for this process. 
 We have discussed one-loop EW renormalization in chapter~\ref{chap_ren_ol}. In this process, the counterterms are needed for $HZZ$, $Z\nu_\mu \bar{\nu}_\mu$ and $Z\nu_e \bar{\nu}_e$ vertices; and for $Z{\text{-}}Z$ self-energies. 
 There are $5$ CT diagrams in this process. We have shown three of them in Fig.~\ref{fig_h24nu_ct_dia}. The counterterms for these CT diagrams have been given in the chapter~\ref{chap_ren_ol}. The counterterms given in Fig.~\ref{fig_h24nu_ct_dia} (and other two) remove all UV poles ($\frac{1}{\epsilon}$) from the one-loop virtual amplitudes.
 As discussed in chapter~\ref{chap_ren_ol}, one needs to calculate the self-energy diagrams to get the counterterms. 
 The self-energy diagrams of $Z$ boson have $ZZHH$ and $ZZWW$ coupling dependencies. Similarly, $H$ boson self-energy diagrams can have $HHH$ coupling and $W$ boson self-energy diagrams can have $ZZWW$ coupling dependencies. As our primary goal is to study the effect of these anomalous couplings, we also need to scale these couplings in CT diagrams.

We use complex mass scheme (CMS) \cite{Denner:2006ic} to treat unstable particles in one-loop electroweak corrections. In CMS, the unstable masses are defined with a complex part as 
\begin{eqnarray}
m_V^2\rightarrow\mu_V^2=m_V^2-im_V\Gamma_V,
\label{eq_h24nu_cms_equ_cal}
\end{eqnarray} 
where $V=W,Z$ and $\Gamma_V$ is the corresponding decay width. This treatment also makes Weinberg angle complex as $\cos^2\theta_W=\mu^2_W/\mu^2_Z$.
 For the $t$-quarks, same prescription has been followed. These complex masses and $\cos\theta$ have been used everywhere in the perturbative calculation to maintain the gauge invariance. The renormalization in CMS has been done in a modified version of the on-shell scheme \cite{Denner:2006ic,Denner:2005fg}. In this treatment, the renormalized mass is the pole of the corresponding propagator in the complex plane. When renormalization conditions are imposed, one needs to perform the self-energy computation with complex momenta. This computation can be done with Taylor expansion of self-energies about the real mass and maintaining the one-loop accuracy. We follow the treatment from Ref. \cite{Denner:2005fg} and write the renormalized counterterms as 
 \begin{gather}
\delta \mu_W^2=\Sigma_T^W(M_W^2)+(\mu_W^2-M_W^2)\Sigma^{\prime W}_T(M_W^2)+c_T^W+\mathcal{O}(\alpha^3),\nonumber\\
\delta \mu_Z^2=\Sigma_T^{ZZ}(M_Z^2)+(\mu_Z^2-M_Z^2)\Sigma^{\prime Z}_T(M_Z^2)+\mathcal{O}(\alpha^3),\nonumber\\
\delta \mathcal{Z}_W=-\Sigma^{\prime W}_T(M_W^2),\quad \delta \mathcal{Z}_H =-\Sigma^{\prime H}(M_H^2),\nonumber\\
\delta \mathcal{Z}_{ZZ}=-\Sigma^{\prime ZZ}_T(M_Z^2),\quad \delta\mathcal{Z}_{AZ} = -\frac{2}{M_Z^2}\Sigma_T^{AZ}(M_Z^2)+\Big(\frac{\mu_Z^2}{M_Z^2}-1\Big)\delta \mathcal{Z}_{ZA}\nonumber\\
\quad \delta \mathcal{Z}_{ZA}=\frac{2}{\mu_Z^2}\Sigma_T^{AZ}(0),\quad \delta \mathcal{Z}_{AA}=-\Sigma^{\prime AA}_T(0).
\label{eq_h24nu_ct_terms_cms_cal}
\end{gather}
$\delta Z_e$ follows same equation as given in Eq.~\ref{equ_ren_ren_cnst}. In Eq.~\ref{eq_h24nu_ct_terms_cms_cal}, the `Re' part is taken out in contrast to the Eq.~\ref{equ_ren_ren_cnst} as the self-energies become complex in CMS due to their complex masses and couplings. The extra term $c_T^W$($=\frac{i\alpha}{\pi}M_W\Gamma_W$) in $\delta \mu_W^2$ comes from the extra photon exchange diagram in $W$ boson self-energy that has the branch cut at $k^2=\mu^2_W$. With the counterterms given in Eq.~\ref{eq_h24nu_ct_terms_cms_cal}, we do the renormalization for this process and remove all UV divergences from the one-loop virtual amplitudes in CMS.

\subsection{Input parameter scheme}
\label{subsec:input_ps}
  The input parameters for EW correction should be taken in a consistent way to get the right results. As we do on-shell renormalization, the pole masses of massive fermions and vector bosons have been used in our computation.
  The Weinberg angle is not an independent parameter, but written in terms of $W$ and $Z$ boson masses. The convenient choice for input parameters for EW correction are the electroweak coupling, masses of vector bosons $M_Z$ and $M_W$, Higgs boson mass $M_H$ and the fermion masses. Depending on the choice of the scale of the process, the weak coupling $\alpha$ may differ by a few percent, so the choice of the weak coupling also has an impact on results.

  The charge renormalization constant $\delta Z_e$ is calculated from the photon self-energy renormalization constant $\delta Z_{AA}$ as it can be seen from Eq.~\ref{equ_ren_ren_cnst}. 
  The renormalization constant $\delta Z_{AA}$ contains mass singular terms $\alpha \log m_f$ where $m_f$ is the mass of the fermion. These contribution comes from every light fermion loop in $\delta Z_{AA}$ and remains uncancelled in EW corrections. 
  To renormalize the electric charge, the standard QED on-shell renormalization condition is being imposed in the Thomson limit , where the photon momentum transfer is zero. This renormalizes the QED coupling $\alpha=\alpha(0)$ at $Q^2=0$. To have the weak coupling $\alpha$ at desire scale ($Q^2\sim M^2_Z$) one needs running of $\alpha$ from $Q^2=0$ to $Q^2=M^2_Z$. The running of $\alpha$ remove the mass singular terms from the charge renormalization. 
  
  The choice of the running of the coupling $\alpha$ leads to the notion of the input parameter scheme. In $\alpha(M_Z)$ input parameter scheme, the $\Delta \alpha(M_Z)$ is given by~\cite{Andersen:2014efa}
  \begin{gather}
    \Delta\alpha(M_Z)=\frac{\alpha(0)}{3\pi}\sum_{f\neq t}N_f^cQ^2_f\Big[\text{ln}\Big(\frac{M_Z^2}{m_f^2}\Big)-\frac{5}{3}\Big].
   \label{eq:da_amzs}
  \end{gather}
  The shift in charge renormalization $\delta Z_e |_{\alpha(M_Z^2)}\rightarrow \delta Z_e|_{\alpha(0)}-\frac{1}{2}\Delta \alpha(M_Z^2)$, will remove all mass singularities in $\delta Z_e$. The numerical value of $\alpha(M_Z)$ has been extracted from an experimental analysis of $e^+e^-$ annihilation to hadrons~\cite{Eidelman:1995ny}.
  In $\alpha_{G_F}$ scheme, the electromagnetic coupling is derived from the Fermi constant as 
  \begin{gather}
    \alpha_{G_F}=\frac{\sqrt{2}G_FM_W^2(M_Z^2-M_W^2)}{\pi M_Z^2}.
   \label{equ_h24nu_alp_gms}
  \end{gather}
  In this scheme, the shift in charge renormalization is given by $\delta Z_e |_{G_F}\rightarrow \delta Z_e|_{\alpha(0)}-\frac{1}{2}\Delta r$, where the $\Delta r$ is the radiative correction to muon decay~\cite{PhysRevD.22.971,Denner:1991kt}. The $\Delta r$ is given by
  \begin{eqnarray}
    \Delta r &=& \Sigma^{AA}_T(0)-\frac{c_W^2}{s_W^2}\Big(\frac{\Sigma_T^{ZZ}(M_Z^2)}{M_Z^2}-\frac{\Sigma_T^W(M_W^2)}{M_W^2}\Big)+\frac{\Sigma^W_T(0)-\Sigma^W_T(M_W^2)}{M_W^2}\nonumber \\
    &&\quad \quad +2\:\frac{c_W}{s_W}\frac{\Sigma^{AZ}_T(0)}{M_Z^2}+\frac{\alpha}{4\pi s_W^2}\Big(6+\frac{7-4s_W^2}{2s_W^2}{\text {log}}\:c_W^2\Big).
   \label{eq:dr_gms}
   \end{eqnarray}
   We calculate EW correction to this process in the both $\alpha(M_Z)$ and $G_F$ schemes. The ``best" scheme will be the one in which the universal correction will be absorbed into the corresponding lower order prediction and leading to smaller perturbative correction.
   We will see in Sec.~\ref{sec_h24nu_numr_res}, the EW correction is smaller in the $G_F$ scheme than $\alpha(M_Z)$ scheme. Hence, the $G_F$ scheme can be regarded as the ``best" input scheme for this process.
\subsection{Anomalous couplings}
\label{subsec_h24nu_renorm_cms}
As discussed in the previous sections, our main goal is to study the  effect of anomalous $HHH$, $ZZHH$ and $ZZWW$ couplings in this process. There are a few virtual diagrams as displayed in Fig.~\ref{fig_h24nu_hhh_dia}, Fig.~\ref{fig_h24nu_zzhh_dia} and  Fig.~\ref{fig_h24nu_wwzz_dia}, where we can introduce such anomalous couplings in the $\kappa$-framework. The $HHH$ coupling is involved in the diagrams shown in the Fig.~\ref{fig_h24nu_hhh_dia} and in the Higgs boson wave function renormalization constant, which has been computed from its self-energy diagrams. The sum of the triangle diagrams shown in Fig.~\ref{fig_h24nu_hhh_dia} is UV finite and the contribution of the related diagrams to the Higgs boson wave function renormalization constant is also UV finite. With this UV pole structure, the renormalizability is sustained even after arbitrary scaling of $HHH$ coupling in this process. The reason behind this UV structure is that the $HHH$ coupling comes from the potential term of the standard model Lagrangian and it does not get coupled with the other terms in the Lagrangian. Therefore, we can vary the $HHH$ coupling within the allowed region and study its effect on the partial decay width of the Higgs boson.

The $ZZHH$ and $ZZWW$ couplings can be scaled in the virtual diagrams displayed in Fig.~\ref{fig_h24nu_zzhh_dia} and Fig.~\ref{fig_h24nu_wwzz_dia}. There are also other places to vary these couplings in various counterterms that involve $W$, $Z$ and Higgs boson self-energies. Scaling these two couplings in the diagrams leads to renormalization problem within the SM as $HVV$ and $VVHH$ ($V=Z,W$) couplings are not independent. In this regard, we recall that in the HEFT framework, 
one can vary $HVV$ and $VVHH$ couplings independently. Furthermore, many beyond-the-standard-model scenarios have different relationship
between HVV and HHVV couplings than that in the standard model. The excess UV pole contribution arising from the scaling of the couplings can be absorbed in the $HZZ$ coupling, as, for example, in the context of HEFT. We adopt $\overline{MS}$ scheme for scaled couplings. Therefore, we only cancel excess UV divergent pieces with appropriate
counterterm. Then, we vary $ZZHH$ and $ZZWW$ couplings and study their effect on the Higgs decay width.
\section{Numerical Results}
\label{sec_h24nu_numr_res}
\subsection{SM prediction}
\label{subsec_h24nu_sm_pdct}
We use the following set of the standard model input parameters
\begin{gather}
 M_W=80.358\: {\text{GeV}},\:\:M_Z=91.153\: {\text{GeV}},\:\:M_t=172.5\: {\text{GeV}}\:\:M_H=125\: {\text{GeV}},\nonumber\\
\Gamma_W=2.0872 \: {\text{GeV}},\:\:\Gamma_Z=2.4944 \: {\text{GeV}},\:\Gamma_H=4.187\: {\text{MeV}} \:{\text{and}}\: \Gamma_t=1.481 \: {\text{GeV}}.
\label{eq_h24nu_sm_param}
\end{gather}
We have taken lepton and light quarks as massless particles.
    Mass parameters in Eq.~\ref{eq_h24nu_sm_param} and cosine of Weinberg angle have been promoted to complex numbers following the Eq.~\ref{eq_h24nu_cms_equ_cal}.
We calculate this process in the both $\alpha(M_Z)$ and $G_F$ input parameter schemes. The value of weak coupling $\alpha=1/128.896$ for the $\alpha(M_Z)$ scheme has been taken from the Ref.~\cite{Eidelman:1995ny}.
   In the $G_F$ scheme, the value of weak coupling can be calculated from the Eq.~\ref{equ_h24nu_alp_gms}. With the above SM parameters, its numerical value is $1/132.36$ in the $G_F$ scheme.
   The standard model prediction for the partial decay width of Higgs boson for this process has been listed in Tab.~\ref{tab_h24nu_sm_rst}.
   The  LO decay widths are $930.71$ eV and $1007.72$ eV in the $G_F$ and $\alpha(M_Z)$ scheme respectively. The NLO corrected decay widths are $959.66$ eV and $948.01$ eV in the $G_F$ and $\alpha(M_Z)$ schemes respectively. We define the relative enhancement as ${\text{RE}}=\frac{\Gamma^{NLO}-\Gamma^{LO}}{\Gamma^{LO}}\; \times 100\%$.
   The RE is $3.11\%$ in the $G_F$ scheme whereas it is $-5.93\%$ in the $\alpha(M_Z)$ scheme. The LO decay widths differ by $\sim 8\%$ but the NLO corrected widths differ by $\sim 1\%$ among the two input parameter schemes.
    Our results agree (differ by $\sim 0.1\%$) with the {\tt Prophecy4f} package~\cite{BREDENSTEIN2006131,Bredenstein:2006nk} where the $H\rightarrow 4l$ has been studied in the $G_F$ scheme.
     As we can see from Tab.~\ref{tab_h24nu_sm_rst}, the relative enhancement is smaller in the $G_F$ scheme i.e., the universal correction has been absorbed in the lower order prediction, so, this scheme can be considered as the ``better" scheme.
\begin{table}[H]
\begin{center}
\begin{tabular}{|c|c|c|c|}
\hline
Input&&&\\
parameter&$\Gamma^{LO}$ (eV)&$\Gamma^{NLO}$ (eV)&RE\\
scheme&&&\\
\hline
&&&\\
$G_F$&$930.71$&$959.66$&$3.11\%$\\
&&&\\

\hline
&&&\\
$\alpha(M_Z)$&$1007.72$&$948.01$&$-5.93\%$\\
&&&\\
\hline
\end{tabular}
\caption{Partial decay widths of Higgs boson in the channel $H\rightarrow \nu_e \bar{\nu}_e \nu_\mu \bar{\nu}_\mu$ in the $G_F$ and $\alpha(M_Z)$ scheme and their relative enhancement.}
\label{tab_h24nu_sm_rst}
\end{center}
\end{table}
\subsection{Anomalous coupling effect}
\label{subsec_h34nu_anoma_cpl_efct}
  We vary $HHH$, $ZZHH$ and $ZZWW$ coupling in the context of kappa ($\kappa$) framework. We define relative increment as ${\text{RI}}=\frac{\Gamma^{NLO}_\kappa-\Gamma^{NLO}_{SM}}{\Gamma^{NLO}_{SM}}\; \times 100\%$.

  We have varied $\kappa_{HHH}$ from $-10$ to $10$ and listed the corresponding RI in the Tab.~\ref{tab_h24nu_hhh_kappa}. In Tab.~\ref{tab_h24nu_hhh_kappa}, we see a significant change in NLO EW decay width ($\Gamma^{NLO}$) with varying $\kappa_{HHH}$.
  The RI varies from $\sim 0.35\%$ to $\sim -23.52\%$ in the $G_F$ scheme and from $\sim 0.40\%$ to $\sim -26.48\%$ in the $\alpha(M_Z)$ scheme depending on
the value of $\kappa_{HHH}$. The RI in two input schemes become positive near $k_{HHH}\sim 2{\text{-}}4$.
\begin{table}[H]
\begin{center}
\begin{tabular}{|c|c|c|}
\hline
\multirow{2}{*}{$\kappa_{HHH}$}&\multicolumn{2}{|c|}{RI}\\
\cline{2-3}
&$G_F$ scheme&$\alpha(M_Z)$ scheme\\
\hline
$10$&$-7.54$&$-8.48$\\
\hline
$8$&$-3.78$&$-4.25$\\
\hline
$6$&$-1.21$&$-1.36$\\
\hline
$4$&$0.17$&$0.19$\\
\hline
$2$&$0.35$&$0.40$\\
\hline
$-1$&$-1.60$&$-1.80$\\
\hline
$-2$&$-2.84$&$-3.20$\\
\hline
$-4$&$-6.23$&$-7.01$\\
\hline
$-6$&$-10.80$&$-12.16$\\
\hline
$-8$&$-16.57$&$-18.65$\\
\hline
$-10$&$-23.52$&$-26.48$\\
\hline
\end{tabular}
\caption{Effect of anomalous $HHH$ coupling on the partial decay width of the process $H\rightarrow\nu_e \bar{\nu}_e \nu_\mu \bar{\nu}_\mu$.}
\label{tab_h24nu_hhh_kappa}
\end{center}
\end{table}
%
%

Next, we examine the effect of scaling the $ZZHH$ coupling. We have listed the RI with different $\kappa_{ZZHH}$ values in Tab.~\ref{tab_h24nu_zzhh_kappa}. As shown in the table, the change in RI due to $\kappa_{ZZHH}$ is hardly visible in the $\alpha(M_Z)$ scheme. It is less than $1 \%$.
 In the $G_F$ scheme, the RI varies from $\sim 5.7\%$ to $\sim -7.0\%$ depending upon $\kappa_{ZZHH}$.
It is positive with positive scaling and negative with negative scaling.
\begin{table}[H]
\begin{center}
\begin{tabular}{|c|c|c|}
\hline
\multirow{2}{*}{$\kappa_{ZZHH}$}&\multicolumn{2}{|c|}{RI}\\
\cline{2-3}
&$G_F$ scheme&$\alpha(M_Z)$ scheme\\
\hline
$10$&$5.74$&$0.29$\\
\hline
$8$&$4.46$&$0.22$\\
\hline
$6$&$3.19$&$0.16$\\
\hline
$4$&$1.91$&$0.10$\\
\hline
$2$&$0.64$&$0.03$\\
\hline
$-1$&$-1.27$&$-0.06$\\
\hline
$-2$&$-1.91$&$-0.09$\\
\hline
$-4$&$-3.19$&$-0.16$\\
\hline
$-6$&$-4.46$&$-0.22$\\
\hline
$-8$&$-5.74$&$-0.29$\\
\hline
$-10$&$-7.01$&$-0.35$\\
\hline
\end{tabular}
\caption{Effect of anomalous $ZZHH$ coupling on the partial decay width of the process $H\rightarrow \nu_e \bar{\nu}_e \nu_\mu \bar{\nu}_\mu$.}
\label{tab_h24nu_zzhh_kappa}
\end{center}
\end{table}
%

  Although it was not our main goal, since the width of the process
  also depends on the $ZZWW$ coupling, we scale it to see the effects.
We have listed the RI by scaling $ZZWW$ coupling in Tab.~\ref{tab_h24nu_zzww_kappa}. As shown in Tab.~\ref{tab_h24nu_zzww_kappa}, we see a significant change in $\Gamma^{NLO}$ with varying $\kappa_{ZZWW}$.
 In the $G_F$ scheme, the RI goes from $\sim 10.5$ to $\sim -12.8\%$, whereas in the $\alpha(M_Z)$ scheme  it goes from $-19.3\%$ to $23.5\%$ with the varying $\kappa_{ZZWW}$ from $10$ to $-10$.
\begin{table}[H]
\begin{center}
\begin{tabular}{|c|c|c|}
\hline
\multirow{2}{*}{$\kappa_{ZZWW}$}&\multicolumn{2}{|c|}{RI}\\
\cline{2-3}
&$G_F$ scheme&$\alpha(M_Z)$ scheme\\
\hline
$10$&$10.45$&$-19.29$\\
\hline
$8$&$8.13$&$-14.97$\\
\hline
$6$&$5.81$&$-10.72$\\
\hline
$4$&$3.48$&$-6.43$\\
\hline
$2$&$1.16$&$-2.14$\\
\hline
$-1$&$-2.32$&$4.29$\\
\hline
$-2$&$-3.48$&$6.43$\\
\hline
$-4$&$-5.80$&$10.72$\\
\hline
$-6$&$-8.13$&$15.00$\\
\hline
$-8$&$-10.45$&$19.29$\\
\hline
$-10$&$-12.78$&$23.58$\\
\hline
\end{tabular}
\caption{Effect of anomalous $ZZWW$ coupling on the partial decay width of the process $H\rightarrow\nu_e \bar{\nu}_e \nu_\mu \bar{\nu}_\mu$.}
\label{tab_h24nu_zzww_kappa}
\end{center}
\end{table}
%
%
%

\section{Conclusion}
\label{sec_h24nu_conclusion}

     We have studied the effect of scaling $HHH$ and $ZZHH$
     couplings, as in $\kappa$-framework, on the decay width of the
     process $H\rightarrow \nu_e\bar{\nu}_e\nu_\mu\bar{\nu}_\mu$.
     As this process also depends on $ZZWW$ coupling, we also investigated
     the effect of its variation. Most interesting thing that we find is
     that the width of this process has significant dependence on the
     $\kappa_{HHH}$. 
    The dependency on $\kappa_{HHH}$ is similar among the two input parameter scheme.
    As the scaling of $HHH$ coupling does not effect the gauge invariance, we see similar behaviour in the two input schemes. 
     A precise measurement of the width may lead to a better bound on the $HHH$ coupling.
     We also examine the dependency of $\kappa_{ZZHH}$ on decay partial width.
      We see very minimal dependency on $\kappa_{ZZHH}$ for the $\alpha(M_Z)$ scheme, but in the $G_F$ scheme, we see a bit stronger dependency on $\kappa_{ZZHH}$. 
      This is due to violating the gauge invariance by scaling $ZZHH$ coupling independently.
      Because of the dependence of
     the process on $\kappa_{ZZWW}$, we also examined the dependence
     on this parameter. However, there are other processes to better
     determine this coupling.

\chapter[Effect of anomalous HHH and ZZHH couplings on
the decay width\\ of $H\rightarrow e^+e^-\mu^+\mu^-$]{Effect of anomalous HHH and ZZHH couplings on
the decay width of $H\rightarrow e^+e^-\mu^+\mu^-$}
\label{chap_h22e2m}
\section{Introduction}
\label{sec:intro}
We have studied one-loop EW correction to $H\rightarrow \nu_e\bar{\nu}_e\nu_\mu\bar{\nu}_\mu$ process in the previous chapter, where we have studied the effects of anomalous $HHH$ and $ZZWW$ couplings. It is difficult to probe this process at colliders as the final state particles are neutrinos. 
  The $H\rightarrow e^+e^-\mu^+\mu^-$ process can be a good process to probe the Higgs couplings for its collider signature at colliders. In this process, again one can probe $HHH$ and $ZZHH$ couplings as they appear in one-loop EW virtual diagrams. 
  In the previous chapter, we have discussed the importance of $HHH$ and $VVHH$ couplings to fully establish the SM. We also discussed the current bounds on these couplings. 
  This process can be a good channel to probe these couplings and can give more significant findings due to its collider signature. We calculate one-loop EW correction to the process $H\rightarrow e^+e^-\mu^+\mu^-$ and study the effect of anomalous $HHH$ and $ZZHH$ couplings on the decay width of Higgs boson in the context of $\kappa$-framework.
  This process is more complicated in the sense of the computation of a larger number of virtual diagrams along with the counterterm and real emission diagrams.
   In the next sections, we have discussed the process in detail. In Sec.~\ref{sec:calc_check}, we discuss the  computation techniques, dipole subtraction, renormalization and the anomalous coupling.
   In Sec.~\ref{sec:numr_res}, we have shown the numerical results for the SM prediction and the results with the anomalous coupling effects. In the last section, we made some conclusions about this study.

\section{The Process}
\label{sec:prcs}
We are interested in studying the width of the decay channel $H\rightarrow e^+e^-\mu^+\mu^-$. We calculate NLO electroweak correction to this process. 
$HHH$, $VVHH$ and $VVVV$ type  couplings appear in the one-loop virtual diagrams of this process. These couplings can also appear in counterterm (CT) diagrams.
These couplings can be varied in virtual and CT diagrams appropriately to study the effects on partial decay width of Higgs boson.

\begin{figure}[!h]
  \begin{center}
\includegraphics [angle=0,width=0.5\linewidth]{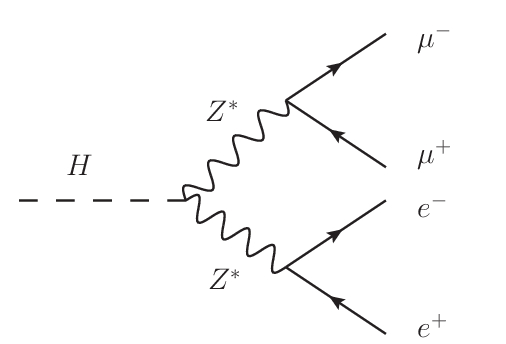}\\
	\caption{The LO Feynman diagram for the decay channel $H \rightarrow e^+e^-\mu^+\mu^-$.  }
	\label{fig:tree_dia}
	\end{center}
\end{figure}

At the leading order, there is only one tree level diagram. The tree-level diagram is shown in Fig.~\ref{fig:tree_dia}. This process only have intermediate $Z$ bosons, one of which decays to $e^+e^-$ and another decays to $\mu^+\mu^-$. We allow off-shell intermediate $Z$ bosons in this process as shown in Fig.~\ref{fig:tree_dia}.

\begin{figure}[!h]
  \begin{center}
\includegraphics [angle=0,width=1\linewidth]{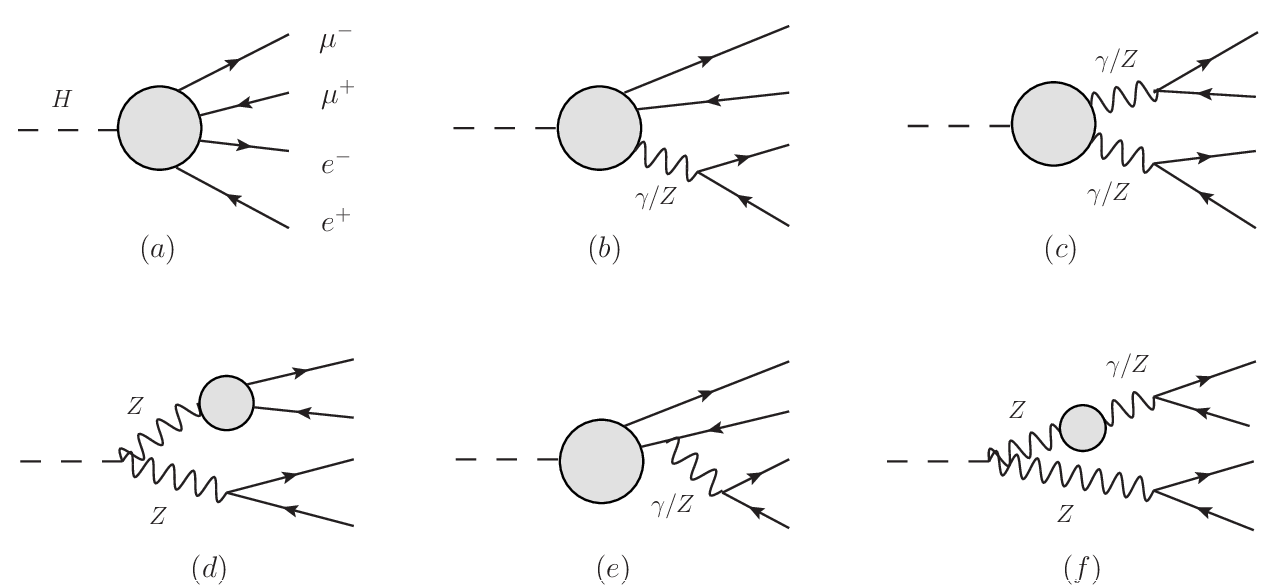}\\
	\caption{Generic class of NLO EW virtual Feynman diagrams for $H\rightarrow e^+e^-\mu^+\mu^-$.  }
	\label{fig:loop_dia}
	\end{center}
\end{figure}

As the process is $1\rightarrow 4$, there are pentagon, box, triangle and bubble-type of virtual diagrams at one loop. There are a total $256$ virtual Feynman diagrams at one loop.
 The generic class of one-loop virtual diagrams are shown in Fig.~\ref{fig:loop_dia}. The generic class of diagrams (a) shown in Fig.~\ref{fig:loop_dia} is the pentagon-type. There are a total of $10$ pentagon-type virtual diagrams in this process. 
 The generic class of diagrams (b) displayed in Fig.~\ref{fig:loop_dia} is the box-type diagram. In this generic diagram, the gauge bosons attached to $e^+$ and $e^-$ can be $\gamma$ or $Z$ boson as depicted in Fig.~\ref{fig:loop_dia}.
  The box loop associated with this generic diagram is attached to Higgs, $\gamma/Z$ bosons and muons. There is also another similar type of generic diagram (which is not displayed in Fig.~\ref{fig:loop_dia}) where the box loop is attached with Higgs, $\gamma/Z$ bosons and electrons.
 There are a total of $22$ box-type diagrams in this process. The generic class of diagrams (c) shown in Fig.~\ref{fig:loop_dia} represents both triangle and bubble-type diagrams. This generic class of diagrams includes the correction to $HZZ$ vertex. It also includes triangle diagrams with loops involving Higgs, $\gamma$, $Z$ and Higgs, $\gamma$, $\gamma$ bosons. 
 There are a total of $122$ virtual diagrams that can be represented by the generic diagrams (c).
  The generic class of diagrams (d) shown in Fig.~\ref{fig:loop_dia} is the correction to $Z\mu^+\mu^-$ vertex and it is a triangle-type diagram. There is also another similar type of diagram for $Ze^+e^-$ vertex which is not shown in the Fig.~\ref{fig:loop_dia}.
   There are a total of $8$ virtual diagrams that are included in this generic diagram.
  The generic class of diagrams (e) in Fig.~\ref{fig:loop_dia} represents a set of triangle-type diagrams associated with Higgs boson and muons. There is another similar type of generic triangle diagram associated with Higgs boson and electrons, which is not shown in the Fig.~\ref{fig:loop_dia}. There are a total of $16$ diagrams that can be represented by this generic diagram.
  The generic class of diagrams (f) in Fig.~\ref{fig:loop_dia} represents the self-energy diagrams. There are $Z\text{-} Z$ and $Z\text{-}\gamma$ type self-energy diagrams as shown in Fig.~\ref{fig:loop_dia}.
   These are the bubble and tadpole-type diagrams. There are also same set of diagrams with the other $Z$ boson propagator which are not shown in Fig.~\ref{fig:loop_dia}.
    There are a total of $78$ (bubble and tadpole-type) diagrams in this generic class.  
   The generic class of diagrams (c) is very important in this study. It includes a few diagrams as shown in Fig.~\ref{fig:hhh_dia} where we can study the effect of anomalous $HHH$, $VVHH$ and  $VVVV$ couplings.
   We can also see the anomalous effect of $VVHH$ coupling in the generic class of diagrams (f). We encounter top-quark loop diagrams in the generic class of diagrams (c).
    The top-quark triangular loop diagrams are shown in Fig.~\ref{fig:top_dia}.

\begin{figure}[!h]
  \begin{center}
\includegraphics [angle=0,width=0.8\linewidth]{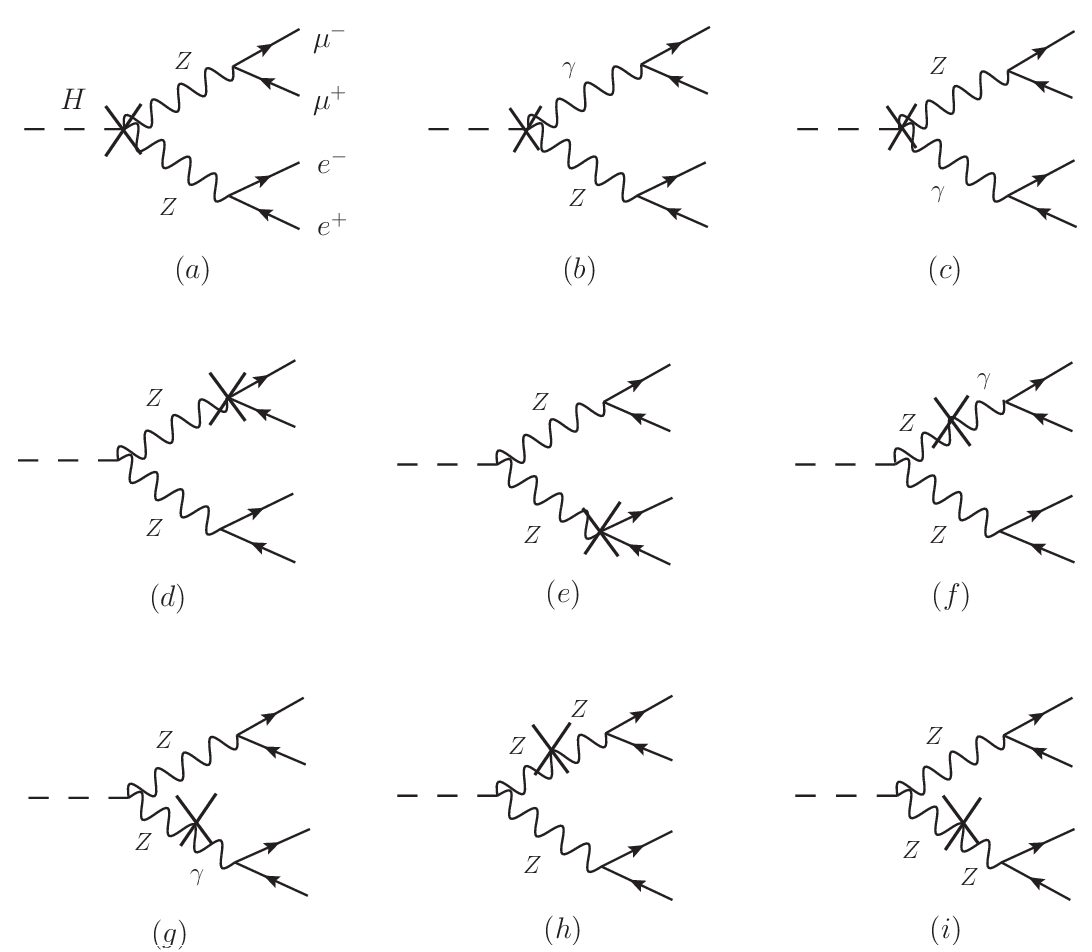}\\
	\caption{Counterterm diagrams for NLO EW correction to $H\rightarrow e^+e^-\mu^+\mu^-$.  }
	\label{fig:ct_dia}
	\end{center}
\end{figure}

We have enlisted the counterterm (CT) diagrams in Fig.~\ref{fig:ct_dia}. As displayed in Fig.~\ref{fig:ct_dia}, there are a total of $9$ CT diagrams in this process. The diagram (a) is the CT diagram for $HZZ$ vertex correction.
  The diagrams (b) and (c) in Fig.~\ref{fig:ct_dia} are the CT diagrams for the generic loop diagram (c) (in Fig.~\ref{fig:loop_dia}) with the loops associated with Higgs, $\gamma$ and $Z$ boson.
  The diagram (d) in Fig.~\ref{fig:ct_dia} is the CT diagram for $Z\mu^+\mu^-$ vertex correction. Similarly, the diagram (e) in Fig.~\ref{fig:ct_dia} is the CT diagram for $Ze^+e^-$ vertex correction.
  The diagrams (f) and (g) in Fig.~\ref{fig:ct_dia} are the CT diagrams for $Z\text{-}\gamma$ type self-energy diagrams. The diagrams (h) and (i) in Fig.~\ref{fig:ct_dia} are the CT diagrams for $Z\text{-}Z$ type self-energy diagrams.
  CT diagrams are also sensitive to the $HHH$, $VVHH$, and $VVVV$ couplings. As counterterms depend on Higgs, $Z$, and $W$ boson self-energies, these couplings appear in the CT diagrams for this process.
  
\begin{figure}[!h]
  \begin{center}
\includegraphics [angle=0,width=0.8\linewidth]{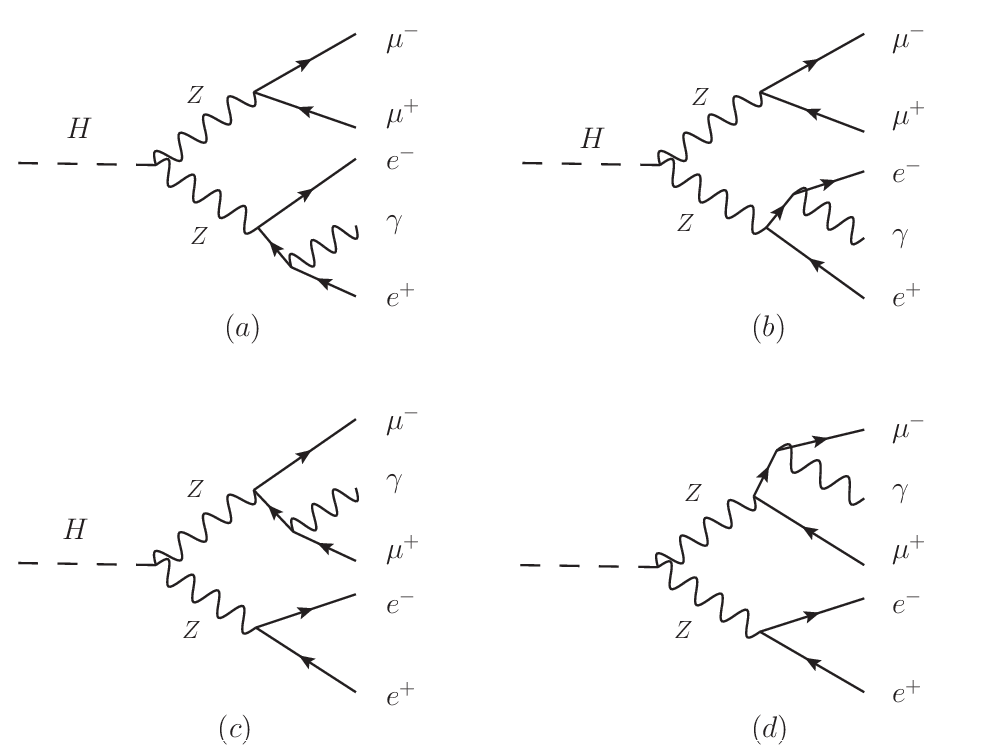}\\
	\caption{Real emission diagrams for NLO EW correction to the decay channel $H\rightarrow e^+e^-\mu^+\mu^-$.  }
	\label{fig:re_dia}
	\end{center}
\end{figure}

   In Fig.~\ref{fig:re_dia}, we have shown the real emission (photon) diagrams for this process. As shown in the Fig.~\ref{fig:re_dia}, there are $4$ real emission diagrams. Each charged lepton can radiate a photon.
    The photon is being emitted from positron and electron as shown in diagrams (a) and (b) respectively in Fig.~\ref{fig:re_dia}. Similarly, photon radiation from anti-muon and muon are shown in diagrams (c) and (d) respectively in Fig.~\ref{fig:re_dia}.
   The diagrams for this process have been generated using a mathematica package, {\tt FeynArts}~\cite{Hahn:2000kx}.

\section{Calculations}
\label{sec:calc_check}
\begin{figure}[!h]
  \begin{center}
\includegraphics [angle=0,width=0.8\linewidth]{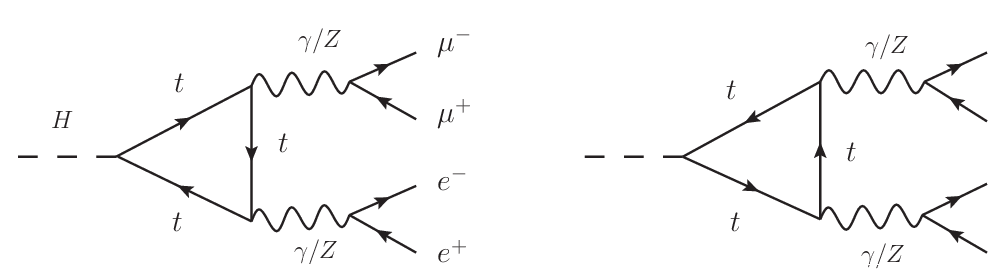}\\
	\caption{ The triangle virtual diagrams with $t$-quark fermion loop.}
	\label{fig:top_dia}
	\end{center}
\end{figure}

  We compute a few hundred one-loop virtual diagrams along with tree-level, CT diagrams and real emission diagrams.
  We treat leptons, light quarks as massless particles. A large set of diagrams are zero due to the vanishing coupling of scalars with the massless leptons and fermions.
   We ignore tadpole diagrams as they can be set to zero with the proper CT diagrams. With these considerations, we compute a total of $270$ Feynman diagrams for this process. 
   There are a few bubble-type diagrams for mixed type propagators between Goldstone boson and $\gamma/Z$ boson as the generic diagram (f) in Fig.~\ref{fig:loop_dia}. These diagrams do not contribute. 
    To calculate total virtual amplitudes, first we calculate the prototype type of diagrams and then we map rest of the diagrams to these diagrams. 
   The mapping is done using the proper choice of couplings, masses and crossings of momenta.
  As discussed in chapter~\ref{chap_shf}, we use helicity formalism to calculate the helicity amplitudes. 
   We use t'Hooft-Veltman (HV) regularization scheme~\cite{THOOFT1972189} for this process to calculate the virtual and CT diagrams.
   There are a few top-quark loop triangle diagrams in this process as shown in the Fig.~\ref{fig:top_dia}. There are a total of $8$ fermion-loop diagrams in this process. We encounter $\gamma^5$ anomaly in these diagrams. 
   We have already discussed about $\gamma^5$ anomaly in the previous chapter. We follow the same KKS scheme to remove the $\gamma^5$ anomaly. We also follow the same scheme in the virtual correction of $Zf\bar{f}$ vertices in this process.

   We use the symbolic manipulation program {\tt FORM}~\cite{Vermaseren:2000nd}, to simplify helicity amplitude with helicity identities. Using {\tt FORM}, the amplitudes have been written in terms of spinor products and scalar products of vector objects with the momenta and polarizations.
    We calculate scalar integrals which appear at one loop using the package {\tt OneLOop}~\cite{vanHameren:2010cp}. The tensor integrals associated with the one-loop diagrams have been computed with a in-house code, {\tt OVReduce}~\cite{Agrawal:2012df,Agrawal:1998ch}.
    For tree and loop-level diagrams, we perform $4$-body phase-space integrals and for radiation diagrams, we perform $5$-body phase-space integral. We use a Monte-Carlo integration package, called {\tt AMCI}~\cite{Veseli:1997hr} to perform the phase-space integrals.

\subsection{Renormalization and CMS scheme }
\label{subsec:renorm_cms}
In this process, one-loop virtual diagrams are UV divergent. Rank-two, rank-three tensors are present in triangle-type diagrams which lead to UV divergences. Bubble diagrams are UV divergent.
 There are no UV singularities from pentagon and box-type diagrams. The UV singularities have been removed with the appropriate CT diagrams shown in Fig.~\ref{fig:ct_dia}.
 As discussed in Sec.~\ref{sec:prcs}, the diagrams (a)-(e) in Fig.~\ref{fig:ct_dia} are the vertex CT diagrams and the diagrams (f)-(i) are the self-energy CT diagrams.
 The CT diagrams given in Fig.~\ref{fig:ct_dia} remove all UV divergent poles from the desired one-loop virtual amplitudes. The counterterms for corresponding diagrams are given in chapter~\ref{chap_ren_ol}.
   
   As we calculate the self-energy diagrams, we can study the effect of anomalous $HHH$, $VVHH$ and $VVVV$ couplings in counterterms. The $Z$ boson self-energy diagrams are sensitive to anomalous $ZZHH$ coupling. The Higgs boson self-energy diagrams are sensitive to anomalous $HHH$ coupling.
    The $W$ and $Z$ boson both self-energy diagrams are sensitive to $ZZWW$ coupling. With appropriate scaling in CT diagrams, we study effect of anomalous $HHH$, $VVHH$ and $VVVV$ couplings in this process.
    We implement CMS scheme for this process to treat unstable particles in the loop. We have discussed in details about the CMS scheme and the one-loop renormalization in this scheme in the last chapter. Here we implement the same.

\subsection{IR divergences and dipole subtractions}
\label{subsec:ir_dp}
Most of the one-loop virtual amplitudes are IR singular. Box and bubble diagrams do not have any IR singularities. IR singularities from virtual diagrams exactly cancel with the IR singularities from real emission diagrams.
 There are four real emission diagrams as shown in the Fig.~\ref{fig:re_dia}, which are IR singular in soft and collinear regions. 
  We have discussed the Catani-Seymour dipole subtraction in chapter~\ref{chap_ir_div_dp_sub}. We have discussed the dipole subtraction only for QCD partons in chapter~\ref{chap_ir_div_dp_sub}.
  Here we are calculating the EW one-loop correction for this process. One needs to implement the Catani-Seymour dipole subtraction for EW correction in order to remove the IR singularities from the desired amplitudes.   
 We follow the Ref.~\cite{Schonherr:2017qcj}, and implement Catani-Seymour dipole subtraction for EW one-loop correction.
  The {\textit{\textbf{I}}}-term in dipole subtraction exactly cancels the IR singularities from virtual amplitudes. The dipole terms $\mathcal{D}_{ij,k}$ exhibits the same behavior as real emission amplitudes in collinear and soft regions.
   There are four leptons in the final state in this process. There are three ${\textit{\textbf{I}}}_{ik}$-terms for each lepton, so total $12$ ${\textit{\textbf{I}}}_{ik}$ terms for this process.
   As they are massless charged particles, their structure are same. For this process, the ${\textit{\textbf{I}}}_{ik}$ term will be
   \begin{eqnarray}
{\textit{\textbf{I}}}_{ik}(\epsilon,\mu^2;\kappa,\{\alpha_{\text{dip}}\}) &=& {\textit{\textbf{Q}}}_{ik}^2\Big[\frac{1}{\epsilon^2}+\frac{1}{\epsilon}\Big(\frac{3}{2}+\log \frac{\mu^2}{s_{ik}}\Big)-\frac{\pi^2}{3}+\frac{3}{2}\Big(1+\log\frac{\mu^2}{s_{ik}}\Big) \nonumber \\
&& \quad \quad +\frac{1}{2}\log^2\Big(\frac{\mu^2}{s_{ij}}\Big)+\Big(\frac{7}{2}-\frac{\pi^2}{6}\Big)+A^I_{ik}\{\alpha_{\text{dip}}\}+\mathcal{O}(\alpha)\Big]
\label{eq:dp_Iik}
\end{eqnarray}
  Here ${\textit{\textbf{Q}}}^2_{ik}$ is the number associated with the charge of the emitter and spectator. It can be calculated from the relation ${\textit{\textbf{Q}}}_{\tilde{ij},\tilde{k}}^2=\frac{Q_{\tilde{ij}}Q_{\tilde{k}}\theta_{\tilde{ij}}\theta_{\tilde{k}}}{Q_{\tilde{ij}}^2}$, where $Q_{\tilde{ij}}$ and $Q_{\tilde{k}}$ are the charges of emitter and spectator and $\theta=\pm 1$ fixed by whether they are in final or initial state.
   $s_{ij}/2$ is the dot product of the momentum of emitter and spectator.
  This {\textit{\textbf{I}}} term removes IR singularities from virtual amplitudes and adds finite contributions to one-loop corrected decay width. We can see from Eq.~\ref{eq:dp_Iik}, the finite contribution of ${\textit{\textbf{I}}}$ term has $log$ and $log^2$ functions.
   The expression of $A^I_{ik}(\{\alpha_\text{dip}\})$ is given in Ref.~\cite{Schonherr:2017qcj}.
   
   In this process, both emitter and spectator are in the final states. The final-final dipole reads as
   \begin{eqnarray}
  \mathcal{D}_{ij,k}=-\frac{1}{(p_i+p_j)^2-m^2_{\tilde{ij}}}{\textit{\textbf{{Q}}}}^2_{\tilde{ij}\tilde{k}}\: {_m}\langle ...,\tilde{ij},...,\tilde{k},...|V_{ij,k}|...,\tilde{ij},...,\tilde{k},...\rangle_m\:.
   \label{eq:dp_dijk}
      \end{eqnarray}
      Here $\tilde{ij}$ is the emitter, $i$ is the emittee, $j$ is the emitted, and $k$ is the spectator. The dipole $\mathcal{D}_{ij,k}$ is being evaluated in $m$ parton phase space with the new set of momenta. In this process fermions splits into photons and fermions. The sigular factor $V_{ij,k}$ for massless fermion is given by 
       \begin{eqnarray}
  \langle s|V_{f\gamma,k}(\tilde{z_i;y_{ij,k}})|s\prime\rangle=8\pi\mu^{2\epsilon}\alpha\Big[\frac{2}{1-\tilde{z_i}(1-y_{ij,k})}-(1+\tilde{z}_i)-\epsilon(1-\tilde{z}_i)\Big]\delta_{ss\prime}.
   \label{eq:dp_vijk}
      \end{eqnarray}
      Here $y_{ijk}$ and $\tilde{z}_i$ are defined as 
      \begin{eqnarray}
  y_{ij,k}=\frac{p_ip_j}{p_ip_j+p_jp_k+p_kp_i}\quad{\text {and}}\quad \tilde{z}_i=\frac{p_ip_k}{p_jp_k+p_ip_k}\:,
   \label{eq:dp_yz}
      \end{eqnarray}
      where $p_i$, $p_j$, and $p_k$ are the momenta of emittee, emitted, and spectator respectively.
     $s$ and $s^\prime$ denotes spin of the emitter $\tilde{ij}$ in $\langle ..,\tilde{ij}, ..|$ and $| ..,\tilde{ij},..\rangle$ respectively.
       The dipole $\mathcal{D}_{ij,k}$ are subtracted from real emission diagrams and it has exactly the same behavior as the real emission diagrams in soft and collinear regions.
\subsection{Anomalous couplings}
\label{subsec:anomalous_cpl}
\begin{figure}[!h]
  \begin{center}
\includegraphics [angle=0,width=1\linewidth]{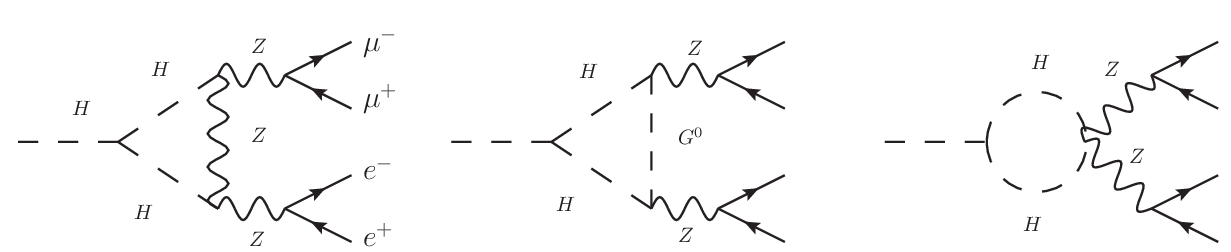}\\
	\caption{ NLO EW virtual diagrams with $HHH$ and $ZZHH$ couplings.  }
	\label{fig:hhh_dia}
	\end{center}
\end{figure}

  As we have discussed, this process is sensitive to the anomalous behavior of $HHH$, $VVHH$ and $VVVV$ couplings. As we have discussed in Sec.~\ref{sec:prcs} and~\ref{subsec:renorm_cms}, there are a few diagrams where we can introduce such anomalous coupling in $\kappa$-framework.
  The one-loop diagrams where $HHH$ coupling is involved have been shown in Fig.~\ref{fig:hhh_dia}. $HHH$ coupling also appears in Higgs wave function renormalization that has been used in $HZZ$ vertex counterterm. Collectively the diagrams given in Fig.~\ref{fig:hhh_dia} are UV finite. 
  The diagrams associated with the $HHH$ couplings in wave function renormalization of Higgs boson do not affect the UV renormalization. With this singular structure, the arbitrary scaling of $HHH$ do not affect the renormalizability in this process.
  The $HHH$ coupling comes from the scalar potential term of the standard model Lagrangian and not from the gauge sector of the model. So, one can scale $HHH$ coupling independently with no loss of renormalizability. The gauge invariance is also preserved.
  We scale $HHH$ coupling within the experimentally allowed region in the context of $\kappa$-framework and study its effect on the partial decay width of Higgs boson for this process.
  
\begin{figure}[!h]
  \begin{center}
\includegraphics [angle=0,width=1\linewidth]{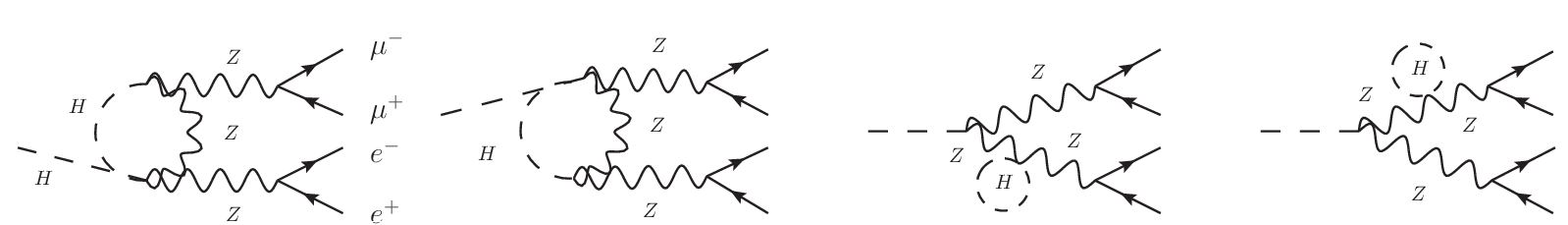}\\
	\caption{ NLO EW virtual diagrams with $ZZHH$ couplings.  }
	\label{fig:zzhh_dia}
	\end{center}
\end{figure}

\begin{figure}[!h]
  \begin{center}
\includegraphics [angle=0,width=1\linewidth]{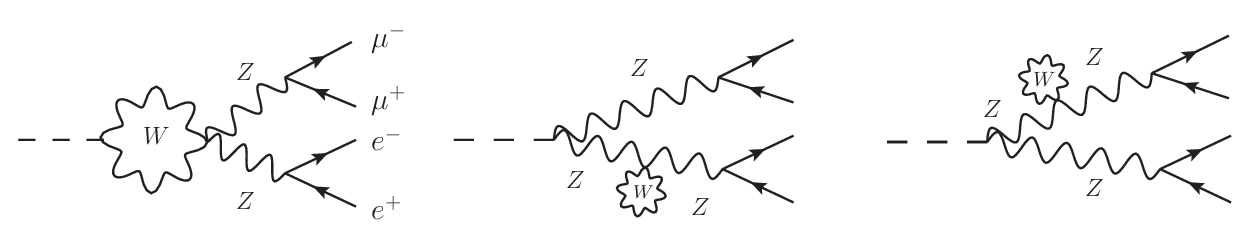}\\
	\caption{ NLO EW virtual diagrams with $ZZWW$ couplings.  }
	\label{fig:wwzz_dia}
	\end{center}
\end{figure}

  We can scale $ZZHH$ coupling in four virtual one-loop diagrams as shown in Fig.~\ref{fig:zzhh_dia}. All are bubble-type diagrams. Apart from these diagrams the counterterms also have this coupling in which the $Z$ and Higgs boson self-energies are involved. We can also vary the $ZZWW$ coupling in this process.
   The corresponding one-loop diagrams are shown in the Fig.~\ref{fig:wwzz_dia}. There are a few counterterms in this process where we can appropriately scale the $ZZWW$ coupling.
   We have discussed the input parameter scheme  in the last chapter. 
   The $\Delta r$ in the $G_F$ input parameter scheme is sensitive to the scaling of $VVHH$ and $VVVV$ couplings. As we can see from Eq.~\ref{eq:dr_gms}, the $\Delta r$ depends on $\Sigma^{ZZ}_T$ and $\Sigma^W_T$ self-energies. These self-energies are sensitive to $VVHH$ and $VVVV$ couplings. 
   In contrast, $\Delta\alpha(M_Z)$ in the $\alpha(M_Z)$ scheme is insensitive to $VVHH$ and $VVVV$ couplings. We also introduce the anomalous couplings in $\Delta r$ in the $G_F$ scheme to see its effects on partial decay width.
    With the scaling of $ZZHH$ and $ZZWW$ coupling in the relevant diagrams, we study the effects on partial decay width of Higgs boson in this process.
   The scaling of $ZZHH$ and $ZZWW$ couplings spoil the renormalization as $VVH$ and $VVHH$ couplings are related, not independent in the standard model. In many beyond-the-standard models, $VVH$ and $VVHH$ have different relationship than the standard model. We may consider that the excess UV pole contribution that came from the scaling of $ZZHH$ and $ZZWW$ couplings can be absorbed in the $ZZH$ coupling in this process.
    The scaling of the anomalous couplings can be done in ${\overline{MS}}$ renormalization scheme where we can set the finite contribution from the counterterms to zero. With this treatment, we vary $ZZHH$ and $ZZWW$ coupling and study the anomalous effects on the Higgs boson decay width.

\section{Numerical Results}
\label{sec:numr_res}
\subsection{SM prediction}
\label{subsec:sm_pdct}
 We use same set of SM parameters as used in the previous chapter. We calculate this process in the both $\alpha(M_Z)$ and $G_F$ input parameter schemes.
   We have listed the standard model prediction for the partial decay width of Higgs boson for this process in Tab.~\ref{table:sm_rst}. The LO partial decay width  are $238.04$ eV and $256.82$ eV; whereas the NLO (EW) corrected decay widths are $241.03$ eV and $237.69$ eV in the $G_F$ and $\alpha(M_Z)$ schemes respectively.
    These results are enlisted in Tab.~\ref{table:sm_rst}. We define relative enhancement as $\text{RE}=\frac{\Gamma^{NLO}-\Gamma^{LO}}{\Gamma^{LO}}\times 100\:\%$.
    The relative enhancement in the $G_F$ scheme is $1.26\%$, whereas in the $\alpha(M_Z)$ scheme the relative enhancement is $-7.45\%$.
     Although the LO results differ quite significantly but the NLO (EW) corrected results differ by only $\sim 1.5\%$ for two input parameter schemes. Our results agree (differ by $\sim 0.2\%$) with the {\tt Prophecy4f} package~\cite{BREDENSTEIN2006131,Bredenstein:2006nk}.
       As we can see from the Tab.~\ref{table:sm_rst}, the relative enhancement is smaller in the $G_F$ scheme i.e., the universal correction has been absorbed in the lower order prediction so, this scheme can be considered as the ``better" scheme.
\begin{table}[H]
\begin{center}
\begin{tabular}{|c|c|c|c|}
\hline
Input&&&\\
parameter&$\Gamma^{LO}$ (eV)&$\Gamma^{NLO}$ (eV)&RE\\
scheme&&&\\
\hline
&&&\\
$G_F$&$238.04$&$241.03$&$1.26\%$\\
&&&\\

\hline
&&&\\
$\alpha(M_Z)$&$256.82$&$237.69$&$-7.45\%$\\
&&&\\
\hline
\end{tabular}
\caption{Partial decay widths of Higgs boson in the channel $H\rightarrow e^+e^-\mu^+\mu^-$ in the $G_F$ and $\alpha(M_Z)$ schemes and their relative enhancements.}
\label{table:sm_rst}
\end{center}
\end{table}
\subsection{Anomalous coupling effect}
\label{subsec:anoma_cpl_efct}
  We vary $\kappa$ of $HHH$, $ZZHH$ and $ZZWW$ couplings within the experimental bounds and study their effects on the decay width $\Gamma^{NLO}$. 
   We define relative increment as ${\text{RI}}=\frac{\Gamma^{NLO}_\kappa-\Gamma^{NLO}_{SM}}{\Gamma^{NLO}_{SM}}\; \times 100\%$.
   
   We vary $\kappa_{HHH}$ from $10$ to $-10$~\cite{TheATLAScollaboration:2014scd} and enlisted corresponding relative increment in Tab.~\ref{table:hhh_kappa}. As we see from the Tab.~\ref{table:hhh_kappa}, the RI goes from $3.6\%$ to $-23.91\%$ in the $G_F$ scheme depending on $\kappa_{HHH}$ value and becomes positive near $\kappa_{HHH}\sim 2\text{-}4$; 
   whereas in the $\alpha(M_Z)$ scheme, the RI goes from $0.40\%$ to $-26.72\%$ and become positive near $\kappa_{HHH}\sim 2\text{-}4$. 
\begin{table}[H]
\begin{center}
\begin{tabular}{|c|c|c|}
\hline
\multirow{2}{*}{$\kappa_{HHH}$}&\multicolumn{2}{|c|}{RI}\\
\cline{2-3}
&$G_F$ scheme&$\alpha(M_Z)$ scheme\\
\hline
$10$&$-7.65$&$-8.62$\\
\hline
$8$&$-3.84$&$-4.32$\\
\hline
$6$&$-1.23$&$-1.39$\\
\hline
$4$&$0.17$&$0.19$\\
\hline
$2$&$0.36$&$0.40$\\
\hline
$-1$&$-1.62$&$-1.83$\\
\hline
$-2$&$-3.17$&$-3.25$\\
\hline
$-4$&$-6.31$&$-7.12$\\
\hline
$-6$&$-10.95$&$-12.41$\\
\hline
$-8$&$-16.82$&$-18.95$\\
\hline
$-10$&$-23.91$&$-26.72$\\
\hline
\end{tabular}
\caption{Effect of anomalous $HHH$ coupling on the partial decay width of the process $H\rightarrow e^+e^-\mu^+\mu^-$.}
\label{table:hhh_kappa}
\end{center}
\end{table}
%
  We vary $\kappa_{ZZHH}$ from $-10$ to $10$~\cite{Aad_2021,Nordstrom:2018ceg} and enlisted the corresponding RI in the Tab.~\ref{table:zzhh_kappa}. The RI goes from $5.13\%$ to $-6.26\%$ for different value of $\kappa_{ZZHH}$ in the $G_F$ scheme whereas there is not significant relative increment in the $\alpha(M_Z)$ scheme with the scaling of $ZZHH$ coupling.
\begin{table}[H]
\begin{center}
\begin{tabular}{|c|c|c|}
\hline
\multirow{2}{*}{$\kappa_{ZZHH}$}&\multicolumn{2}{|c|}{RI}\\
\cline{2-3}
&$G_F$ scheme&$\alpha(M_Z)$ scheme\\
\hline
$10$&$5.13$&$-0.50$\\
\hline
$8$&$3.98$&$-0.38$\\
\hline
$6$&$2.85$&$-0.28$\\
\hline
$4$&$1.71$&$-0.16$\\
\hline
$2$&$0.57$&$-0.06$\\
\hline
$-1$&$-1.14$&$0.11$\\
\hline
$-2$&$-1.71$&$0.16$\\
\hline
$-4$&$-2.85$&$0.27$\\
\hline
$-6$&$-3.91$&$0.38$\\
\hline
$-8$&$-5.12$&$0.50$\\
\hline
$-10$&$-6.26$&$0.88$\\
\hline
\end{tabular}
\caption{Effect of anomalous $ZZHH$ coupling on the partial decay width of the process $H\rightarrow e^+e^-\mu^+\mu^-$.}
\label{table:zzhh_kappa}
\end{center}
\end{table}
  
  In the same way, we also vary $\kappa_{ZZWW}$ to study its impact on the partial decay width $\Gamma^{NLO}$. We vary $\kappa_{ZZWW}$ from $-10$ to $10$ and enlist the corresponding RI in Tab.~\ref{table:zzww_kappa}. As we can see in Tab.~\ref{table:zzww_kappa}, RI goes from $6.56\%$ to $-8.56\%$ and $-23.68\%$ to $28.90\%$ in the $G_F$ and $\alpha(M_Z)$ schemes respectively for different $\kappa_{ZZWW}$ values.
\begin{table}[H]
\begin{center}
\begin{tabular}{|c|c|c|}
\hline
\multirow{2}{*}{$\kappa_{ZZWW}$}&\multicolumn{2}{|c|}{RI}\\
\cline{2-3}
&$G_F$ scheme&$\alpha(M_Z)$ scheme\\
\hline
$10$&$6.56$&$-23.68$\\
\hline
$8$&$5.51$&$-18.45$\\
\hline
$6$&$3.95$&$-13.15$\\
\hline
$4$&$2.47$&$-7.93$\\
\hline
$2$&$0.78$&$-2.79$\\
\hline
$-1$&$-1.57$&$5.23$\\
\hline
$-2$&$-2.35$&$7.85$\\
\hline
$-4$&$-3.88$&$13.15$\\
\hline
$-6$&$-5.48$&$18.38$\\
\hline
$-8$&$-7.03$&$23.64$\\
\hline
$-10$&$-8.59$&$28.90$\\
\hline
\end{tabular}
\caption{Effect of anomalous $ZZWW$ coupling on the partial decay width of the process $H\rightarrow e^+e^-\mu^+\mu^-$.}
\label{table:zzww_kappa}
\end{center}
\end{table}
\section{Conclusion}
 We have studied one-loop EW correction to the process $H\rightarrow e^+e^-\mu^+\mu^-$. We investigate effect of the anomalous $HHH$ and $ZZHH$ coupling on the partial decay width of the Higgs boson.
 We varied the $\kappa_{HHH}$ within the experimental bound and saw the significant change in the decay width. 
 The behavior of RI on varying $\kappa_{HHH}$ is same for two input schemes as the scaling of $HHH$ coupling does not affect gauge invariance.
 We also varied $\kappa_{ZZHH}$ and studied the effects on relative enhancement of the decay width. For two input parameter schemes, the RI are different for $ZZHH$ scaling.
  As we know, the naive scaling of $ZZHH$ coupling breaks gauge invariance. This discrepancy for two input parameter schemes came from different shift in charge renormalization in the two schemes. The $\Delta r$ in the $G_F$ scheme contains $\Sigma^{ZZ}_T$ self-energy where the $\kappa_{ZZHH}$ appears whereas in $\Delta \alpha(M_Z)$ in the $\alpha(M_Z)$ scheme there is no such term.
  As there is the dependency of decay width on $\kappa_{ZZWW}$, we scale $ZZWW$ coupling and study its effects. Due to the same reason, we see different relative increments in two input parameter schemes. 
 
\label{sec:conclusion}

%
%

\chapter{Conclusions}
\label{chap_con}

  In this thesis, we mainly focus on the Higgs sector of the standard model, especially the Higgs boson self-couplings and its couplings with the vector bosons.
  We have considered a few processes in which we have studied the anomalous effects of $HHH$ and $VVHH$ couplings.
  These couplings are involved at tree level for the process $b\bar{b}\rightarrow W^+W^-H$, for which we have computed one-loop QCD correction and have studied the effects of anomalous $HHH$ and $ZZHH$ couplings.
  We have studied the anomalous effects in the context of the $\kappa$-framework.
  In the Higgs boson decay processes, these couplings do not appear in the LO amplitudes but appear in the EW NLO amplitudes.
  In the first few chapters of this thesis, we describe on computing the one-loop Feynman amplitudes, the challenges in computing and the techniques to resolve the challenges. Then in the last few chapters, we have studied a few processes at the NLO level in detail.
  
  In the chapter~\ref{chap_shf}, we have described spinor helicity formalism.
  We have derived and listed a set of spinor helicity identities that are relevant to compute the helicity amplitudes.
   We have shown that not only the tree-level amplitudes but also loop-level amplitudes can be computed with the help of this technique.
   We obtained the functional forms of the spinor product and each component of the vector current, which is necessary to compute the loop-level amplitudes.
  In the chapter~\ref{chap_ren_ol}, we have discussed the one-loop integrals, the source of UV divergences and the renormalization at the one-loop.  
 First, we have discussed the structure of  
generic one-loop integral, then tensor and scalar integrals, and their decomposition into lower point scalar integrals.
 Then we showed the source of UV singularities from tadpole and bubble scalar integrals. Our main focus in this chapter is one-loop EW renormalization.
 We have listed the expressions for the renormalization constants.
  Then we have given the expressions for self-energies that are needed to compute the renormalization constants.
 These self-energies are computed in both HV and FDH regularization schemes.
 We have also given the Feynman rules for a few counterterm diagrams, which are relevant for the EW corrections to the Higgs boson decay processes discussed in this thesis.
   
   In the chapter~\ref{chap_ir_div_dp_sub}, we have discussed the IR divergences and the dipole subtraction technique. In this chapter, first, we discussed the sources of IR singularities in virtual amplitudes.
   Then we described the Catani-Saymour dipole subtraction method to construct IR-safe amplitudes. In this chapter, we have described this scheme for one-loop QCD correction.
   We have discussed how one introduces the local counter terms in perturbative computation and make the results finite and integrable so that a Monte-Carlo integration can be performed with the finite amplitudes.
 We gave the explicit factorization formulas for amplitude-square in collinear and soft regions.
   We discussed a few dipole configurations which are useful for calculating one-loop correction for the processes discussed in this thesis.
    We have given the expressions for the insertion operators, which are useful for one-loop perturbative computation. The insertion operator \textit{\textbf{I}} will make virtual amplitude IR finite and will add some finite pieces.
    The insertion operators \textit{\textbf{P}} and \textit{\textbf{K}} are the finite reminders after collinear singularities factorize into the PDFs. These operators are also universal, i.e., they depend only on the identified partons, not on the given process.
    
    In the chapter~\ref{chap_wwh_bb_fus}, we calculated the one-loop QCD correction to the $b\bar{b}\rightarrow W^+W^-H$ process.
  The NLO cross section for this process are $289$, $1559$ and $23097$ $ab$ at $14$, $27$ and $100$ TeV CME respectively.
    The RE are $33.2\%$, $43.6\%$ and $51.4\%$ for $14$, $27$ and $100$ TeV CME respectively. This channel contributes $\sim 2\%$ and $\sim 14\%$ to $pp\rightarrow W^+W^-H$ process at $14$ and $100$ TeV CME respectively.
    We also computed the cross sections with the possible polarization states of the $W^+$ and $W^-$ bosons. We have seen that the contributions are significantly large when both $W$ bosons are longitudinaly polarized.
    The contribution from this polarization state are $\sim 23\%$ and $\sim 42\%$ at $14$ and $100$ TeV CME at the NLO respectively.
    The QCD correction in this polarization state is also higher compared to other polarization states. The RE in this polarization states are $\sim 31\%$ and $\sim 117\%$  at $14$ and $100$ TeV CME respectively.
    There are top mediated diagrams in this process and the pseudo Goldstone bosons couple to top quarks with the coupling proportional to its mass.
    This leads to a larger contribution when both $W$ bosons are in longitudinal polarization modes. This mode (both $W$ bosons are longitudinaly polarized) is also useful for background suppression as the background contributions come from the processes with gauge bosons or gluons or photons couplings with massless fermions.
    We also examined the anomalous coupling effects for this process. We do not see any significant change with the scaling of $HHH$ coupling, but this process has a significant dependency on $ZZHH$ coupling.
    The dependency is even stronger in longitudinal modes. The RI are $\sim 2\%$ and $\sim 10\%$ at $14$ and $100$ TeV CME when we have set $\kappa_{V_2H_2}=-2$ and there are marginal changes in RI for positive scaling of $ZZHH$.
    But in longitudinal mode, the RI goes up to $\sim 23\%$ at $100$ TeV CME. We also find that $p_T^W$ and invariant distributions are considerably harder for negative scaling of $VVHH$ coupling. One can put stronger bounds on the coupling from this.
    
    In the chapter~\ref{chap_h24nu}, we studied the partial decay width of the Higgs boson in the channel $H\rightarrow \nu_e\bar{\nu}_e\nu_\mu\bar{\nu}_\mu$.
    In this process, we studied the one-loop EW correction and the anomalous effects of $HHH$ and $VVHH$ couplings. In this process, these couplings appear in NLO level amplitudes.
  Few of the virtual and CT diagrams are sensitive to $HHH$ and $ZZHH$ couplings.
    We adopted complex mass scheme for the EW correction to this process due to the involvement of heavy particles in loop-level amplitudes. We upgraded the EW renormalization to complex mass scheme.
    We have discussed the $\gamma^5$ anomaly related to this process. We have implemented the $KKS$ scheme to resolve this issue.
 We have implemented both $\alpha(M_Z)$ and $G_F$ schemes for this process.
    The EW-corrected partial decay widths of the Higgs boson in this channel are $959.7$ eV and $948.0$ eV in the $G_F$ and $\alpha(M_Z)$ schemes respectively.
    Although the LO decay widths differ a lot depending on the input parameter schemes, the NLO corrected decay widths are nearly the same for the two input parameter schemes. Our results are in agreement with the {\tt Prophecy4f} package where the $H\rightarrow 4l$ has been studied in the $G_F$ scheme.
    We vary the $\kappa_{HHH}$ and $\kappa_{ZZHH}$ within the experimental bounds. We see that there are significant changes in RI when we vary $\kappa_{HHH}$. The RI goes from $\sim 0.5\%$ to $\sim -25\%$ when we vary $\kappa_{HHH}$ from $10$ to $-10$. The change in RI are nearly the same for the two input schemes, as the scaling of $HHH$ does not disturb the gauge invariance.
    There are significant changes in RI in the $G_F$ scheme but a very marginal change in RI in the $\alpha(M_Z)$ scheme when we vary $\kappa_{ZZHH}$ within the experimental bounds. This discrepancy is due to disturbing the gauge invariance by varying $VVHH$ coupling independently.
    We see the same behavior in RI due to the same reason when we vary $\kappa_{ZZWW}$ coupling. A better theory is needed to restore the gauge invariance with independently varying $VVHH$ and $VVVV$ couplings.
    
    In the chapter~\ref{chap_h22e2m}, we studied the Higgs boson partial decay widths in the channel $H\rightarrow e^+e^-\mu^+\mu^-$. We have calculated one-loop EW correction to this process and have studied the effect of anomalous $HHH$ and $ZZHH$ couplings.
     We have implemented the dipole subtraction scheme for QED to get the IR-safe amplitudes.
     The NLO corrected decay widths are $241.0$ eV and $237.7$ eV in the $G_F$ and $\alpha(M_Z)$ schemes respectively.
     Our results agree with the {\tt Prophecy4f} package for this process also.
     In this process, similarly, we scaled the $HHH$ and $ZZHH$ couplings to study the anomalous effects on Higgs decay width.    
    We see the same behavior in RI for this process as for the process $H\rightarrow \nu_e\bar{\nu}_e\nu_\mu\bar{\nu}_\mu$ with varying $\kappa_{HHH}$, $\kappa_{ZZHH}$ and $\kappa_{ZZWW}$.
    In this process, again, the gauge invariance is maintained when varying $HHH$ coupling, but the same is not maintained when varying $VVHH$ and $VVVV$ couplings independently. One needs a better theory to vary the $VVHH$ and $VVVV$ couplings independently for this process.
    
    In brief, we have studied the Higgs anomalous couplings for a few processes which can be probed in colliders.
    One can make use of the techniques given in this thesis to compute one-loop amplitudes for any process at colliders.
     Our future goal is to implement SMEFT, HEFT for these processes, especially for the Higgs decay processes, so that we can study the anomalous effect of Higgs couplings without affecting the gauge invariance.
    We are also computing the other Higgs decay modes in four leptonic channel for a complete study.

\newpage

%
%
%
%

\phantomsection
\addcontentsline{toc}{chapter}{Bibliography}

\bibliographystyle{JHEP}
\bibliography{references}

\providecommand{\href}[2]{#2}\begingroup\raggedright\begin{thebibliography}{10}

\bibitem{conference1}
K.~Monig, \emph{{Highlights and perspectives from the ATLAS experiment}},
  {\emph{The Large Hadron Collider Physics Conference, 25-30 May 2020} }.

\bibitem{conference2}
P.~Mcbride, \emph{{Highlights and perspectives from the CMS experiment}},
  {\emph{The Large Hadron Collider Physics Conference, 25-30 May 2020} }.

\bibitem{Agrawal:2019bpm}
P.~Agrawal, D.~Saha, L.-X. Xu, J.-H. Yu and C.~P. Yuan, \emph{{Determining the
  shape of the Higgs potential at future colliders}},
  \href{https://doi.org/10.1103/PhysRevD.101.075023}{\emph{Phys. Rev. D}
  {\bfseries 101} (2020) 075023}
  [\href{https://arxiv.org/abs/1907.02078}{{\ttfamily 1907.02078}}].

\bibitem{Bishara_2017}
F.~Bishara, R.~Contino and J.~Rojo, \emph{Higgs pair production in vector-boson
  fusion at the lhc and beyond},
  \href{https://doi.org/10.1140/epjc/s10052-017-5037-9}{\emph{The European
  Physical Journal C} {\bfseries 77} (2017) }.

\bibitem{Aad_2021}
G.~Aad, B.~Abbott, D.~C. Abbott, A.~Abed~Abud, K.~Abeling, D.~K. Abhayasinghe
  et~al., \emph{Erratum to: Search for the $hh \to b\overline{b}b\overline{b} $
  process via vector-boson fusion production using proton-proton collisions at
  $ \sqrt{s} = 13$ tev with the atlas detector},
  \href{https://doi.org/10.1007/jhep01(2021)145}{\emph{Journal of High Energy
  Physics} {\bfseries 2021} (2021) }.

\bibitem{Nordstrom:2018ceg}
K.~Nordstr\"om and A.~Papaefstathiou, \emph{{$VHH$ production at the
  High-Luminosity LHC}},
  \href{https://doi.org/10.1140/epjp/i2019-12614-2}{\emph{Eur. Phys. J. Plus}
  {\bfseries 134} (2019) 288}
  [\href{https://arxiv.org/abs/1807.01571}{{\ttfamily 1807.01571}}].

\bibitem{Baglio:2015eon}
J.~Baglio, \emph{{Next-To-Leading Order QCD Corrections to Associated
  Production of a SM Higgs Boson with a Pair of Weak Bosons in the
  POWHEG-BOX}}, \href{https://doi.org/10.1103/PhysRevD.93.054010}{\emph{Phys.
  Rev. D} {\bfseries 93} (2016) 054010}
  [\href{https://arxiv.org/abs/1512.05787}{{\ttfamily 1512.05787}}].

\bibitem{Agrawal:2019ffb}
P.~Agrawal, D.~Saha and A.~Shivaji, \emph{{Di-vector boson production in
  association with a Higgs boson at hadron colliders}},
  \href{https://arxiv.org/abs/1907.13168}{{\ttfamily 1907.13168}}.

\bibitem{Chatrchyan:2011ig}
{\scshape CMS} collaboration, S.~Chatrchyan et~al., \emph{{Measurement of the
  Polarization of W Bosons with Large Transverse Momenta in W+Jets Events at
  the LHC}}, \href{https://doi.org/10.1103/PhysRevLett.107.021802}{\emph{Phys.
  Rev. Lett.} {\bfseries 107} (2011) 021802}
  [\href{https://arxiv.org/abs/1104.3829}{{\ttfamily 1104.3829}}].

\bibitem{Aad:2012ky}
{\scshape ATLAS} collaboration, G.~Aad et~al., \emph{{Measurement of the W
  boson polarization in top quark decays with the ATLAS detector}},
  \href{https://doi.org/10.1007/JHEP06(2012)088}{\emph{JHEP} {\bfseries 06}
  (2012) 088} [\href{https://arxiv.org/abs/1205.2484}{{\ttfamily 1205.2484}}].

\bibitem{Aaboud:2019gxl}
{\scshape ATLAS} collaboration, M.~Aaboud et~al., \emph{{Measurement of
  $W^{\pm}Z$ production cross sections and gauge boson polarisation in $pp$
  collisions at $\sqrt{s} = 13$ TeV with the ATLAS detector}},
  \href{https://doi.org/10.1140/epjc/s10052-019-7027-6}{\emph{Eur. Phys. J. C}
  {\bfseries 79} (2019) 535}
  [\href{https://arxiv.org/abs/1902.05759}{{\ttfamily 1902.05759}}].

\bibitem{Baglio:2016ofi}
J.~Baglio, \emph{{Gluon fusion and $b\bar{b}$ corrections to $H W^+ W^- / H Z
  Z$ production in the POWHEG-BOX}},
  \href{https://doi.org/10.1016/j.physletb.2016.10.066}{\emph{Phys. Lett. B}
  {\bfseries 764} (2017) 54}
  [\href{https://arxiv.org/abs/1609.05907}{{\ttfamily 1609.05907}}].

\bibitem{Hahn:2000kx}
T.~Hahn, \emph{{Generating Feynman diagrams and amplitudes with FeynArts 3}},
  \href{https://doi.org/10.1016/S0010-4655(01)00290-9}{\emph{Comput. Phys.
  Commun.} {\bfseries 140} (2001) 418}
  [\href{https://arxiv.org/abs/hep-ph/0012260}{{\ttfamily hep-ph/0012260}}].

\bibitem{Peskin:2011in}
M.~E. Peskin, \emph{{Simplifying Multi-Jet QCD Computation}},  in \emph{{13th
  Mexican School of Particles and Fields}}, 1, 2011,
  \href{https://arxiv.org/abs/1101.2414}{{\ttfamily 1101.2414}}.

\bibitem{Kleiss:1986qc}
R.~Kleiss and W.~Stirling, \emph{{Cross-sections for the Production of an
  Arbitrary Number of Photons in Electron - Positron Annihilation}},
  \href{https://doi.org/10.1016/0370-2693(86)90454-5}{\emph{Phys. Lett. B}
  {\bfseries 179} (1986) 159}.

\bibitem{BERN1992451}
Z.~Bern and D.~A. Kosower, \emph{The computation of loop amplitudes in gauge
  theories},
  \href{https://doi.org/https://doi.org/10.1016/0550-3213(92)90134-W}{\emph{Nuclear
  Physics B} {\bfseries 379} (1992) 451 }.

\bibitem{Gnendiger:2017pys}
C.~Gnendiger et~al., \emph{{To ${d}$, or not to ${d}$: recent developments and
  comparisons of regularization schemes}},
  \href{https://doi.org/10.1140/epjc/s10052-017-5023-2}{\emph{Eur. Phys. J. C}
  {\bfseries 77} (2017) 471}
  [\href{https://arxiv.org/abs/1705.01827}{{\ttfamily 1705.01827}}].

\bibitem{Vermaseren:2000nd}
J.~Vermaseren, \emph{{New features of FORM}},
  \href{https://arxiv.org/abs/math-ph/0010025}{{\ttfamily math-ph/0010025}}.

\bibitem{Shao:2011zza}
H.-S. Shao, Y.-J. Zhang and K.-T. Chao, \emph{{Dijet Invariant Mass
  Distribution in Top Quark Hadronic Decay with QCD Corrections}},
  \href{https://doi.org/10.1103/PhysRevD.84.094021}{\emph{Phys. Rev. D}
  {\bfseries 84} (2011) 094021}
  [\href{https://arxiv.org/abs/1106.5483}{{\ttfamily 1106.5483}}].

\bibitem{Gnendiger:2017rfh}
C.~Gnendiger and A.~Signer, \emph{{$\gamma_{5}$ in the four-dimensional
  helicity scheme}},
  \href{https://doi.org/10.1103/PhysRevD.97.096006}{\emph{Phys. Rev. D}
  {\bfseries 97} (2018) 096006}
  [\href{https://arxiv.org/abs/1710.09231}{{\ttfamily 1710.09231}}].

\bibitem{vanHameren:2010cp}
A.~van Hameren, \emph{{OneLOop: For the evaluation of one-loop scalar
  functions}}, \href{https://doi.org/10.1016/j.cpc.2011.06.011}{\emph{Comput.
  Phys. Commun.} {\bfseries 182} (2011) 2427}
  [\href{https://arxiv.org/abs/1007.4716}{{\ttfamily 1007.4716}}].

\bibitem{Agrawal:2012df}
P.~Agrawal and A.~Shivaji, \emph{{Di-Vector Boson + Jet Production via Gluon
  Fusion at Hadron Colliders}},
  \href{https://doi.org/10.1103/PhysRevD.86.073013}{\emph{Phys. Rev. D}
  {\bfseries 86} (2012) 073013}
  [\href{https://arxiv.org/abs/1207.2927}{{\ttfamily 1207.2927}}].

\bibitem{Agrawal:1998ch}
P.~Agrawal and G.~Ladinsky, \emph{{Production of two photons and a jet through
  gluon fusion}}, \href{https://doi.org/10.1103/PhysRevD.63.117504}{\emph{Phys.
  Rev. D} {\bfseries 63} (2001) 117504}
  [\href{https://arxiv.org/abs/hep-ph/0011346}{{\ttfamily hep-ph/0011346}}].

\bibitem{Veseli:1997hr}
S.~Veseli, \emph{{Multidimensional integration in a heterogeneous network
  environment}},
  \href{https://doi.org/10.1016/S0010-4655(97)00120-3}{\emph{Comput. Phys.
  Commun.} {\bfseries 108} (1998) 9}
  [\href{https://arxiv.org/abs/physics/9710017}{{\ttfamily physics/9710017}}].

\bibitem{Lepage:1977sw}
G.~Lepage, \emph{{A New Algorithm for Adaptive Multidimensional Integration}},
  \href{https://doi.org/10.1016/0021-9991(78)90004-9}{\emph{J. Comput. Phys.}
  {\bfseries 27} (1978) 192}.

\bibitem{10.7551/mitpress/5712.001.0001}
A.~Geist, A.~Beguelin, J.~Dongarra, W.~Jiang, R.~Manchek and V.~S. Sunderam,
  \emph{{PVM: A Users' Guide and Tutorial for Network Parallel Computing}}. The
  MIT Press, 11, 1994,
  \href{https://doi.org/10.7551/mitpress/5712.001.0001}{10.7551/mitpress/5712.001.0001}.

\bibitem{Catani:1996vz}
S.~Catani and M.~Seymour, \emph{{A General algorithm for calculating jet
  cross-sections in NLO QCD}},
  \href{https://doi.org/10.1016/S0550-3213(96)00589-5}{\emph{Nucl. Phys. B}
  {\bfseries 485} (1997) 291}
  [\href{https://arxiv.org/abs/hep-ph/9605323}{{\ttfamily hep-ph/9605323}}].

\bibitem{Catani:1996pk}
S.~Catani, M.~Seymour and Z.~Trocsanyi, \emph{{Regularization scheme
  independence and unitarity in QCD cross-sections}},
  \href{https://doi.org/10.1103/PhysRevD.55.6819}{\emph{Phys. Rev. D}
  {\bfseries 55} (1997) 6819}
  [\href{https://arxiv.org/abs/hep-ph/9610553}{{\ttfamily hep-ph/9610553}}].

\bibitem{Grazzini:2016ctr}
M.~Grazzini, S.~Kallweit, S.~Pozzorini, D.~Rathlev and M.~Wiesemann,
  \emph{{$W^{+}W^{-}$ production at the LHC: fiducial cross sections and
  distributions in NNLO QCD}},
  \href{https://doi.org/10.1007/JHEP08(2016)140}{\emph{JHEP} {\bfseries 08}
  (2016) 140} [\href{https://arxiv.org/abs/1605.02716}{{\ttfamily
  1605.02716}}].

\bibitem{Denner:2012yc}
A.~Denner, S.~Dittmaier, S.~Kallweit and S.~Pozzorini, \emph{{NLO QCD
  corrections to off-shell top-antitop production with leptonic decays at
  hadron colliders}},
  \href{https://doi.org/10.1007/JHEP10(2012)110}{\emph{JHEP} {\bfseries 10}
  (2012) 110} [\href{https://arxiv.org/abs/1207.5018}{{\ttfamily 1207.5018}}].

\bibitem{Cascioli:2013wga}
F.~Cascioli, S.~Kallweit, P.~Maierh\"ofer and S.~Pozzorini, \emph{{A unified
  NLO description of top-pair and associated Wt production}},
  \href{https://doi.org/10.1140/epjc/s10052-014-2783-9}{\emph{Eur. Phys. J. C}
  {\bfseries 74} (2014) 2783}
  [\href{https://arxiv.org/abs/1312.0546}{{\ttfamily 1312.0546}}].

\bibitem{Gehrmann:2014fva}
T.~Gehrmann, M.~Grazzini, S.~Kallweit, P.~Maierh\"ofer, A.~von Manteuffel,
  S.~Pozzorini et~al., \emph{{$W^+W^-$ Production at Hadron Colliders in Next
  to Next to Leading Order QCD}},
  \href{https://doi.org/10.1103/PhysRevLett.113.212001}{\emph{Phys. Rev. Lett.}
  {\bfseries 113} (2014) 212001}
  [\href{https://arxiv.org/abs/1408.5243}{{\ttfamily 1408.5243}}].

\bibitem{Bierweiler:2012kw}
A.~Bierweiler, T.~Kasprzik, J.~H. K\"uhn and S.~Uccirati, \emph{{Electroweak
  corrections to W-boson pair production at the LHC}},
  \href{https://doi.org/10.1007/JHEP11(2012)093}{\emph{JHEP} {\bfseries 11}
  (2012) 093} [\href{https://arxiv.org/abs/1208.3147}{{\ttfamily 1208.3147}}].

\bibitem{DENNER200622}
A.~Denner and S.~Dittmaier, \emph{The complex-mass scheme for perturbative
  calculations with unstable particles},
  \href{https://doi.org/https://doi.org/10.1016/j.nuclphysbps.2006.09.025}{\emph{Nuclear
  Physics B - Proceedings Supplements} {\bfseries 160} (2006) 22 }.

\bibitem{GRAZZINI2020135399}
M.~Grazzini, S.~Kallweit, M.~Wiesemann and J.~Y. Yook, \emph{{$W^{+}W^{-}$
  production at the LHC: NLO QCD corrections to the loop-induced gluon fusion
  channel}},
  \href{https://doi.org/https://doi.org/10.1016/j.physletb.2020.135399}{\emph{Physics
  Letters B} {\bfseries 804} (2020) 135399}.

\bibitem{Harlander:2003ai}
R.~V. Harlander and W.~B. Kilgore, \emph{{Higgs boson production in bottom
  quark fusion at next-to-next-to leading order}},
  \href{https://doi.org/10.1103/PhysRevD.68.013001}{\emph{Phys. Rev. D}
  {\bfseries 68} (2003) 013001}
  [\href{https://arxiv.org/abs/hep-ph/0304035}{{\ttfamily hep-ph/0304035}}].

\bibitem{Dulat:2015mca}
S.~Dulat, T.-J. Hou, J.~Gao, M.~Guzzi, J.~Huston, P.~Nadolsky et~al.,
  \emph{{New parton distribution functions from a global analysis of quantum
  chromodynamics}},
  \href{https://doi.org/10.1103/PhysRevD.93.033006}{\emph{Phys. Rev. D}
  {\bfseries 93} (2016) 033006}
  [\href{https://arxiv.org/abs/1506.07443}{{\ttfamily 1506.07443}}].

\bibitem{Whalley:2005nh}
M.~Whalley, D.~Bourilkov and R.~Group, \emph{{The Les Houches accord PDFs
  (LHAPDF) and LHAGLUE}},  in \emph{{HERA and the LHC: A Workshop on the
  Implications of HERA and LHC Physics (Startup Meeting, CERN, 26-27 March
  2004; Midterm Meeting, CERN, 11-13 October 2004)}}, pp.~575--581, 8, 2005,
  \href{https://arxiv.org/abs/hep-ph/0508110}{{\ttfamily hep-ph/0508110}}.

\bibitem{Alwall:2014hca}
J.~Alwall, R.~Frederix, S.~Frixione, V.~Hirschi, F.~Maltoni, O.~Mattelaer
  et~al., \emph{{The automated computation of tree-level and next-to-leading
  order differential cross sections, and their matching to parton shower
  simulations}}, \href{https://doi.org/10.1007/JHEP07(2014)079}{\emph{JHEP}
  {\bfseries 07} (2014) 079} [\href{https://arxiv.org/abs/1405.0301}{{\ttfamily
  1405.0301}}].

\bibitem{LHCHiggsCrossSectionWorkingGroup:2012nn}
{\scshape LHC Higgs Cross Section Working Group} collaboration, A.~David,
  A.~Denner, M.~Duehrssen, M.~Grazzini, C.~Grojean, G.~Passarino et~al.,
  \emph{{LHC HXSWG interim recommendations to explore the coupling structure of
  a Higgs-like particle}},  \href{https://arxiv.org/abs/1209.0040}{{\ttfamily
  1209.0040}}.

\bibitem{Ghezzi:2015vva}
M.~Ghezzi, R.~Gomez-Ambrosio, G.~Passarino and S.~Uccirati, \emph{{NLO Higgs
  effective field theory and \ensuremath{\kappa}-framework}},
  \href{https://doi.org/10.1007/JHEP07(2015)175}{\emph{JHEP} {\bfseries 07}
  (2015) 175} [\href{https://arxiv.org/abs/1505.03706}{{\ttfamily
  1505.03706}}].

\end{thebibliography}\endgroup



\providecommand{\href}[2]{#2}\begingroup\raggedright\begin{thebibliography}{10}

\bibitem{Agrawal:2019bpm}
P.~Agrawal, D.~Saha, L.-X. Xu, J.-H. Yu and C.~P. Yuan, \emph{{Determining the
  shape of the Higgs potential at future colliders}},
  \href{https://doi.org/10.1103/PhysRevD.101.075023}{\emph{Phys. Rev. D}
  {\bfseries 101} (2020) 075023}
  [\href{https://arxiv.org/abs/1907.02078}{{\ttfamily 1907.02078}}].

\bibitem{Bishara_2017}
F.~Bishara, R.~Contino and J.~Rojo, \emph{Higgs pair production in vector-boson
  fusion at the lhc and beyond},
  \href{https://doi.org/10.1140/epjc/s10052-017-5037-9}{\emph{The European
  Physical Journal C} {\bfseries 77} (2017) }.

\bibitem{Aad_2021}
G.~Aad, B.~Abbott, D.~C. Abbott, A.~Abed~Abud, K.~Abeling, D.~K. Abhayasinghe
  et~al., \emph{Erratum to: Search for the $hh \to b\overline{b}b\overline{b} $
  process via vector-boson fusion production using proton-proton collisions at
  $ \sqrt{s} = 13$ tev with the atlas detector},
  \href{https://doi.org/10.1007/jhep01(2021)145}{\emph{Journal of High Energy
  Physics} {\bfseries 2021} (2021) }.

\bibitem{Nordstrom:2018ceg}
K.~Nordstr\"om and A.~Papaefstathiou, \emph{{$VHH$ production at the
  High-Luminosity LHC}},
  \href{https://doi.org/10.1140/epjp/i2019-12614-2}{\emph{Eur. Phys. J. Plus}
  {\bfseries 134} (2019) 288}
  [\href{https://arxiv.org/abs/1807.01571}{{\ttfamily 1807.01571}}].

\bibitem{LHCHiggsCrossSectionWorkingGroup:2012nn}
{\scshape LHC Higgs Cross Section Working Group} collaboration, A.~David,
  A.~Denner, M.~Duehrssen, M.~Grazzini, C.~Grojean, G.~Passarino et~al.,
  \emph{{LHC HXSWG interim recommendations to explore the coupling structure of
  a Higgs-like particle}},  \href{https://arxiv.org/abs/1209.0040}{{\ttfamily
  1209.0040}}.

\bibitem{Ghezzi:2015vva}
M.~Ghezzi, R.~Gomez-Ambrosio, G.~Passarino and S.~Uccirati, \emph{{NLO Higgs
  effective field theory and \ensuremath{\kappa}-framework}},
  \href{https://doi.org/10.1007/JHEP07(2015)175}{\emph{JHEP} {\bfseries 07}
  (2015) 175} [\href{https://arxiv.org/abs/1505.03706}{{\ttfamily
  1505.03706}}].

\bibitem{Hahn:2000kx}
T.~Hahn, \emph{{Generating Feynman diagrams and amplitudes with FeynArts 3}},
  \href{https://doi.org/10.1016/S0010-4655(01)00290-9}{\emph{Comput. Phys.
  Commun.} {\bfseries 140} (2001) 418}
  [\href{https://arxiv.org/abs/hep-ph/0012260}{{\ttfamily hep-ph/0012260}}].

\bibitem{Peskin:2011in}
M.~E. Peskin, \emph{{Simplifying Multi-Jet QCD Computation}},  in \emph{{13th
  Mexican School of Particles and Fields}}, 1, 2011,
  \href{https://arxiv.org/abs/1101.2414}{{\ttfamily 1101.2414}}.

\bibitem{Vermaseren:2000nd}
J.~Vermaseren, \emph{{New features of FORM}},
  \href{https://arxiv.org/abs/math-ph/0010025}{{\ttfamily math-ph/0010025}}.

\bibitem{vanHameren:2010cp}
A.~van Hameren, \emph{{OneLOop: For the evaluation of one-loop scalar
  functions}}, \href{https://doi.org/10.1016/j.cpc.2011.06.011}{\emph{Comput.
  Phys. Commun.} {\bfseries 182} (2011) 2427}
  [\href{https://arxiv.org/abs/1007.4716}{{\ttfamily 1007.4716}}].

\bibitem{vanNeerven:1983vr}
W.~L. van Neerven and J.~A.~M. Vermaseren, \emph{{LARGE LOOP INTEGRALS}},
  \href{https://doi.org/10.1016/0370-2693(84)90237-5}{\emph{Phys. Lett. B}
  {\bfseries 137} (1984) 241}.

\bibitem{Agrawal:2012df}
P.~Agrawal and A.~Shivaji, \emph{{Di-Vector Boson + Jet Production via Gluon
  Fusion at Hadron Colliders}},
  \href{https://doi.org/10.1103/PhysRevD.86.073013}{\emph{Phys. Rev. D}
  {\bfseries 86} (2012) 073013}
  [\href{https://arxiv.org/abs/1207.2927}{{\ttfamily 1207.2927}}].

\bibitem{Agrawal:1998ch}
P.~Agrawal and G.~Ladinsky, \emph{{Production of two photons and a jet through
  gluon fusion}}, \href{https://doi.org/10.1103/PhysRevD.63.117504}{\emph{Phys.
  Rev. D} {\bfseries 63} (2001) 117504}
  [\href{https://arxiv.org/abs/hep-ph/0011346}{{\ttfamily hep-ph/0011346}}].

\bibitem{BERN1992451}
Z.~Bern and D.~A. Kosower, \emph{The computation of loop amplitudes in gauge
  theories},
  \href{https://doi.org/https://doi.org/10.1016/0550-3213(92)90134-W}{\emph{Nuclear
  Physics B} {\bfseries 379} (1992) 451 }.

\bibitem{Gnendiger:2017pys}
C.~Gnendiger et~al., \emph{{To ${d}$, or not to ${d}$: recent developments and
  comparisons of regularization schemes}},
  \href{https://doi.org/10.1140/epjc/s10052-017-5023-2}{\emph{Eur. Phys. J. C}
  {\bfseries 77} (2017) 471}
  [\href{https://arxiv.org/abs/1705.01827}{{\ttfamily 1705.01827}}].

\bibitem{THOOFT1972189}
G.~{'t Hooft} and M.~Veltman, \emph{Regularization and renormalization of gauge
  fields},
  \href{https://doi.org/https://doi.org/10.1016/0550-3213(72)90279-9}{\emph{Nuclear
  Physics B} {\bfseries 44} (1972) 189}.

\bibitem{Veseli:1997hr}
S.~Veseli, \emph{{Multidimensional integration in a heterogeneous network
  environment}},
  \href{https://doi.org/10.1016/S0010-4655(97)00120-3}{\emph{Comput. Phys.
  Commun.} {\bfseries 108} (1998) 9}
  [\href{https://arxiv.org/abs/physics/9710017}{{\ttfamily physics/9710017}}].

\bibitem{Lepage:1977sw}
G.~Lepage, \emph{{A New Algorithm for Adaptive Multidimensional Integration}},
  \href{https://doi.org/10.1016/0021-9991(78)90004-9}{\emph{J. Comput. Phys.}
  {\bfseries 27} (1978) 192}.

\bibitem{10.7551/mitpress/5712.001.0001}
A.~Geist, A.~Beguelin, J.~Dongarra, W.~Jiang, R.~Manchek and V.~S. Sunderam,
  \emph{{PVM: A Users' Guide and Tutorial for Network Parallel Computing}}. The
  MIT Press, 11, 1994,
  \href{https://doi.org/10.7551/mitpress/5712.001.0001}{10.7551/mitpress/5712.001.0001}.

\bibitem{Berger:2009zb}
C.~F. Berger and D.~Forde, \emph{{Multi-Parton Scattering Amplitudes via
  On-Shell Methods}},
  \href{https://doi.org/10.1146/annurev.nucl.012809.104538}{\emph{Ann. Rev.
  Nucl. Part. Sci.} {\bfseries 60} (2010) 181}
  [\href{https://arxiv.org/abs/0912.3534}{{\ttfamily 0912.3534}}].

\bibitem{ELLIS2012141}
R.~K. Ellis, Z.~Kunszt, K.~Melnikov and G.~Zanderighi, \emph{One-loop
  calculations in quantum field theory: From feynman diagrams to unitarity
  cuts},
  \href{https://doi.org/https://doi.org/10.1016/j.physrep.2012.01.008}{\emph{Physics
  Reports} {\bfseries 518} (2012) 141}.

\bibitem{Kleiss:1985yh}
R.~Kleiss and W.~J. Stirling, \emph{{Spinor Techniques for Calculating p anti-p
  ---\ensuremath{>} W+- / Z0 + Jets}},
  \href{https://doi.org/10.1016/0550-3213(85)90285-8}{\emph{Nucl. Phys. B}
  {\bfseries 262} (1985) 235}.

\bibitem{Kleiss:1986qc}
R.~Kleiss and W.~Stirling, \emph{{Cross-sections for the Production of an
  Arbitrary Number of Photons in Electron - Positron Annihilation}},
  \href{https://doi.org/10.1016/0370-2693(86)90454-5}{\emph{Phys. Lett. B}
  {\bfseries 179} (1986) 159}.

\bibitem{Passarino:1978jh}
G.~Passarino and M.~J.~G. Veltman, \emph{{One Loop Corrections for e+ e-
  Annihilation Into mu+ mu- in the Weinberg Model}},
  \href{https://doi.org/10.1016/0550-3213(79)90234-7}{\emph{Nucl. Phys. B}
  {\bfseries 160} (1979) 151}.

\bibitem{Denner:1991kt}
A.~Denner, \emph{{Techniques for calculation of electroweak radiative
  corrections at the one loop level and results for W physics at LEP-200}},
  \href{https://doi.org/10.1002/prop.2190410402}{\emph{Fortsch. Phys.}
  {\bfseries 41} (1993) 307} [\href{https://arxiv.org/abs/0709.1075}{{\ttfamily
  0709.1075}}].

\bibitem{Kinoshita:1962ur}
T.~Kinoshita, \emph{{Mass singularities of Feynman amplitudes}},
  \href{https://doi.org/10.1063/1.1724268}{\emph{J. Math. Phys.} {\bfseries 3}
  (1962) 650}.

\bibitem{Lee:1964is}
T.~D. Lee and M.~Nauenberg, \emph{{Degenerate Systems and Mass Singularities}},
  \href{https://doi.org/10.1103/PhysRev.133.B1549}{\emph{Phys. Rev.} {\bfseries
  133} (1964) B1549}.

\bibitem{Harris:2001sx}
B.~W. Harris and J.~F. Owens, \emph{{The Two cutoff phase space slicing
  method}}, \href{https://doi.org/10.1103/PhysRevD.65.094032}{\emph{Phys. Rev.
  D} {\bfseries 65} (2002) 094032}
  [\href{https://arxiv.org/abs/hep-ph/0102128}{{\ttfamily hep-ph/0102128}}].

\bibitem{Frederix:2009yq}
R.~Frederix, S.~Frixione, F.~Maltoni and T.~Stelzer, \emph{{Automation of
  next-to-leading order computations in QCD: The FKS subtraction}},
  \href{https://doi.org/10.1088/1126-6708/2009/10/003}{\emph{JHEP} {\bfseries
  10} (2009) 003} [\href{https://arxiv.org/abs/0908.4272}{{\ttfamily
  0908.4272}}].

\bibitem{Catani:1996vz}
S.~Catani and M.~Seymour, \emph{{A General algorithm for calculating jet
  cross-sections in NLO QCD}},
  \href{https://doi.org/10.1016/S0550-3213(96)00589-5}{\emph{Nucl. Phys. B}
  {\bfseries 485} (1997) 291}
  [\href{https://arxiv.org/abs/hep-ph/9605323}{{\ttfamily hep-ph/9605323}}].

\bibitem{Schonherr:2017qcj}
M.~Sch\"onherr, \emph{{An automated subtraction of NLO EW infrared
  divergences}},
  \href{https://doi.org/10.1140/epjc/s10052-018-5600-z}{\emph{Eur. Phys. J. C}
  {\bfseries 78} (2018) 119}
  [\href{https://arxiv.org/abs/1712.07975}{{\ttfamily 1712.07975}}].

\bibitem{Baglio:2015eon}
J.~Baglio, \emph{{Next-To-Leading Order QCD Corrections to Associated
  Production of a SM Higgs Boson with a Pair of Weak Bosons in the
  POWHEG-BOX}}, \href{https://doi.org/10.1103/PhysRevD.93.054010}{\emph{Phys.
  Rev. D} {\bfseries 93} (2016) 054010}
  [\href{https://arxiv.org/abs/1512.05787}{{\ttfamily 1512.05787}}].

\bibitem{Agrawal:2019ffb}
P.~Agrawal, D.~Saha and A.~Shivaji, \emph{{Di-vector boson production in
  association with a Higgs boson at hadron colliders}},
  \href{https://arxiv.org/abs/1907.13168}{{\ttfamily 1907.13168}}.

\bibitem{Chatrchyan:2011ig}
{\scshape CMS} collaboration, S.~Chatrchyan et~al., \emph{{Measurement of the
  Polarization of W Bosons with Large Transverse Momenta in W+Jets Events at
  the LHC}}, \href{https://doi.org/10.1103/PhysRevLett.107.021802}{\emph{Phys.
  Rev. Lett.} {\bfseries 107} (2011) 021802}
  [\href{https://arxiv.org/abs/1104.3829}{{\ttfamily 1104.3829}}].

\bibitem{Aad:2012ky}
{\scshape ATLAS} collaboration, G.~Aad et~al., \emph{{Measurement of the W
  boson polarization in top quark decays with the ATLAS detector}},
  \href{https://doi.org/10.1007/JHEP06(2012)088}{\emph{JHEP} {\bfseries 06}
  (2012) 088} [\href{https://arxiv.org/abs/1205.2484}{{\ttfamily 1205.2484}}].

\bibitem{Aaboud:2019gxl}
{\scshape ATLAS} collaboration, M.~Aaboud et~al., \emph{{Measurement of
  $W^{\pm}Z$ production cross sections and gauge boson polarisation in $pp$
  collisions at $\sqrt{s} = 13$ TeV with the ATLAS detector}},
  \href{https://doi.org/10.1140/epjc/s10052-019-7027-6}{\emph{Eur. Phys. J. C}
  {\bfseries 79} (2019) 535}
  [\href{https://arxiv.org/abs/1902.05759}{{\ttfamily 1902.05759}}].

\bibitem{Baglio:2016ofi}
J.~Baglio, \emph{{Gluon fusion and $b\bar{b}$ corrections to $H W^+ W^- / H Z
  Z$ production in the POWHEG-BOX}},
  \href{https://doi.org/10.1016/j.physletb.2016.10.066}{\emph{Phys. Lett. B}
  {\bfseries 764} (2017) 54}
  [\href{https://arxiv.org/abs/1609.05907}{{\ttfamily 1609.05907}}].

\bibitem{Shao:2011zza}
H.-S. Shao, Y.-J. Zhang and K.-T. Chao, \emph{{Dijet Invariant Mass
  Distribution in Top Quark Hadronic Decay with QCD Corrections}},
  \href{https://doi.org/10.1103/PhysRevD.84.094021}{\emph{Phys. Rev. D}
  {\bfseries 84} (2011) 094021}
  [\href{https://arxiv.org/abs/1106.5483}{{\ttfamily 1106.5483}}].

\bibitem{Gnendiger:2017rfh}
C.~Gnendiger and A.~Signer, \emph{{$\gamma_{5}$ in the four-dimensional
  helicity scheme}},
  \href{https://doi.org/10.1103/PhysRevD.97.096006}{\emph{Phys. Rev. D}
  {\bfseries 97} (2018) 096006}
  [\href{https://arxiv.org/abs/1710.09231}{{\ttfamily 1710.09231}}].

\bibitem{Catani:1996pk}
S.~Catani, M.~Seymour and Z.~Trocsanyi, \emph{{Regularization scheme
  independence and unitarity in QCD cross-sections}},
  \href{https://doi.org/10.1103/PhysRevD.55.6819}{\emph{Phys. Rev. D}
  {\bfseries 55} (1997) 6819}
  [\href{https://arxiv.org/abs/hep-ph/9610553}{{\ttfamily hep-ph/9610553}}].

\bibitem{Grazzini:2016ctr}
M.~Grazzini, S.~Kallweit, S.~Pozzorini, D.~Rathlev and M.~Wiesemann,
  \emph{{$W^{+}W^{-}$ production at the LHC: fiducial cross sections and
  distributions in NNLO QCD}},
  \href{https://doi.org/10.1007/JHEP08(2016)140}{\emph{JHEP} {\bfseries 08}
  (2016) 140} [\href{https://arxiv.org/abs/1605.02716}{{\ttfamily
  1605.02716}}].

\bibitem{Denner:2012yc}
A.~Denner, S.~Dittmaier, S.~Kallweit and S.~Pozzorini, \emph{{NLO QCD
  corrections to off-shell top-antitop production with leptonic decays at
  hadron colliders}},
  \href{https://doi.org/10.1007/JHEP10(2012)110}{\emph{JHEP} {\bfseries 10}
  (2012) 110} [\href{https://arxiv.org/abs/1207.5018}{{\ttfamily 1207.5018}}].

\bibitem{Cascioli:2013wga}
F.~Cascioli, S.~Kallweit, P.~Maierh\"ofer and S.~Pozzorini, \emph{{A unified
  NLO description of top-pair and associated Wt production}},
  \href{https://doi.org/10.1140/epjc/s10052-014-2783-9}{\emph{Eur. Phys. J. C}
  {\bfseries 74} (2014) 2783}
  [\href{https://arxiv.org/abs/1312.0546}{{\ttfamily 1312.0546}}].

\bibitem{Gehrmann:2014fva}
T.~Gehrmann, M.~Grazzini, S.~Kallweit, P.~Maierh\"ofer, A.~von Manteuffel,
  S.~Pozzorini et~al., \emph{{$W^+W^-$ Production at Hadron Colliders in Next
  to Next to Leading Order QCD}},
  \href{https://doi.org/10.1103/PhysRevLett.113.212001}{\emph{Phys. Rev. Lett.}
  {\bfseries 113} (2014) 212001}
  [\href{https://arxiv.org/abs/1408.5243}{{\ttfamily 1408.5243}}].

\bibitem{Bierweiler:2012kw}
A.~Bierweiler, T.~Kasprzik, J.~H. K\"uhn and S.~Uccirati, \emph{{Electroweak
  corrections to W-boson pair production at the LHC}},
  \href{https://doi.org/10.1007/JHEP11(2012)093}{\emph{JHEP} {\bfseries 11}
  (2012) 093} [\href{https://arxiv.org/abs/1208.3147}{{\ttfamily 1208.3147}}].

\bibitem{DENNER200622}
A.~Denner and S.~Dittmaier, \emph{The complex-mass scheme for perturbative
  calculations with unstable particles},
  \href{https://doi.org/https://doi.org/10.1016/j.nuclphysbps.2006.09.025}{\emph{Nuclear
  Physics B - Proceedings Supplements} {\bfseries 160} (2006) 22 }.

\bibitem{GRAZZINI2020135399}
M.~Grazzini, S.~Kallweit, M.~Wiesemann and J.~Y. Yook, \emph{{$W^{+}W^{-}$
  production at the LHC: NLO QCD corrections to the loop-induced gluon fusion
  channel}},
  \href{https://doi.org/https://doi.org/10.1016/j.physletb.2020.135399}{\emph{Physics
  Letters B} {\bfseries 804} (2020) 135399}.

\bibitem{Harlander:2003ai}
R.~V. Harlander and W.~B. Kilgore, \emph{{Higgs boson production in bottom
  quark fusion at next-to-next-to leading order}},
  \href{https://doi.org/10.1103/PhysRevD.68.013001}{\emph{Phys. Rev. D}
  {\bfseries 68} (2003) 013001}
  [\href{https://arxiv.org/abs/hep-ph/0304035}{{\ttfamily hep-ph/0304035}}].

\bibitem{Dulat:2015mca}
S.~Dulat, T.-J. Hou, J.~Gao, M.~Guzzi, J.~Huston, P.~Nadolsky et~al.,
  \emph{{New parton distribution functions from a global analysis of quantum
  chromodynamics}},
  \href{https://doi.org/10.1103/PhysRevD.93.033006}{\emph{Phys. Rev. D}
  {\bfseries 93} (2016) 033006}
  [\href{https://arxiv.org/abs/1506.07443}{{\ttfamily 1506.07443}}].

\bibitem{Whalley:2005nh}
M.~Whalley, D.~Bourilkov and R.~Group, \emph{{The Les Houches accord PDFs
  (LHAPDF) and LHAGLUE}},  in \emph{{HERA and the LHC: A Workshop on the
  Implications of HERA and LHC Physics (Startup Meeting, CERN, 26-27 March
  2004; Midterm Meeting, CERN, 11-13 October 2004)}}, pp.~575--581, 8, 2005,
  \href{https://arxiv.org/abs/hep-ph/0508110}{{\ttfamily hep-ph/0508110}}.

\bibitem{Alwall:2014hca}
J.~Alwall, R.~Frederix, S.~Frixione, V.~Hirschi, F.~Maltoni, O.~Mattelaer
  et~al., \emph{{The automated computation of tree-level and next-to-leading
  order differential cross sections, and their matching to parton shower
  simulations}}, \href{https://doi.org/10.1007/JHEP07(2014)079}{\emph{JHEP}
  {\bfseries 07} (2014) 079} [\href{https://arxiv.org/abs/1405.0301}{{\ttfamily
  1405.0301}}].

\bibitem{BREDENSTEIN2006131}
A.~Bredenstein, A.~Denner, S.~Dittmaier and M.~Weber, \emph{Precision
  calculations for the higgs decays h → zz/ww → 4 leptons},
  \href{https://doi.org/https://doi.org/10.1016/j.nuclphysbps.2006.09.104}{\emph{Nuclear
  Physics B - Proceedings Supplements} {\bfseries 160} (2006) 131}.

\bibitem{Boselli:2015aha}
S.~Boselli, C.~M. Carloni~Calame, G.~Montagna, O.~Nicrosini and F.~Piccinini,
  \emph{{Higgs boson decay into four leptons at NLOPS electroweak accuracy}},
  \href{https://doi.org/10.1007/JHEP06(2015)023}{\emph{JHEP} {\bfseries 06}
  (2015) 023} [\href{https://arxiv.org/abs/1503.07394}{{\ttfamily
  1503.07394}}].

\bibitem{TheATLAScollaboration:2014scd}
\emph{{Prospects for measuring Higgs pair production in the channel
  $H(\rightarrow\gamma\gamma)H(\rightarrow b\overline{b}) $ using the ATLAS
  detector at the HL-LHC, ATL-PHYS-PUB-2014-019}}, .

\bibitem{PhysRevD.84.094021}
H.-S. Shao, Y.-J. Zhang and K.-T. Chao, \emph{Dijet invariant mass distribution
  in top quark hadronic decay with qcd corrections},
  \href{https://doi.org/10.1103/PhysRevD.84.094021}{\emph{Phys. Rev. D}
  {\bfseries 84} (2011) 094021}.

\bibitem{Korner:1991sx}
J.~G. Korner, D.~Kreimer and K.~Schilcher, \emph{{A Practicable gamma(5) scheme
  in dimensional regularization}},
  \href{https://doi.org/10.1007/BF01559471}{\emph{Z. Phys. C} {\bfseries 54}
  (1992) 503}.

\bibitem{Garzelli:2009is}
M.~V. Garzelli, I.~Malamos and R.~Pittau, \emph{{Feynman rules for the rational
  part of the Electroweak 1-loop amplitudes}},
  \href{https://doi.org/10.1007/JHEP10(2010)097}{\emph{JHEP} {\bfseries 01}
  (2010) 040} [\href{https://arxiv.org/abs/0910.3130}{{\ttfamily 0910.3130}}].

\bibitem{Denner:2006ic}
A.~Denner and S.~Dittmaier, \emph{{The Complex-mass scheme for perturbative
  calculations with unstable particles}},
  \href{https://doi.org/10.1016/j.nuclphysbps.2006.09.025}{\emph{Nucl. Phys. B
  Proc. Suppl.} {\bfseries 160} (2006) 22}
  [\href{https://arxiv.org/abs/hep-ph/0605312}{{\ttfamily hep-ph/0605312}}].

\bibitem{Denner:2005fg}
A.~Denner, S.~Dittmaier, M.~Roth and L.~H. Wieders, \emph{{Electroweak
  corrections to charged-current e+ e- ---\ensuremath{>} 4 fermion processes:
  Technical details and further results}},
  \href{https://doi.org/10.1016/j.nuclphysb.2011.09.001}{\emph{Nucl. Phys. B}
  {\bfseries 724} (2005) 247}
  [\href{https://arxiv.org/abs/hep-ph/0505042}{{\ttfamily hep-ph/0505042}}].

\bibitem{Andersen:2014efa}
J.~R. Andersen et~al., \emph{{Les Houches 2013: Physics at TeV Colliders:
  Standard Model Working Group Report}},
  \href{https://arxiv.org/abs/1405.1067}{{\ttfamily 1405.1067}}.

\bibitem{Eidelman:1995ny}
S.~Eidelman and F.~Jegerlehner, \emph{{Hadronic contributions to g-2 of the
  leptons and to the effective fine structure constant alpha (M(z)**2)}},
  \href{https://doi.org/10.1007/BF01553984}{\emph{Z. Phys. C} {\bfseries 67}
  (1995) 585} [\href{https://arxiv.org/abs/hep-ph/9502298}{{\ttfamily
  hep-ph/9502298}}].

\bibitem{PhysRevD.22.971}
A.~Sirlin, \emph{Radiative corrections in the
  $\mathrm{SU}{(2)}_{L}\ifmmode\times\else\texttimes\fi{}\mathrm{U}(1)$ theory:
  A simple renormalization framework},
  \href{https://doi.org/10.1103/PhysRevD.22.971}{\emph{Phys. Rev. D} {\bfseries
  22} (1980) 971}.

\bibitem{Bredenstein:2006nk}
A.~Bredenstein, A.~Denner, S.~Dittmaier and M.~M. Weber, \emph{{Precision
  calculations for the Higgs decays H ---\ensuremath{>} ZZ/WW ---\ensuremath{>}
  4 leptons}},
  \href{https://doi.org/10.1016/j.nuclphysbps.2006.09.104}{\emph{Nucl. Phys. B
  Proc. Suppl.} {\bfseries 160} (2006) 131}
  [\href{https://arxiv.org/abs/hep-ph/0607060}{{\ttfamily hep-ph/0607060}}].

\end{thebibliography}\endgroup

\end{document}